\documentclass[twocolumn,showpacs,preprintnumbers,eqsecnum,amsmath,amssymb]{revtex4}

\newcommand{\sgn}{{\mathrm{sgn}}}

\newcommand{\be}{\begin{equation}}
\newcommand{\ee}{\end{equation}}
\newcommand{\bea}{\begin{eqnarray}}
\newcommand{\eea}{\end{eqnarray}}

\newcommand{\cl}{{\hbox{\scriptsize cl}}}
\newcommand{\DN}{{D$_{\hbox{\scriptsize N}}$} {}}
\newcommand{\DP}{{D$_{\hbox{\scriptsize P}}$} {}}

\newcommand{\shbox}[1]{{\hbox{\scriptsize #1}}}

\usepackage{graphicx}
\usepackage{dcolumn}
\usepackage{bm}

\begin{document}

\preprint{}

\title{Junctions of three quantum wires}

\author{Masaki Oshikawa}
\affiliation {Department of Physics, Tokyo Institute of Technology,
Oh-okayama, Meguro-ku, Tokyo 152-8551 JAPAN}

\author{Claudio Chamon}%
\affiliation{Department of Physics, Boston
University, Boston, MA 02215, U.S.A.}

\author{Ian Affleck}
\affiliation{Dept. of Physics and Astronomy, University of British
Columbia, Vancouver, BC, Canada V6T 1Z1}

\date{\today}

\begin{abstract}
We study a junction of three quantum wires enclosing a magnetic flux.
The wires are modeled as single-channel spinless Tomonaga-Luttinger
liquids. This is the simplest problem of a quantum junction between
Tomonaga-Luttinger liquids in which Fermi statistics enter in a
non-trivial way. We study the problem using a mapping onto the
dissipative Hofstadter model, describing a single particle moving on a
plane in a magnetic field and a periodic potential coupled to a
harmonic oscillator bath. Alternatively we study the problem by
identifying boundary conditions corresponding to the low energy fixed
points.  We obtain a rich phase diagram including a chiral fixed point
in which the asymmetric current flow is highly sensitive to the sign of
the flux and a phase in which electron pair tunneling dominates. We
also study the effects 
on the conductance tensor of the junction
of contacting the three quantum wires to Fermi
liquid reservoirs.
\end{abstract}

\maketitle

\section{Introduction}
\label{sec:intro}
While electron-electron interactions can, in many cases, be adequately
described by Fermi liquid theory in two- or three-dimensional systems, they
lead to more exotic effects in one-dimensional Tomonaga-Luttinger liquids
(TLL). One context in which striking TLL behavior occurs is in the response
of a quantum wire~\cite{Kane} or a spin chain~\cite{Eggert} to a single
constriction or weak link. The conductance through the constriction scales to
zero at low temperatures in the case of repulsive interactions, corresponding
to a TLL parameter $g<1$ or to the ideal value of $ge^2/h$ for attractive
interactions, $g>1$.  TLL behavior has recently been studied experimentally
in carbon nano-tubes.~\cite{Bockrath,Yao}

In order to construct a useful circuit out of quantum wires it will be
necessary to incorporate junctions of three or more wires. These and other
closely related problems exhibit rather rich TLL effects which have been the
subject of a number of recent
papers.~\cite{Nayak,Rao,Rao-SenPRB2004,Egger,Egger2,Lederer,Safi,Moore,Yi,KVF2004,Akira2004,Sodano2005,Meden1,Meden2,Meden3,Benoit2005,short-paper1}
Much of the work in this field has used bosonization. As emphasized by Nayak
et al.~\cite{Nayak}, when the number of wires meeting at a junction exceeds
two, the Klein factors which give bosonized operators Fermi statistics play a
crucial role. We recently introduced~\cite{short-paper1} a new method to study
this problem, mapping it into the dissipative Hofstadter model (DHM), which
describes a single particle moving in a uniform magnetic field and a periodic
potential in two dimensions and coupled to a bath of harmonic
oscillators. This mapping is useful because it allows us to take advantage of
earlier results of Callan and Freed on the DHM~\cite{CF}. When the three
quantum wires enclose a magnetic flux, the mapping to the DHM also allows us
to identify a new low energy chiral fixed point with an asymmetric flow of
current that is highly sensitive to the sign of the flux.

The purpose of this paper is to present a comprehensive study of the
three-wire junction (or Y-junction) problem. We describe the mapping to the DHM
and its implications in more detail, and we further study the problem using a
complimentary method which involves identifying boundary conditions on the
boson fields which correspond to low energy fixed points. In combination,
these methods provide a coherent picture of the possible stable fixed point
conductances of the three-wire junction, and this picture shows a regime with
chiral current flow and a regime in which electron pair tunneling
dominates. This paper also addresses the problem of how the fixed point
conductance tensor of the three-wire junction is modified if the
TLL wires are connected to Fermi liquid leads far from the junction.

The paper is organized as follows.
In Sec.~\ref{sec:results} we provide a brief summary of the results for the
conductance tensor of the Y-junction and for the Renormalization Group (RG)
flow diagrams for different ranges of the Luttinger parameter $g$.
In Sec.~\ref{sec:model} we present our effective model for the three-wire
junction. In particular, we show explicitly how the magnetic flux
dependence comes into the problem in which the three wires are connected to a
circular ring by taking into account the discrete energy levels on the ring.
In Sec.~\ref{sec:set-up} we review the bosonization of the Y-junction model.
In Sec.~\ref{sec:potential} we discuss the strong coupling limit of the
junction that is obtained by simply taking the hopping between the three wires
to infinity. We show that such a simple argument does not provide information
on the nature of the fixed points for a range of the parameter $g$, and that
different methods that we describe in the paper are needed to shed light onto
the nature of the strong coupling fixed points.
In Sec.~\ref{sec:DHM} we establish the mapping of the Y-junction model onto
the DHM and deduce its consequences for the phase diagram of the Y-junction
model. 
In Sec.~\ref{sec:approaches} we introduce the 
study of the system by looking at the
boundary conditions satisfied by the bosonic fields at the
junction. 
In Sec.~\ref{sec:twist} we discuss the twisted structure of the
Hilbert space that is manifest in the compactification of the bosons
so as to take proper care of the fermionic statistics of the electrons
in the wires.
In Sec.~\ref{sec:DEBC} we introduce what we refer to as the method of
delayed evaluation of boundary conditions (DEBC), which is less
reliant on the apparatus of BCFT.
In Sec.~\ref{sec:fpts} we apply all the results from the previous
sections to analyze the RG fixed points in the junction problem.  We
also show in this section that asymmetries in the couplings between
the three wires are irrelevant at low energies and temperatures, so
the simple Z$_3$ symmetric model of the system may indeed be a good
description of realistic junctions.
In Sec.~\ref{sec:Econservation} we use simple arguments based on energy
conservation to obtain constraints on the conductance tensor. We show that
all the fixed points that we are able to understand in this paper correspond
to physical situations where no energy is transfered to neutral (non-charge
carrying) modes in the quantum wires.
In Sec.~\ref{sec:FLL} we discuss how the conductance tensors are modified if
the Luttinger liquid wires are connected to Fermi liquid leads far from the
junction. Depending on details of the potential experiment, this may be a
better model. 
In Sec.~\ref{sec:open} we present and discuss open problems. 
In Appendix~\ref{sec:free-fermion} we give results on the non-interacting
version of our model.  
In Appendix~\ref{sec:BCFT} we review  boundary conformal field 
theory.
In Appendix~\ref{sec:refboson} we review the boundary 
conformal field theory of standard free bosons.
In Appendix~\ref{sec:BC_cond} we review the boundary conformal 
field theory approach to calculating the conductance.
In Appendix~\ref{sec:fqh} we discuss the related problem of a Y-junction of
quantum Hall edge states.

\section{Summary of Results}
\label{sec:results}
We summarize in this section our results for the conductance tensor
for the three wire junction, as well as our conjectured RG flow
diagrams for several different ranges of the interaction parameter
$g$. These results are based on our findings from three different
methods discussed in the paper: mapping to the dissipative Hofstadter
model, boundary conformal field theory, and delayed evaluation of
boundary conditions.
Our results rely on various assumptions
and arguments which we believe to be reasonable
in the light of several supporting evidences.
The details of our analysis and assumptions
are given in later sections.

\begin{figure}
\includegraphics[width=0.60\linewidth]{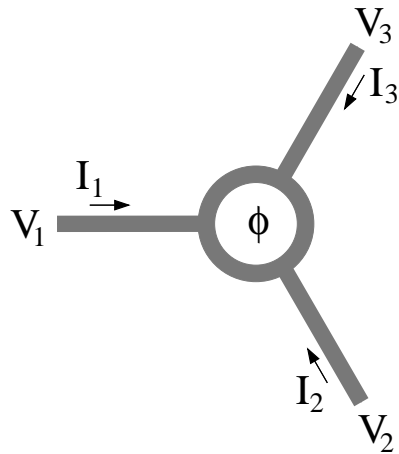}
\caption{Junction of three quantum wires with a magnetic flux threading 
the ring. The $V_{1,2,3}$ are the voltages applied on each wire, and
the $I_{1,2,3}$ the currents arriving at the junction from each of the
three wires.}
\label{fig:deviceA}
\end{figure}

The device we consider is shown in Fig.~\ref{fig:deviceA}, where the
three quantum wires are connected to a ring which can be threaded by a
magnetic flux.
The electron transfer processes between the three wires meeting at the
Y-junction is modeled by the effective Hamiltonian
\begin{equation}
H_{\cal B}=
-\sum_{j=1}^3 [\Gamma \;e^{i \phi/3} \psi^{\dagger}_{j}(0) 
\psi_{j-1}(0)
 + \mbox{h.c.}]
\;, 
\label{H_B-summary}
\end{equation}
where $\psi_j(x=0)$ is the electron operator at the endpoint of the
$j$-th ($j=1,2,3$, and we identify $j\equiv j+3$) quantum wire that connects
to the junction ($x=0$). $\Gamma$ is the tunneling amplitude between the
leads, and $\phi$ is a phase that depends on the flux that threads the
ring. In addition to the boundary term Eq.~(\ref{H_B-summary}), there is a
bulk Hamiltonian $H_{\rm bulk}$ that models the Luttinger liquid wires
characterized by the interaction parameter $g$.
The derivation of this effective Hamiltonian at the junction,
and the effective Hamiltonian for the wire part,
is given in section~\ref{sec:model}.

The presence of a magnetic flux breaks generically
time-reversal invariance, with the exception of $\phi=0,\pi$, in which
case time-reversal symmetry exists.
In this paper, we assume the three wires to be identical.
Nevertheless, the junction can be asymmetric if the wires are attached to the
ring in different ways.
If all three wires are attached
identically to the ring, the junction is symmetric under cyclic
permutations of the wires; this is the Z$_3$ symmetric case.

The important physical quantity for the three-wire junction problem is the
conductance tensor.
Within the linear response theory, the total current
$I_j$ flowing into the junction from wire $j$ is related
to the voltage $V_k$ applied to wire $k$ by
\begin{equation}
 I_j = \sum_k G_{jk} V_k,
\label{eq:def-of-conduc-tensor}
\end{equation}
where $j,k = 1,2,3$ and
$G_{jk}$ is the $3 \times 3$ conductance tensor.
Note that current conservation implies that:
\be \sum_jI_j=0.\ee
Furthermore, a common voltage applied to all three wires results
in zero current.
Thus:
\be 
\sum_jG_{jk}=\sum_kG_{jk}=0
\;.
\label{eq:conduc-aspen}
\ee

For a Z$_3$ symmetric junction, the conductance tensor takes the form
\begin{equation}
 G_{jk} = 
\frac{G_S}{2}
\, (3\delta_{jk}-1)+\frac{G_A}{2}\,\epsilon_{jk}
\;,
\label{eq:Z3tensor}
\end{equation}
where we separate the symmetric and anti-symmetric components of the tensor,
and $G_S$ and $G_A$ are scalar conductances. (The $\epsilon_{ij}$ are defined
as follows: $\epsilon_{12}=\epsilon_{23}=\epsilon_{31}=1$, $\epsilon_{21}
=\epsilon_{32}=\epsilon_{13}=-1$ and $\epsilon_{jj}=0$.) The anti-symmetric
component, controlled by $G_A$, is present only when time-reversal symmetry
is broken by a magnetic flux. 
Namely, $G_A$ should vanish in the time-reversal symmetric case $\phi=0,\pi$.
Even when  time-reversal symmetry is broken in the microscopic model,
$G_A$ might vanish in the low-energy fixed point
if the RG flow restores the time-reversal symmetry.

{}From the tensor conductance
Eq.~(\ref{eq:Z3tensor}), one reads
\begin{equation}
G_S \;=\;G_{11} = G_{22} = G_{33}.
\end{equation}
This represents the conductance of each wire when zero voltage is applied to
the other two wires. For this reason, we shall sometimes refer to $G_S$ in
the text as the single terminal conductance.

\begin{figure}
\includegraphics[width=0.60\linewidth]{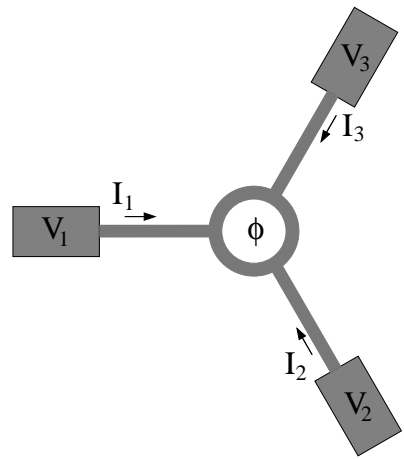}
\caption{The same three-wire junction as in Fig.~\ref{fig:deviceA}, but now 
with the Fermi liquid reservoirs attached at the ends of the three
wires. The $V_{1,2,3}$ are now the voltages as measured at the three
reservoirs.}
\label{fig:deviceB}
\end{figure}

We also consider the same three-wire junction device when connected to Fermi
liquid reservoirs, as depicted in Fig.~\ref{fig:deviceB} (the wide regions at
the ends of the wires represent the Fermi liquid contacts). It is known that
the contact to the reservoirs affect the two-terminal conductance of single
quantum wires~\cite{Tarucha,Maslov-Stone,Safi-Schulz}. This renormalization
of the conductance also takes place in the three-wire junctions, and we
calculate these renormalization effects in this paper. In the presence of the
Fermi liquid leads, we obtain that the ``dressed'' $3\times 3$ conductance
tensor $\bar{\bm G}\equiv{\bm G}_{\rm w/\;leads}$ of the system is related to
the ``bare'' conductance tensor $\bm G\equiv{\bm G}_{\rm w/o\;leads}$ by
\begin{equation}
{{\bm G}^{-1}_{\rm w/\;leads}}=
{{\bm G}^{-1}_{\rm w/o\;leads}}+ G_c^{-1} \openone
\; ,
\label{eq:RbarR-results}
\end{equation}
where $G_c=2g/(g-1)\;e^2/h$ is effectively a contact conductance between the leads
and the quantum wires. This relationship in Eq.~(\ref{eq:RbarR-results}) can
be interpreted physically as the addition in series of the lead/wire
interface resistances to the Y-junction resistance tensor. (Notice that,
because of the $\sum_k G_{jk} =0$ condition, the 
inverse $\bm G^{-1}$ does not exist.
Hence one must work directly with the conductance instead of the
resistance tensor as discussed in section~\ref{sec:FLL}.)

We are mostly interested in the stable RG fixed points of the problem.
They would govern the physics in the low-energy limit
-- namely in the limit of low bias voltages and low temperature.
Below we list our results for the low-energy fixed point conductances for the
Y-junction devices. Which fixed point is the stable one is determined by the
Luttinger parameter $g$ that depends on the strength of local
electron-electron interactions.
We find that the stable fixed point has the Z$_3$ symmetry in most cases.
Namely, even if the Z$_3$ symmetry is broken at the junction,
the Z$_3$ symmetry would be restored in the low-energy limit,
except for a few special cases.
Thus we primarily focus on the RG flow in the presence of the Z$_3$ symmetry.
The most important parameters of the junction are then
the hopping strength $\Gamma$ and the enclosed magnetic flux $\phi$.
If we understand the RG flow most simply by the effective values of
$\Gamma$ and $\phi$ varying as functions of the energy scale,
the RG flow diagram could be drawn on the $\Gamma$--$\phi$ plane
as shown in the following.

As in the case of junctions of two wires, the interaction parameter $g$
controls the RG flow and dictates the phase diagram. Now we present our
results for several different ranges of $g$.

\subsection{$g<1$}

\begin{figure}
\includegraphics[width=0.45\linewidth]{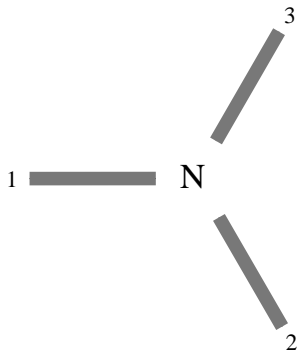}
\caption{Pictorial representation of the N fixed point.
It just corresponds to three decoupled wires.}
\label{fig:Npicture}
\end{figure}

When the interaction in the quantum wires is repulsive ($g<1$), the hopping
amplitude $\Gamma$ decreases along the RG flow.  The stable fixed point
corresponds to $\Gamma=0$, namely to completely decoupled wires, as
pictorially illustrated in Fig.~\ref{fig:Npicture}. We will call this fixed
point the N fixed point, where N stands for Neumann boundary condition. We
note that the value of the effective flux $\phi$ does not matter in the
decoupled limit, so in the N fixed point there is no breaking of time
reversal.  Moreover, even if the junction is Z$_3$ asymmetric, the system is
generically renormalized into the N fixed point; the Z$_3$ asymmetry turns
out to be irrelevant.  The leading irrelevant perturbation to the N fixed
point is the electron hopping between two wires, which has scaling dimension
$(1/g)>1$.

The conductance tensor at the N fixed point is obviously
\begin{equation}
 G_{jk} = \bar{G}_{jk} =0.
\end{equation}
Of course there is no difference between the ``bare'' conductance $\bm G$ and
the ``dressed'' (by the reservoirs) conductance $\bar{\bm G}$ at the N fixed
point.

\begin{figure}
\includegraphics[width=0.75\linewidth]{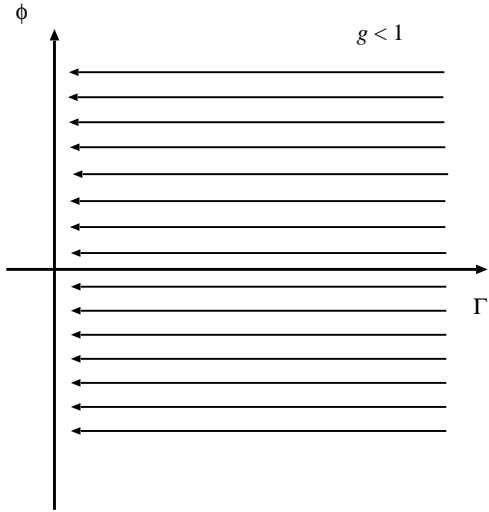}
\caption{RG flow diagram for $g<1$.
The junction is renormalized toward the decoupled N fixed point.}
\label{fig:Nflow}
\end{figure}

The corresponding flow diagram in the $\Gamma$--$\phi$ plane is shown in
Fig.~\ref{fig:Nflow}.

\subsection{$g=1$}

For the noninteracting case $g=1$, one can solve the original electron model
as a single particle problem.  The junction is then characterized by the $3
\times 3$ scattering matrix of a single free electron. Thus there is a
continuous manifold of RG fixed points.
In other words, electron hopping is exactly marginal.

\begin{figure}
\includegraphics[width=0.45\linewidth]{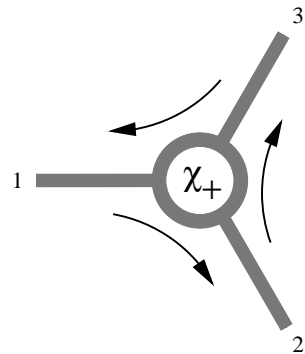}
\caption{Pictorial representation of the chiral fixed point.
The incoming electrons from one wire is diverted to one of the
other wires.}
\label{fig:Chipicture}
\end{figure}

Several special cases on the ``free electron''
manifold are worth mentioning.
Let us consider the Z$_3$ symmetric case first.
The N fixed point, as introduced above for $g<1$, corresponds to
a complete
reflection for each wire and thus is a special point on the RG fixed
manifold. 
If we further impose the time reversal symmetry,
we always have $G_A=0$ while
the maximal single-terminal conductance on the
``free electron'' manifold is $G_S=8/9 \; e^2/h$.
The constraint on the maximum of $G_S$ in the time reversal symmetric 
case 
is due to the unitarity of the single electron scattering matrix
as discussed in Ref.~\cite{Nayak}.
On the other
hand, if the time reversal symmetry is broken by the flux $\phi$, the maximal
single-terminal conductance $G_S=e^2/h$ can be realized by a complete
transmission of electrons from wire $j$ to wire $j+1$ (or $j-1$), where again
we identify $j\equiv j+3$. We call them chiral fixed points $\chi_{\pm}$;
they are pictorially represented in Fig.~\ref{fig:Chipicture}, and they are
also on the RG fixed manifold. At the $\chi_{\pm}$ points $G_A=\pm G_S$. The
conductance tensor at $\chi_{\pm}$ reads
\begin{equation}
G_{jk} = \frac{e^2}{2h}\;
\left[(3\delta_{jk}-1)\pm\,\epsilon_{jk}\right]=
\frac{e^2}{h} \big( \delta_{jk} - \delta_{j,k \pm 1} \big).
\label{eq:Gchifree}
\end{equation}

\begin{figure}
\includegraphics[width=0.45\linewidth]{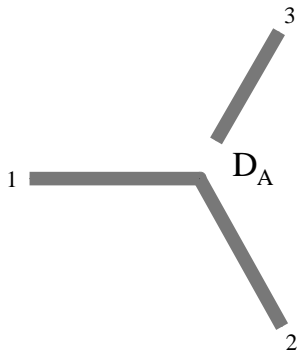}
\caption{Pictorial representation of the asymmetric fixed point D$_\shbox{A}$, 
in which two of the wires (for example wires 1 and 2 in the figure) are
perfectly connected while the other (wire 3) is left decoupled. This fixed
point is unstable for all values of $g$ with the exception of $g=1,3$, in
which case D$_\shbox{A}$ may belong to the continuous
manifold of fixed points.}
\label{fig:DAnew}
\end{figure}

If we allow breaking of the Z$_3$ symmetry,
a simple fixed point is given by 
connecting perfectly two of the wires, while the other is
left decoupled completely, as depicted in Fig.~\ref{fig:DAnew}.
This fixed point, which we call D$_\shbox{A}$, is also described by a
single electron scattering matrix and thus a special point
on the ``free electron'' manifold.
It is clear that there are other fixed points on the manifold
with varying degrees of Z$_3$ asymmetry. 

\begin{figure}
\includegraphics[width=0.75\linewidth]{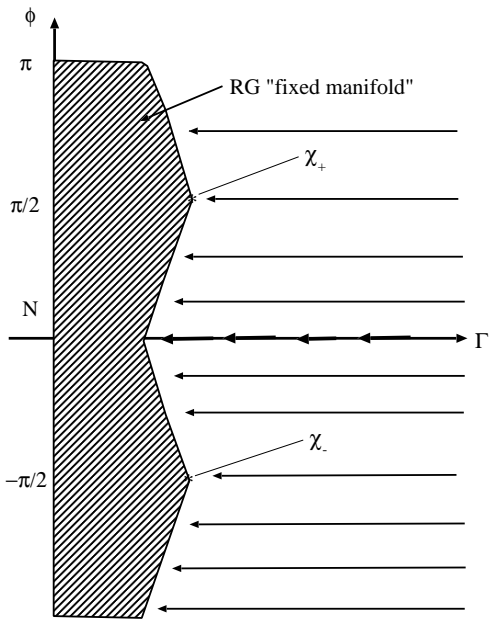}
\caption{RG flow diagram for $g=1$.
There is a stable RG fixed manifold which includes N and $\chi_{\pm}$
as special points.}
\label{fig:g=1flow}
\end{figure}

In fact, even for noninteracting wires or $g=1$, other RG fixed points, which
cannot be characterized by the single electron scattering matrix, exist in
the presence of the interaction at the junction. However, these fixed points
are more unstable, and the RG flows (shown in Fig.~\ref{fig:g=1flow}) are
towards the ``free electron'' manifold.

\subsection{$1<g<3$}

The N fixed point is now unstable against the inclusion of electron hoppings,
which have scaling dimension $1/g < 1$.  For the time-reversal symmetric case
$\phi=0,\pi$, the system is renormalized into the M fixed point.  The M fixed
point should have the single-terminal conductance $G_S$ in the range
$0<G_S<(4g/3)\;(e^2/h)$, as shown in Section~\ref{sec:Econservation}. The M fixed point
is unstable against turning on the flux $\phi$. Unfortunately, we do not have
an exact solution corresponding to the M fixed point, and other properties
are not known.

For $0 < \phi < \pi$ and $-\pi < \phi < 0$, the RG flows are towards the
stable, chiral fixed points $\chi_+$ and $\chi_-$, respectively.  The chiral
fixed points exhibit the asymmetric conductance tensor
\begin{equation}
  G_{jk}^{\pm} = \frac{G_{\chi}}{2} \;
  \left[(3\delta_{jk}-1)\pm g\,\epsilon_{jk}\right]
\label{eq:Gchi}
\end{equation}
with the single terminal conductance or symmetric component
\begin{equation}
 G_S=G_{\chi} = \frac{4g}{3+ g^2} \frac{e^2}{h},
\end{equation}
and anti-symmetric component $G_A=\pm g\;G_S$. We note that, in the limit
$g\rightarrow 1$, the chiral $\chi_{\pm}$ fixed points reduce to the complete
transmission of single electron from wire $j$ to wire $j \pm 1$ discussed
previously.

\begin{figure}
\includegraphics[width=0.65\linewidth]{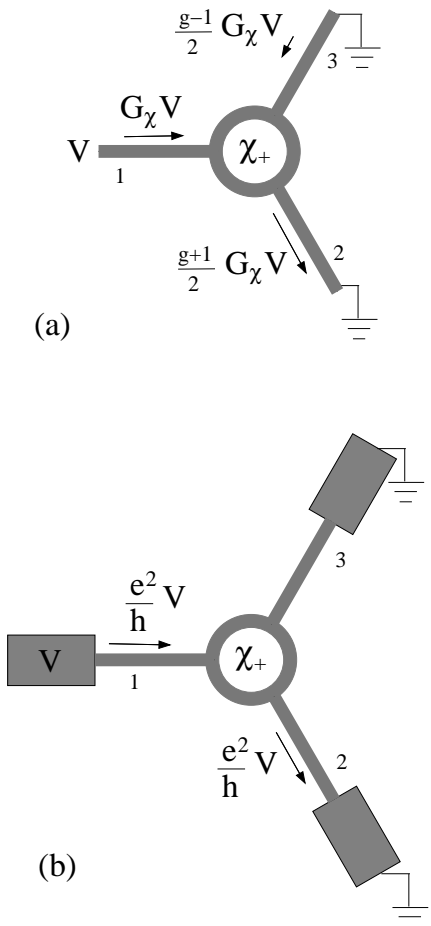}
\caption{Response of the Y-junction system at the $\chi_+$ fixed point 
when a voltage $V$ is applied to wire 1, while wires 2 and 3 are
grounded. Case (a) is when wire 1 is directly at a voltage $V$, and case (b)
is when wire 1 is in contact with a reservoir at voltage $V$. Notice that in
case (a) some current is sucked into the junction region from lead 3, while
in case (b) this effect goes away and the junction behaves as a perfect
``circulator'', transmitting all incoming current from lead 1 to lead 2. The
response for the $\chi_-$ fixed point is similar, with the current flowing
from 1 to 3 instead.}
\label{fig:Chi1V2ground}
\end{figure}

An intriguing aspect of the conductance~(\ref{eq:Gchi}) is that, when a
voltage is applied to one of the wires while the other two are kept at zero
voltage (as illustrated in Fig.~\ref{fig:Chi1V2ground}a), the current is
rather ``sucked in'' from one of the wires with the zero voltage. However, if
we include the non-interacting leads to the system (as shown now in
Fig.~\ref{fig:Chi1V2ground}b), the effective conductance tensor for
$\chi_{\pm}$ are given by the conductance tensor Eq.~(\ref{eq:Gchifree})
obtained for $\chi_{\pm}$ at $g=1$ (in which $\bar G_A=\pm \bar G_S$ with
$\bar G_S={e^2}/{h}$). Namely, the ``sucking'' effect disappears in the
presence of non-interacting leads, while the asymmetry holds.

\begin{figure}
\includegraphics[width=0.75\linewidth]{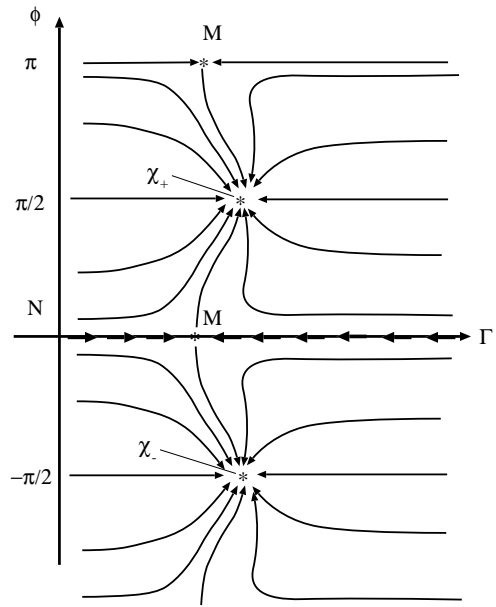}
\caption{RG flow diagram for $1<g<3$.
For $\phi \neq 0,\pi$ the system is renormalized toward the chiral fixed
point $\chi_{\pm}$.  For the time-reversal symmetric case $\phi=0,\pi$, the
infrared fixed point is the nontrivial M fixed point.  It is unstable against
the addition of flux $\phi$.}
\label{fig:Chiflow}
\end{figure}

The RG flow diagram (shown in Fig.~\ref{fig:Chiflow}) implies that even a
tiny flux $\phi$ brings the junction to the completely asymmetric conductance
in the low-energy limit. The leading irrelevant perturbation to the
$\chi_{\pm}$ fixed points has scaling dimension $4g/(3+g^2) >1$. A remarkable
aspect of the fixed points $\chi_\pm$ is that the conductance, as well as the
scaling dimension, exhibit {\em non-monotonic} dependence on the interaction
parameter $g$, unlike in other known applications of TLL. Making the electron
interaction more attractive can decrease the conductance!

We note that the D$_\shbox{A}$ fixed point, which breaks the Z$_3$
symmetry, is
unstable for this range of $g$. Hence, small asymmetries in the physical
device do not affect the conductance of the $\chi_\pm$ fixed points at low
voltages and temperatures for this range of the $g$ parameter.

\subsection{$g=3$}

\begin{figure}
\includegraphics[width=0.75\linewidth]{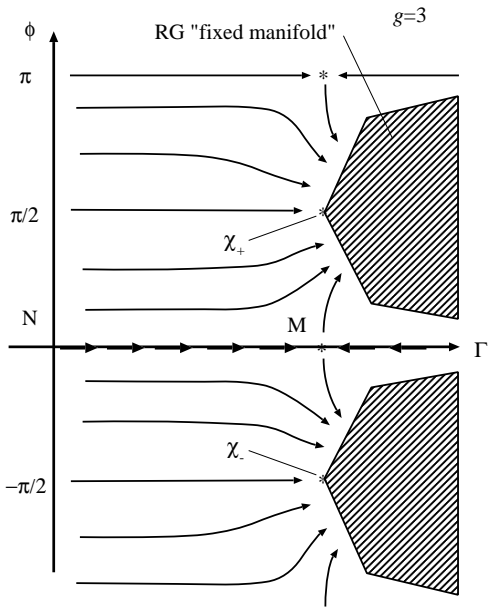}
\caption{RG flow diagram for $g=3$.
For $\phi \neq 0,\pi$ the system is renormalized toward the RG fixed
manifold, which includes $\chi_{\pm}$ as special points.  For the
time-reversal symmetric case $\phi=0,\pi$, the infrared fixed point is the
nontrivial M fixed point.  It is unstable against the addition of flux
$\phi$.}
\label{fig:g=3flow}
\end{figure}

The N fixed point is again unstable. On the other hand, there is an RG fixed
manifold (shown in Fig.~\ref{fig:g=3flow}), as we have seen for $g=1$.  For
$g=3$, the manifold is again characterized by the scattering matrix of a free
fermion, which is {\em not} the original electron. The chiral fixed points
$\chi_{\pm}$ become the special points on the manifold as $g \rightarrow 3$.
The largest single-terminal conductance is realized on another special point
\DP on the manifold, which is explained in the next subsection for $3<g<9$.

The system with $\phi \neq 0,\pi$ is renormalized into the fixed manifold.
On the other hand, for the time-reversal symmetric case
$\phi =0, \pi$, the infrared M fixed point is {\em not} on the manifold,
and it is unstable against the inclusion of the flux $\phi$.

\subsection{$3<g<9$}

There is a stable fixed point D$_\shbox{P}$, which becomes a special
point on the
fixed manifold in the limit $g \rightarrow 3$ as discussed above. The
conductance tensor at \DP takes the symmetric form ($G_A=0$) with
$G_S=(4g/3)\; (e^2/h)$.  Including the non-interacting leads, the effective
conductance tensor is again symmetric ($\bar G_A=0$), but the effective
single-terminal conductance is $\bar{G}_S=(4/3)\; (e^2/h)$.

\begin{figure}
\includegraphics[width=0.45\linewidth]{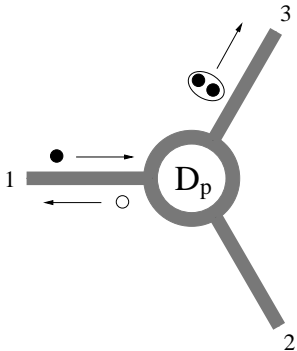}
\caption{Pictorial representation of the \DP fixed point.
The conductance is enhanced by the Andreev reflection process. We only
illustrate the process where the electron pair exits through lead 3; but
notice that, since the \DP fixed point is time-reversal symmetric, the pair
has exactly the same amplitude for exiting through lead 2.}
\label{fig:Dppicture}
\end{figure}

Notice that the maximum values $G_S=(4g/3)\;(e^2/h)$ or 
$\bar G_S=(4/3)\;(e^2/h$) is
larger than the maximum conductance $G_S=(8/9) \; e^2/h$ for unitary scattering
of single particle states; this shows that multiparticle scattering at the
junction takes place for interacting quantum wires. Indeed, the enhanced
conductance at \DP may be understood as a consequence of pair tunneling,
and this is pictorially illustrated in Fig.~\ref{fig:Dppicture}.

\begin{figure}
\includegraphics[width=0.75\linewidth]{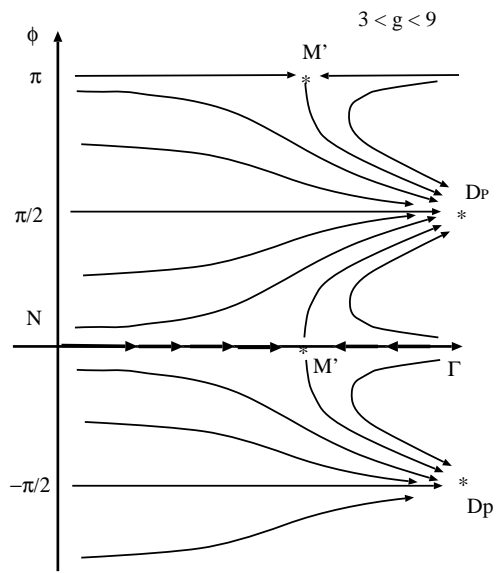}
\caption{RG flow diagram for $3 < g <9$.
For $\phi \neq 0,\pi$ the system is renormalized to the \DP fixed point
exhibiting Andreev reflection.  For the time-reversal symmetric case
$\phi=0,\pi$, the infrared fixed point is the nontrivial M fixed point. It is
unstable against the addition of flux $\phi$.}
\label{fig:Dpflow}
\end{figure}

The RG flow diagram for this range of $g$'s is shown in
Fig.~\ref{fig:Dpflow}. For $\phi\neq 0, \pi$, the system is generically
renormalized into the \DP fixed point.  Although the \DP fixed point
itself does not break the time-reversal symmetry, the system does not reach
there with $\phi=0,\pi$.  In the time-reversal symmetric case $\phi=0,\pi$,
the property of the infrared M fixed point is not known precisely. It should
be unstable against inclusion of the flux.

\subsection{$9 < g$}

\begin{figure}
\includegraphics[width=0.75\linewidth]{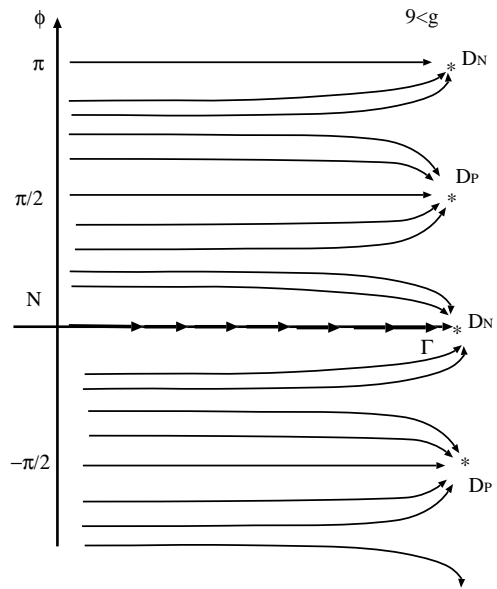}
\caption{RG flow diagram for $9 < g$.
For $\phi \neq 0,\pi$ and $\phi=0, \pi$, the system is renormalized to \DP
and \DN fixed points respectively.  Both exhibits the enhanced conductance
due to the Andreev reflection.}
\label{DNflow}
\end{figure}

The time-reversal invariant system with $\phi=0,\pi$ is now renormalized into
the stable \DN fixed point. For $\phi \neq 0,\pi$, the infrared stable
fixed point is still the \DP fixed point as in the case $3<g<9$. The flow
diagram for $g>9$ is shown in Fig.~\ref{DNflow}.

As far as the conductance is concerned, the \DN fixed point is identical to
the \DP fixed point. The difference is in the scaling dimension of the
leading irrelevant operator; it is $g/3$ for \DP but
is $g/9$ for D$_\shbox{N}$.
As both \DN and \DP are stable, there must be another critical fixed
point between these two; its properties however are not known.

\section{Model}
\label{sec:model}
In this Section, we define the model Hamiltonian precisely,
for the junction introduced in Fig.~\ref{fig:deviceA}.

We study only the simplest model of a Y-junction which includes TLL effects.
Thus we consider only a single channel of electrons in each wire, ignore
electron spin, assume that the interactions are short-ranged, and ignore
phonons and impurities.  We also assume that all three wires are equivalent
in the bulk, and consider the Z$_3$ symmetric junction. (The generalization
to the Z$_3$ asymmetric junctions is straightforward, and we find in
Sec.~\ref{sec:asymmetric-FP} that the Z$_3$ asymmetry turns out to be
irrelevant.)

It is convenient to start by defining a tight-binding version of our
model. Let $\psi_{n,i}$ annihilate an electron on site $n$ on wire $i$. Here
$n=0,1,2,3,\ldots \infty$ and $j=1,2$ or $3$. The Hamiltonian is:
\begin{equation} 
H=H_0+H_{\cal B}+H_{\rm int},
\label{eq:eHam}
\end{equation}
where:
\begin{equation} 
H_0=-t\sum_{n=0}^\infty\sum_{j=1}^3(\psi^\dagger_{n,j}\psi_{n+1,j}+h.c.)
\label{H_0},
\end{equation}
\begin{equation}
H_{\cal B}=-
\sum_{j=1}^3 \left[(\tilde \Gamma /2)\;e^{i \phi/3}\; \psi^{\dagger}_{0,j} 
\psi_{0,j-1}
 + \mbox{h.c.}\right] \label{H_B}
\end{equation} and 
\begin{equation} H_{\rm int}=\tilde V\sum_{n=0}^\infty\sum_{j=1}^3
\hat n_{n,j}\hat n_{n+1,j}.
\end{equation} 
Here $\hat n_{n,j}=\psi^\dagger_{n,j}\psi_{n,j}$ is the number density and
the wire index $j=0$ is identified with $j=3$.  Note that we have allowed for
a non-zero phase $\phi$ in the tunneling matrix element $\Gamma e^{i\phi
/3}$. By making redefinitions of the phases of the fields $\psi_{n,i}$,
independently on the different wires, we can make the phase of each
tunneling matrix element equal.  However, we cannot transform away a uniform
phase, $\phi /3$.  We may think of the phase $\phi$ as being the flux through
the junction. It must vanish if the system is time-reversal invariant but
will generally be non-zero otherwise, for example if a magnetic field is
applied to the junction.

While it seems intuitively reasonable that $\phi $ should become non-zero in
the presence of a magnetic field, it is not so obvious how $\phi$ actually
depends on the field.  To study this issue it is convenient to consider a
slightly more realistic model in which each lead is connected to a circular
conducting ring, which may contain a flux.  In particular, if the ring is
very small and the Fermi energy of the leads is in resonance with a single
unoccupied energy level in the ring, so that we may ignore all other energy
levels of the ring, then $\phi$ actually vanishes.  This follows since the
tunneling Hamiltonian then takes the form:
\begin{equation} 
H_{\cal B}=\sum_{j=1}^3 
\;[a_j\;\psi_{0,j}^\dagger d+ h.c.],
\label{1level}
\end{equation}
where $d$ annihilates an electron at the single level in the ring. We are now
free to redefine the phases of $\psi_{n,j}$ independently on each lead,
making all the $a_j$'s real.  This is nothing but the junction constructed
with a ``quantum dot'' as introduced in Ref.~\cite{Nayak}.

On the other hand, if two or more energy
levels in the ring are considered, $\phi$ will generally be non-zero. More
generally, if the operators $d_l$ annihilate electrons in the ring in
eigenstates with energies $E_l$, then we may write the tunneling Hamiltonian
as:
\begin{equation}  
H_{\cal B}=
\sum_{j=1}^3 \,\sum_l\;
[a_{j,l}\;\psi_{0,j}^\dagger d_l+ h.c.].
\label{mlevel}
\end{equation}
The tunneling amplitudes, $a_{j,l}$ from the $j^{\hbox{th}}$ lead to the
$l^{\hbox{th}}$ ring state may be assumed to be $j$-independent constants
multiplied by the wave-functions of the ring at the locations of the
$j^{\hbox{th}}$ wire, at angle $\theta = 2\pi j/3$.  We may take the ring
wave-functions to be $f_l(\theta )=e^{il\theta}$ for integer $l$, with
energies $E_l=Il^2/2$. A non-zero dimensionless flux, $\tilde \phi$, changes
the energies to $I(l-\tilde\phi /2\pi )^2/2$ so that the states with angular
quantum numbers $\pm l$ do not have the same energy. We may assume that the
phases of the tunneling amplitudes $a_{j,l}$ are given by the phases of the
ring states at the locations $\theta =2\pi j/3$ where the $j^{\hbox{th}}$
wire joins the ring:
\begin{equation} a_{j,l}=a_l\;e^{i2\pi jl/3}.\end{equation}
To get a tunneling Hamiltonian of the form of Eq.~(\ref{H_B}) we assume that
$|a_l|<<\Delta E$, the level spacing in the ring.  Then we consider processes
of second order in the $a_l$'s: 1) in which an electron tunnels from a lead
to an unoccupied ring state $l$ and then tunnels off to a different lead, and
2) in which an electron in an occupied ring state $l$ first tunnels off to a
lead and then an electron from a different lead tunnels into the ring to fill
the hole. This gives a low-energy effective tunneling Hamiltonian of the
form of Eq. (\ref{H_B}) with:
\begin{equation} 
\Gamma e^{i\phi /3}=\sum_{l}{|a_l|^2e^{i2\pi l/3}\over E_F-E_l}
\label{phi_eff}
\end{equation}
We see that if only one (occupied or unoccupied) level $l$ dominates the sum,
then: $\phi = 2\pi l$, equivalent to zero.  If 2 levels dominate the sum, say
$l_1$ and $l_2$, with the contribution from the ${l_1}^{\hbox{th}}$ level
much bigger than the contribution from the ${l_2}^{\hbox{th}}$, then
\begin{equation} 
\phi \approx 3 {|a_{l_2}|^2\over |a_{l_1}|^2}\;{E_{l_1}-E_F\over E_{l_2}-E_F}
\;\sin [2\pi (l_2-l_1)/3],
\end{equation}
with $|\phi|<<1$.  So, $\phi$ becomes small near a resonance. Of course, when
there is zero flux ($\tilde\phi=0$) in the ring, $\phi $ also vanishes due to
the cancellation of the ring levels with quantum numbers $\pm l$. In general,
$\phi$ will be some periodic odd function of the flux $\tilde\phi$ in the
ring.

We now consider the continuum limit of the tight-binding model of
Eq. (\ref{eq:eHam}). We introduce left and right movers as usual, keeping only
narrow bands of wave-vectors near the Fermi ``surface'' at $\pm k_F$:
\begin{equation} 
\psi_{n,j}\approx e^{ik_Fna}\;\psi_{Rj}(na)+e^{-ik_Fna}\;\psi_{Lj}(na).
\end{equation}
($a$ is the lattice spacing.)  Linearizing the dispersion relation, $H_0$
then becomes:
\begin{equation} 
H_0\approx iv_F\sum_j \int_0^\infty \!\!\!dx\;
\left[ \psi^\dagger_{Lj}{d\over dx}\psi_{Lj}
-\psi^\dagger_{Rj}{d\over dx}\psi_{Rj}\right] ,
\end{equation}
where $v_F=2t\sin k_F$. The interaction Hamiltonian $H_{\rm int}$ takes the
form
\begin{eqnarray}
H_{\rm int}=&&{\tilde V}\;\sum_j\int_0^\infty \!\!\!dx\;
\left[
J_{L,j}^2(x)+J_{R,j}^2(x)\right]
\nonumber
\\
&+&{\tilde V}_{RL}\;\sum_j\int_0^\infty \!\!\!dx\;
\;J_{L,j}(x)J_{R,j}(x)
\;,
\label{eq:HintJRL}
\end{eqnarray}
where
\begin{equation}
J_{L,j}(x)=\psi^\dagger_{L,j}\psi_{L,j}
\quad
{\rm and}
\quad
J_{R,j}(x)=\psi^\dagger_{R,j}\psi_{R,j}
\end{equation}
and ${\tilde V}_{RL}=2 {\tilde V}\left[1-\cos(2k_Fa)\right]$. The first term
in Eq.~(\ref{eq:HintJRL}) can be absorbed into a renormalization of the Fermi
velocity. It is the second term, the one proportional to ${\tilde V}_{RL}$,
that controls the Luttinger parameter $g$. For repulsive interactions
(${\tilde V_{RL}}>0$), $g<1$, while for attractive interactions
 (${\tilde V_{RL}}<0$),
$g>1$.

The open boundary conditions in the tight binding model are equivalent to a
vanishing boundary condition at $x=-a$, i.e.
\begin{equation} 
e^{-ik_Fa}\psi_R(-a)+e^{ik_Fa}\psi_L(-a)=0.
\label{L-R}
\end{equation}
Since $\psi_{Rj}$ is a function only of $(x-v_Ft)$ and $\psi_{Lj}$ of
$(x+v_Ft)$, Eq. (\ref{L-R}) implies that we may regard the right-movers as
the analytic continuation of the left-movers to the negative $x$-axis:
\begin{equation} 
\psi_{Rj}(x)\approx -e^{2ik_Fa}\psi_{Lj}(-x),\  \ (x>0).
\end{equation}
Thus we can apparently write $H_0$ entirely in terms of left-movers:
\begin{equation} 
H_0\approx iv_F\sum_j\int_{-\infty}^\infty
\!\!\!dx\;
\psi^\dagger_{Lj}{d\over dx}
\psi_{Lj}.\label{H_0c}
\end{equation}
However, treating the boundary condition of Eq. (\ref{L-R}) 
more carefully, we see that $H$ actually contains an extra 
term when $k_F\neq \pi /2a$.  To obtain this boundary condition 
on the left-movers we must add a term to the Hamiltonian:
\begin{equation}
\delta H_{\cal{B}}=-r\sum_j\psi_{Lj}^\dagger (0)\psi_{Lj}(0).
\end{equation}
The corresponding Schr\"odinger equation (which does not mix the leads at this
point) is:
\begin{equation} 
iv_F{d\over dx}\Psi_j -r\delta (x) \Psi_j =E\Psi_j .
\label{SEmodel}
\end{equation} The
energy eigenvalues are $E=v_Fk$, and the wave-functions are:
\begin{eqnarray}
\Psi_j (x) &=& A_{{\rm in},j}\;e^{-ikx}, \ \  (x>0)\\
&=& A_{{\rm out},j}\;e^{-ikx},\ \  (x<0).
\end{eqnarray}
Interpreting $\Psi_j (0)$ in the second term of Eq. (\ref{SEmodel}) as
$[\Psi_j (0^+)+\Psi_j (0^-)]/2$, this implies:
\begin{equation} 
A_{{\rm out},j}={iv_F-r/2\over iv_F+r/2}\; A_{{\rm in},j}.
\end{equation}
From Eq. (\ref{L-R}) we see that the phase of the wave-function advances by
$e^{-2ik_Fa}$ upon scattering so that:
\begin{equation} 
-e^{2ik_Fa}={iv_F-r/2\over iv_F+r/2},
\end{equation}
and hence:
\begin{equation} 
r=-2v_F\cot (k_Fa).
\label{r}
\end{equation}

In the continuum limit, the total boundary term in $H$ is:
\begin{widetext}
\begin{equation} 
H_{\cal B}\approx 
\sum_{j=1}^3
\left\{
[\Gamma e^{i\phi /3}\;\psi^\dagger_{L,j}(0)\psi_{L,j-1}(0)+h.c.]
+r\;\psi^\dagger_{L,j}(0) \psi_{L,j}(0)
\right\} ,
\label{H_Bc}
\end{equation}
\end{widetext}
where the relation between the coupling $\Gamma$ in the continuum limit and
the $\tilde
\Gamma$ in the tight-binding model is obtained in 
Appendix~\ref{sec:free-fermion} by matching the scattering matrices for
non-interacting electrons calculated in both cases:
\begin{equation}
\Gamma=
|r-2iv_F| \; \tilde\Gamma/2 ,
\end{equation}
with $r$ given by Eq. (\ref{r}). For interacting systems, $r$ has no effect
on the conductance (in the zero frequency, zero temperature limit), so we
henceforth ignore it.  It is important to note that the continuum form of
$H_0$ and $H_{\cal B}$ in Eqs.~(\ref{H_0c}) and (\ref{H_Bc}) is very general
and not restricted to the underlying tight-binding model.

\section{Bosonization}
\label{sec:set-up}
The effects of electron-electron interactions on the three 1D quantum
wires is treated using bosonization methods.
(For a review, see e.g.~\cite{FisherGlazman}).
Here we shall focus on the simpler case of spinless electrons.
In the low-energy limit, the effective theory for
each quantum wire is given by the free boson field theory,
which is nothing but the Tomonaga-Luttinger liquid.
In formulating the junction of three wires, we can generalize
the ``folding'' trick used in Ref.~\cite{WongAffleck} and
define the three wires on the half-line $x>0$, so that
the boundary at $x=0$ represents the junction.
The (imaginary-time) low-energy effective action of the wires is
thus given as 
\begin{equation}
S = \int_{-\infty}^{\infty} d\tau \;
 \int_0^{\infty}dx \; \sum_{j=1}^3
          \frac{g}{4\pi} (\partial_{\mu} \varphi_j)^2,
\label{eq:S0}
\end{equation}
where $j=1,2,3$ labels each wire.
The electron-electron interaction strength is essentially contained in
a single parameter $g$, which determines various physical quantities
and exponents, as we will see later. 
The cases $g<1$ and $g>1$ correspond, respectively, to
repulsive and attractive interactions, while $g=1$ is the Fermi
liquid.

The same theory can be also described in terms of
the dual field $\theta_j$ for each wire, with the similar action
\begin{equation}
S = \int d\tau \; dx \; \sum_{j=1}^3 \frac{1}{4\pi g} (\partial_{\mu} 
\theta_j)^2 .
\label{eq:S0-dual}
\end{equation}
The precise definition of the dual field $\theta_j$
and the derivation of the action~(\ref{eq:S0-dual})
is discussed in Sections~\ref{sec:approaches} and \ref{sec:twist}.
Here we just note that, while $\varphi_j$ represents the
quantum mechanical phase of the electron, $\theta_j$
represents the phase of the charge-density wave.
In the bulk, both phases $\varphi_j$ and $\theta_j$ are not
fixed to a particular value, reflecting the strong fluctuation
in one dimension.

Of course, the effective action~(\ref{eq:S0}) by itself
does not give an answer to the problem of our interest.
In fact, the theory defined on the half-line is not completely
well defined without specifying the {\em boundary condition} at $x=0$.
In the present formalism, it is the boundary condition
that represents the physics of the junction.
In particular, the renormalization-group (RG) fixed points
of the problem correspond to conformally invariant
boundary conditions.
It turns out that there is a rich variety of RG fixed points
(conformally invariant boundary conditions)
for the present problem, as we presented in Sec.~\ref{sec:results}
and will discuss in detail in Sec.~\ref{sec:fpts}.

In this Section,
we begin our analysis starting from
the simplest RG fixed point of the problem:
the limit of three disconnected  wires without any transfer of
electrons between different wires.
It is clear that this limit should exist as a RG fixed point.
The corresponding boundary condition is known from the
previous studies on the open end of a single quantum wire:
each boson field $\phi_j$ obeys the Neumann boundary
condition, $\partial \phi_j / \partial x =0$ at the
boundary $x=0$.
In terms of the dual field $\theta_j$, it is equivalent to the
Dirichlet boundary condition $\theta_j = \mbox{const.}$
at $x=0$.
Thus, the disconnected limit can be described by the
effective action~(\ref{eq:S0}) with the
Neumann boundary condition on $\varphi_j$ imposed.

We defer the technical discussion of the boundary condition
to Sections~\ref{sec:approaches} and \ref{sec:twist}.
Instead, now we introduce the electron hopping term~(\ref{H_B})
as a perturbation.
In the low-energy effective theory, the perturbation should
also be bosonized.

The electron annihilation operators {\em at the open ends} ($x=0$)
of the three 1D wires can be written in terms of bosonic
fields $\varphi_j$, $j=1,2,3$~\cite{FisherGlazman,Nayak,short-paper1}:
\begin{equation}
  \psi_j \sim \eta_j \;e^{ i \varphi_j/\sqrt{2}},
\label{bosend}\end{equation}
where the $\eta_j$ are the so-called Klein factors satisfying $ \{ \eta_j,
\eta_k \} = 2\delta_{jk}, $ which are necessary to ensure the
anticommutation relations of the fermion operators in different wires.
Notice that for electron operators on the same wire $j$, the correct
anticommutation relations follow, via the bosonization scheme, from
the commutation relations of the boson field $\varphi_j$, and the
Klein factor $\eta_j$ plays no role. However, for $i\ne j$,
$[\varphi_i,\varphi_j]=0$; hence, for electron operators on different
leads, it is only the Klein factors that are responsible for the
correct anticommutation relations. The Klein factors may be
represented by the Pauli matrices. An alternative representation using
the boson zero modes is described in Appendix~\ref{sec:fqh} when we
discuss tunneling in junctions of fractional quantum Hall liquids.

The boundary action Eq.~(\ref{H_B}), when rewritten in terms of the
boson fields, is thus equivalent to
\begin{widetext}
\begin{eqnarray}
H_{\cal B}&&=-  \sum_{j=1}^3 \left[\left(\Gamma e^{i \phi/3} \eta_j\eta_{j-1}
\;e^{-\frac{i}{\sqrt{2}}(\varphi_j-\varphi_{j-1})}
 + \mbox{h.c.}\right)+r\; \partial_x \varphi_j\right]\Big{|}_{x=0}
\;.
\label{H_B-bosonic}
\end{eqnarray}
\end{widetext}

It is useful to define the rotated basis
\begin{eqnarray}
\Phi_0 &=& \frac{1}{\sqrt{3}}(\varphi_1 + \varphi_2 + \varphi_3)
\nonumber \\
\Phi_1  &=& \frac{1}{\sqrt{2}}(\varphi_1 - \varphi_2) 
\label{eq:newbasisphi} \\
\Phi_2  &=& \frac{1}{\sqrt{6}}(\varphi_1 + \varphi_2 - 2\varphi_3)
\nonumber
\end{eqnarray}
and similarly
\begin{eqnarray}
\Theta_0 &=& \frac{1}{\sqrt{3}}(\theta_1 + \theta_2 + \theta_3)
\nonumber \\
\Theta_1  &=& \frac{1}{\sqrt{2}}(\theta_1 - \theta_2) 
\label{eq:newbasistheta} \\
\Theta_2  &=& \frac{1}{\sqrt{6}}(\theta_1 + \theta_2 - 2\theta_3)
\; .
\nonumber \label{Phidef}
\end{eqnarray}
The $x=0$ tunneling operators between the three wires are more easily
expressible in terms of these rotated fields. Also, the product of two Klein
factors is simplified by using $\eta_j\eta_{j-1}=-i\eta_{j+1}$ (recall that
lead $0\equiv 3$), we can write
\begin{equation}
H_{\cal B}=\sum_{a=1}^3 \left(i\Gamma e^{i \phi/3} \eta_{a}
\;e^{{i}{\vec K}_a\cdot{\vec \Phi}}
 + \mbox{h.c.}\right)+r\sqrt{3}\; \partial_x \Phi_0\Big{|}_{x=0}
\;,
\label{H_B-bosonic2}
\end{equation}
where 
\be \vec \Phi=(\Phi_1,\Phi_2),\label{vecPhidef}\ee
 and 
\begin{eqnarray}
\vec{K}_1 &&= (-1/2,\sqrt{3}/2)
\nonumber
\\
\vec{K}_2 &&= (-1/2, -\sqrt{3}/2)
\label{eq:Kvec}
\\
\vec{K}_3 &&= (1,0)
\; .
\label{Kdef}\nonumber
\end{eqnarray}

The boundary interactions lead to renormalization to an infrared fixed
point with a different boundary condition from Neumann on the two
fields $\Phi_{1,2}$. However, current conservation at the junction
requires the ``center of mass'' field $\Phi_0$ to always obey Neumann
boundary conditions.  Thus the remaining degrees of freedom at the
boundary comprise the two component boson field
$\vec{\Phi}=(\Phi_1,\Phi_2)$. Notice that the term proportional to $r$
in Eq.~(\ref{H_B-bosonic2}) is a constant due to the boundary
condition on $\Phi_0$, and hence it can be dropped.
Hence, the boundary action can finally be expressed as
\begin{equation}
S_{\cal B}=
i\Gamma e^{i\phi/3}
\int d\tau \sum_{a=1}^3 \eta_a \;e^{i \vec{K}_a \cdot \vec{\Phi}}
+ \mbox{h.c.}
\label{eq:SB-set-up}
\end{equation}
The scaling dimension of the hopping term in the disconnected limit
($\Gamma=0$) is calculated by standard methods~\cite{Kane} as $1/g$,
which is not affected by the Klein factors nor the magnetic flux. The
hopping term is thus irrelevant for $g<1$,
and the N fixed point corresponding to disconnected wires
is stable. 

When $g>1$, the hopping term is relevant and the N fixed point is unstable.
The system is renormalized into an infrared fixed point different from the N
fixed point.  The infrared fixed point should correspond to some conformally
invariant boundary condition.  Identification of such a fixed point in
general is a rather difficult problem, even in the pure bosonic problem
without the Klein factors as discussed in
Refs.~\cite{Kane,Akira2004,YiKane,QBM}.

In the present case, moreover, additional complicacy arises from the Klein
factors, namely due to the Fermi statistics of the electrons.  In the
following Sections, using various approaches, we study possible infrared
fixed points that are reached when the effect of the Klein factors is
included.

\section{Strong hopping limit}
\label{sec:potential}

As we have discussed in the previous section, for $g>1$ the hopping term is a
relevant perturbation to the N (disconnected) fixed point. We would like to
know where the system flows into, once the relevant hopping term is added to
the N fixed point. Tracing the RG flow exactly, starting from the N fixed
point down to the infrared fixed point is generally difficult, except in
fortunate cases where the flow is integrable. Even when we do not have the
exact solution of the integrable RG flow, it often happens that some of the
fixed points can be given exactly. Lacking the solution of the RG flow
itself, we generally resort to conjecturing the RG flow, based on the known
fixed points.

Thus the first step to construct the phase diagram is to
construct the candidates for the infrared fixed points.
Since the small hopping is relevant and grows under the RG
transformation,
a naive guess is that the hopping strength ``goes to infinity'' in the
infrared limit.
This simple guess turns out to be rather useful in many cases.
However, as it will become clear later in this paper,
the limit of the ``infinite hopping strength'' actually depends
on the representation of the given problem.

A natural idea may be to take
the hopping amplitude $\tilde{\Gamma}$ to infinity in the
original electron representation Eq.~(\ref{eq:eHam}).
However, in the presence of interaction, it is difficult to know
what happens in the $\tilde{\Gamma} \rightarrow \infty$.
Instead, we could discuss the strong tunneling limit
by taking the hopping strength $\Gamma$ to infinity,
{\em in the bosonized representation} Eq.~(\ref{eq:SB-set-up}).
This is generally {\em different} from taking the hopping amplitude
to infinity in the original electron models.

This approach was taken in Ref.~\cite{Nayak} for the resonant dot model of
the junction. Here we apply a similar argument to our model, namely the
junction with direct hoppings between the wires, or equivalently the ``ring''
junction.
The hopping term in the bosonized representation
corresponds to the boundary potential for $\vec{\Phi}$.
Thus, the field $\vec{\Phi}$
would be pinned to the potential minimum at the boundary
in the limit $\Gamma \rightarrow \infty$.
Namely, the strong hopping limit corresponds to the
Dirichlet boundary condition on $\vec{\Phi}$.
However, in fact, there are different types of Dirichlet boundary
conditions (or more precisely, boundary states) with different
stability.
All the Dirichlet boundary conditions, on the other hand,
give rise to the same conductance tensor.
As was discussed by Nayak et al.~\cite{Nayak},
the Dirichlet boundary conditions
exhibits an enhanced conductivity which may be understood as
a result of Andreev reflection (see Ref.~\cite{Sandler}).

In order to analyze the stability of the Dirichlet boundary
condition, we have to identify the potential minima
of the hopping action Eq.~(\ref{eq:SB-set-up}) at the boundary.
As was emphasized by Nayak et al.~\cite{Nayak},
the stability of the Dirichlet boundary condition
is indeed affected by the presence of the Klein factors.

To demonstrate the importance of the Klein factors, let us first consider the
fictitious problem without them. If there were no Klein factors
in Eq.~(\ref{H_B-bosonic}) to begin with, the boundary action would read
\begin{equation}
S_{\cal B} = - 2 \Gamma \int d\tau
\sum_{a=1}^3 \cos{\left(\vec{K}_a \cdot \vec{\Phi} + \phi/3\right)} .
\label{eq:SBnoKlein}\end{equation}
This is equivalent to the model studied by Yi and Kane~\cite{YiKane} (see
also Ref.~\cite{QBM}) in the context of the quantum Brownian motion on a
triangular/honeycomb lattice. The minima of this boundary potential is given
by a honeycomb lattice with lattice constant (distance between nearest
neighbors) $4 \pi /3$ if $\phi/3 \equiv \pi ({\rm mod} 2\pi)$, and a
triangular lattice with lattice constant $4 \pi/\sqrt{3}$ otherwise. 

The stability of the $\Gamma \to \infty$ fixed point is 
conveniently studied after 
integrating out the fields $\vec \Phi (\tau ,x)$ for $x>0$, 
thus deriving an effective boundary action. [A more detailed 
discussion in a related context is given in 
Subsubsection (\ref{subsubsec:rel_DHM}).]  Consider a general 
boundary field $\vec \Phi (\tau)$, in the imaginary 
time formalism at temperature $1=1/\beta$, with Fourier expansion:
\be \vec \Phi_b (\tau ) = 
{1\over \beta}\sum_n \vec \Phi (\omega_n) e^{i\omega_n\tau}.\ee
(Here $\omega_n\equiv 2\pi n/\beta$.)  The solution of the 
classical equations of motion, $\partial^2\vec \Phi (\tau ,x)=0$ 
with boundary condition $\vec \Phi (\tau ,0)=\Phi_b(\tau)$,
(defined for $x>0$ only) is 
\be \vec{\Phi}^\cl (\tau ,x)=
{1\over \beta}\sum_n \vec \Phi_n e^{i\omega_n\tau -|\omega_n|x}.\ee
Substituting into Eq. (\ref{eq:S0}) gives the 
non-interaction term in the boundary action (at $\beta \to \infty$):
\be S_0=g\int {d\omega \over (2\pi )^2}|\omega |\vec \Phi^*(\omega )
\cdot \vec \Phi (\omega ).\ee
The full boundary action also includes the potential term of Eq. 
(\ref{eq:SBnoKlein}), which can be written entirely in terms of
 the boundary field $\Phi_b(\tau )$, the Fourier transform of 
$\vec \Phi (\omega )$. When $\Gamma \to \infty$ the path integral 
is dominated by configurations where the fields $\Phi_b(\tau )$
stay close to one of the minima of the potential.  Rare tunneling
events (instantons) occur in which $\Phi_b (\tau )$ goes 
from one minimum to a neighboring one. 
In general, multi-instanton configurations must be considered in 
the dilute gas approximation. This is reviewed in some 
detail in a related context in Subsec. (\ref{subsec:mapping}).
 The conclusion is that, for a multi-instanton configuration 
in which $\vec \Phi$ tunnels between minima of $V$ separated 
by $\vec M_i$ at time $\tau_i$, the instanton interaction term 
in the action, for large time separations, $|\tau_i-\tau_j|$, 
takes the form:
\be S_{\hbox{\scriptsize int}}=-{g\over (2\pi )^2}\sum_{i>j}\vec M_i\cdot 
\vec M_j\ln (\tau_i-\tau_j)^2.\ee 
This implies that a single tunneling process corresponds 
to an operator in the effective Hamiltonian, 
at the infinite $\Gamma$ fixed point, with a RG scaling 
dimension:
\be \Delta = {g\over (2\pi )^2}|\vec M_i|^2.\ee
This is $4g/9$ and $4g/3$ respectively for the
honeycomb and triangular cases. Thus these Dirichlet boundary conditions are
stable for $g > 9/4$ and $g > 3/4$ respectively. This would be troublesome
especially in the noninteracting case $g=1$, as the ``triangular'' Dirichlet
boundary condition is stable. Because the Dirichlet boundary condition
corresponds to the enhanced conductance due to the Andreev reflection, it
should not be realized if there is no interaction.

However, the correct bosonized representation for the junction of electron
systems should include the Klein factors (as Pauli matrices) as in
Eq.~(\ref{eq:SB-set-up}), to reproduce 
the effect of the Fermi statistics of electrons.
Following Ref.~\cite{Nayak}, the classical value for the boundary field
$\vec{\Phi}$ is obtained by diagonalizing the boundary term in the auxiliary
two-dimensional Hilbert space on which the Pauli matrices act. The resulting
potential  is
\begin{equation}
 V(\vec{\Phi}) = \pm \sqrt{ \sum_a
  \sin^2{\left(\vec{K}_a \cdot \vec{\Phi}+ {\phi}/{3}\right)}
  } ,
\end{equation}
with two branches corresponding to $\pm $ for a given boundary value of
$\vec{\Phi} $. The potential minima of $\vec{\Phi}$, in the $-$branch, are
those which make the argument of the square-root maximum. As in the case
without the Klein factors, the structure of the potential minima depends on
the flux $\phi$. For the time-reversal symmetric case $\phi=0, \pi$, the
potential minima form a honeycomb lattice with the nearest neighbor distance
$2 \pi / 3$. For other cases ($\phi \neq 0, \pi$) the potential minima is
instead a triangular lattice with the nearest neighbor distance $2 \pi /
\sqrt{3}$. The scaling dimension of the leading perturbation, which is
proportional to the nearest neighbor distance squared, is $g/9$ and $g/3$
respectively for the honeycomb and triangular cases.

As  was mentioned above, the ``strong hopping'' limit in the bosonized
representation corresponds to Dirichlet boundary condition for the boson
field $\vec{\Phi}$ in either case. However, the operator content (and thus
the stability) differs in these cases. Thus we distinguish them by referring
to the honeycomb ($\phi=0, \pi$) and the triangular ($\phi \neq 0, \pi$) cases
as \DN and \DP fixed points respectively. N of \DN
stands for Nayak et al., as it is identical to the one discussed in
Ref.~\cite{Nayak} for the
junction with the quantum dot. P of \DP stands for pair-tunneling; let us
now show that the \DP fixed point may be regarded as the limit of strong
tunneling of electron pairs.

We emphasize again that the strong hopping limit $\Gamma \to \infty$ in the
bosonized representation is different from the $\tilde \Gamma \to \infty$
limit in the original electron model. This is evident in the non-interacting
case $g=1$, where the original electron model should be always described by a
free electron scattering matrix. In this case, the strong hopping limit in
the bosonized representation gives the \DP or \DN fixed point, which
exhibits the Andreev reflection and thus cannot be described by free
electron scattering. For $g=1$ both \DP and \DN fixed points are
unstable. Thus, as far as the infrared stable fixed point is concerned, the
original electron model and the bosonized representation give consistent
conclusions.

If we introduce the hopping of electron pairs instead of
electrons, the boundary action should be given by
\begin{equation}
S'_{\cal B}=
 2 \Gamma_P
\int d\tau \sum_{a=1}^3 \;
\cos{\left( 2 \vec{K}_a \cdot \vec{\Phi} + {2\phi}/{3} \right)}
+ \mbox{h.c.} .
\label{eq:SB-pair}
\end{equation}
The sign of the pair-tunneling parameter $\Gamma_P$, may be deduced 
from second order perturbation theory in the single-electron 
tunneling parameter $\Gamma$. From the operator product expansion:
\be e^{i\vec K_a\cdot \vec \Phi (z)}e^{i\vec K_a\cdot \vec \Phi (0)}
\to C|z|^{2/g}e^{2i\vec K_a\cdot \vec \Phi (0)},\ee
for a positive constant, $C$, and taking into account that 
$\eta_i^2=-1$, we conclude that $\Gamma_P\propto \Gamma^2>0$. 
We note that the Klein factors are absent from the action.
This is natural, considering that the electron pair is a bosonic object.
For $\phi \neq 0,\pi$, 
the potential minima form a triangular lattice
with the nearest neighbor distance $2\pi/\sqrt{3}$.
Thus the boundary condition in the strong pair-tunneling limit
$\Gamma_P \rightarrow \infty$ is identified with \DP.
For $\phi=0,\pi$,
the potential minima form a honeycomb lattice
with the nearest neighbor distance $2 \pi / 3$.
Thus the strong pair-tunneling limit is again identified with
the more unstable \DN, in the time-reversal invariant cases
$\phi=0,\pi$.

Therefore, the strong pair-tunneling limit $\Gamma_P \rightarrow \infty$
gives the same boundary conditions as the strong hopping
limit $\Gamma \rightarrow \infty$.
As the pair-tunneling term would be generated through the
RG transformation even if the microscopic model only contains
the single electron hopping, this serves as a consistency check.

We note that
even though the junction with $\phi \neq 0, \pi$ is not
time reversal symmetric, the strong hopping limit for $\phi \neq 0,\pi$
is the \DP fixed point which is time-reversal symmetric in itself.
This is of course not a contradiction because
the symmetry may be higher at the infrared fixed point than in the
original microscopic model.
This symmetry restoration may be understood naturally by
considering the maximally time-reversal breaking
flux $\phi = \pm \pi/2$ for the electron hopping model.
At this value of the flux, the pair-tunneling model Eq.~(\ref{eq:SB-pair})
is actually time-reversal invariant, as the Aharonov-Bohm
phase for the electron pair is $\pi$ instead of $\pm \pi/2$.

What is more surprising is that the model with $\phi = 0, \pi$
apparently cannot reach the \DP fixed point, although
there is no symmetry forbidding it.
This was deduced from the structure of the potential minima
in the $\Gamma_P \rightarrow \infty$ limit, as
discussed above.
We note that this result relies on the assumption that
the sign of the coefficient $\Gamma_P$ of the potential does
not change under the RG flow.
A similar assumption was used in Ref.~\cite{YiKane} for the
related model of the quantum Brownian motion.
Of course, if $\Gamma_P$ is the only coupling constant in the
theory, it should not change sign because the
RG flow cannot cross the fixed point corresponding to the
Neumann boundary condition.
However, in the present problem, the validity of the assumption
is subtle. This is because one should, in principle, consider
both the single-electron hopping $\Gamma$ and
the pair hopping.$\Gamma_P$ simultaneously in
discussing the RG flow.

While we have not yet resolved this question, 
the above arguments lead to the following result on the
phase diagram of the system.

\subsection{time-reversal symmetric $\phi = 0, \pi$ case}

The N fixed point (decoupled wire) is stable for $g<1$ and
unstable for $g>1$.
The \DN fixed point is unstable for $g<9$ and stable for $g>9$.
Therefore, the system should flow to some ``intermediate'' stable
fixed point for $1<g<9$. 
The intermediate fixed point corresponds to neither Dirichlet nor
Neumann boundary conditions.

This conclusion is similar to that in Ref.~\cite{Nayak}
for a different setup of the junction.
The only difference is that the N fixed point in their model
is unstable for $g>1/2$ because of the resonant transmission
through the dot.

\subsection{time-reversal breaking $\phi \neq 0, \pi$ case}

In our model, we have the flux $\phi$ as a new parameter which
breaks the time-reversal symmetry, and can lead to new physics.

In fact, as it is seen above, the new Dirichlet fixed point
\DP can be realized for $\phi \neq 0, \pi$.
It is stable for $g>3$, so we expect a generic junction
without time-reversal symmetry to flow to \DP for $g>3$.
We note that although the \DP fixed point itself is time-reversal
invariant, it seems that the time-reversal symmetry in
the original junction model must be broken to reach \DP.

For $1<g<3$, as both the N and \DP fixed points are unstable,
there must be an intermediate stable fixed point to which the
generic system flows.
However, the present argument 
does not shed light on the nature of the stable fixed
point.
In the following sections, we will argue that the
stable fixed point for $1<g<3$ is solvable with
different techniques that we present in this paper.

\section{Connection with the dissipative Hofstadter model}
\label{sec:DHM}
As we have seen in the last section, there are infrared stable fixed
points that cannot be identified with either weak- or strong- hopping
limit in the bosonized representation Eq.~(\ref{eq:SB-set-up})
with the Klein factors.

The presence of the phase factors due to the fermionic statistics,
encoded in the Klein factors, makes the
analysis more difficult than that for the standard problems with only
bosonic terms in the boundary action. The extra phases due to the
fermionic statistics appear, for example, if one attempts a
perturbative expansion in the hopping amplitude $\Gamma$. The
coefficients appearing in each order in perturbation theory are
different than those obtained without the phase factors, and hence
they alter the nature of the strong hopping limit.

In this section, we develop an alternative representation,
in which the fermionic statistics of electrons are encoded
without using the Klein factors.
Namely, we will show that the perturbative series in the
hopping amplitude $\Gamma$ for the Y-junction problem is identical to
the series obtained for the dissipative Hofstadter model (quantum
motion of a single particle under a magnetic field and a periodic
potential, subject to dissipation) studied by Callan and Freed in
Ref.~\cite{CF}. In the dissipative Hofstadter action, phase factors
arise from the motion in a magnetic field, but the action contains
solely bosonic terms, and hence it is amenable to standard methods
for determining the strong coupling behavior, such as instanton
expansions.

\subsection{Generalized Coulomb gas from the Y-junction}

We consider the perturbation theory in the hopping parameter
$\Gamma$, starting from the bosonized
representation Eq.~(\ref{eq:SB-set-up}) with the Klein factor.
The partition function $Z_Y$ of the system can be expanded as 
\begin{widetext}
\begin{eqnarray}
Z_Y=\sum_{n}
\;\Gamma^n
\int_{\tau_{n-1}}^\infty \!\!\!\!\!  d\tau_{n}\;
\int_{\tau_{n-2}}^{\tau_{n}} \!\!\!\!\!  d\tau_{n-1}\;
\cdots
\int_{-\infty}^{\tau_{2}} \!\!\!\! d\tau_{1}\;
\sum_{\{L_j\}}
\left[
\zeta_{\vec L_n}
\cdots
\zeta_{\vec L_2}\;
\zeta_{\vec L_1}
\right]
\;
\langle
e^{i \vec{L}_{n} \cdot \vec{\Phi}(\tau_n)}
\ldots
e^{i \vec{L}_{2} \cdot \vec{\Phi}(\tau_2)}
e^{i \vec{L}_{1} \cdot \vec{\Phi}(\tau_1)}
\rangle_0
\;,
\label{eq:Z_Y}
\end{eqnarray}
\end{widetext}
where ${\vec L}_j=q_j{\vec K}_{a_j}$ is one of the six vectors $\pm
\vec{K}_{1,2,3}$ ($q_j=\pm 1$ and $a_j=1,2,3$), and $\langle \ 
\rangle_0$ denotes the expectation value with the Neumann boundary
condition on $\vec{\Phi}$. The $\zeta_{\vec
L_j}=(-i)^{q_j}\,e^{iq_j\phi/3}\,\eta_{a_j}$ in the $[\ldots]$ term in
Eq.~(\ref{eq:Z_Y}) keep track of the phase factors.

\begin{figure}
\begin{center}
\includegraphics[width=0.45\linewidth]{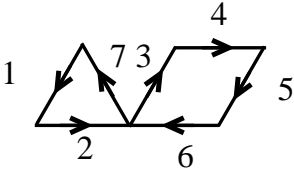}
\caption{A typical closed path occurring in perturbative expansion
of partition function}
\label{fig:pathex}
\end{center}
\end{figure}

Now, each term in the expansion Eq.~(\ref{eq:Z_Y}) corresponds to a
closed directed path
formed by the vectors
${\vec L}_1,{\vec L}_2,\ldots, {\vec L}_n$,
on the triangular lattice
spanned by the primitive vectors ${\vec K}_1$ and ${\vec K}_2$.
An
example of a closed path is shown in Fig.~\ref{fig:pathex}. A charge vector
${\vec L}_j$ is associated to each vertex operator $e^{i{\vec L}_j\cdot
\vec\Phi}$, and the condition that the sequence of vectors must form a
closed path is a consequence of the charge neutrality condition of the
Coulomb gas expansion.

\begin{figure}
\begin{center}
\includegraphics[width=0.4\linewidth]{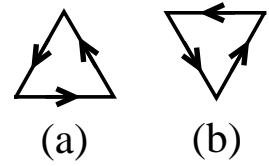}
\caption{(a) a counter-clockwise loop around an ``up triangle''. 
(b) A counter-clockwise loop around a ``down triangle''.}
\label{fig:updown}
\end{center}
\end{figure}

Let us now argue that the total phase factor for a given path ${\cal
L}=({\vec L}_1,{\vec L}_2,\ldots, {\vec L}_n)$,
\begin{equation}
e^{i\Upsilon_{\cal L}}=
\left[
\zeta_{\vec L_n}
\cdots
\zeta_{\vec L_2}\;
\zeta_{\vec L_1}
\right]
\; ,
\end{equation}
is the sum of the phase contributions from all the elementary
triangles enclosed by the loop. The phase contributions from ``up''
and ``down'' triangles, however, are different. The phase factor for a
counterclockwise loop on the up triangle [see part (a) of 
Fig.~\ref{fig:updown}]  is  determined to be
\begin{equation}
 e^{i \upsilon_{\bigtriangleup}}=
(-i \eta_3) e^{i \phi/3} (-i \eta_2) e^{i \phi/3} (-i \eta_1) e^{i \phi/3}
 = e^{i \phi}\label{up}
\end{equation}
while that for the counterclockwise loop on the down triangle
[see part (b) of Fig.~\ref{fig:updown}] is
\begin{equation}
e^{i \upsilon_{\bigtriangledown}}=
(i \eta_3) e^{- i \phi/3} (i \eta_2) e^{- i \phi/3}
(i \eta_1) e^{- i \phi/3}  = e^{i (\pi - \phi)}.
\label{down}\end{equation}
Namely, loops on the triangular lattice pick up a phase as if there is
a staggered magnetic flux of $\phi$ and $\pi - \phi$ in each
elementary triangle. The total phase accumulated due to the loop
${\cal L}$ is thus
\begin{equation}
\Upsilon_{\cal L}=
N_\bigtriangleup({\cal L})\;\upsilon_\bigtriangleup+
N_\bigtriangledown({\cal L})\;\upsilon_\bigtriangledown
\; ,
\label{phaseY}\end{equation}
where $N_\bigtriangleup({\cal L})$ and $N_\bigtriangledown({\cal L})$
are the {\it net} numbers of up and down triangles enclosed by the path 
(a triangle contributes $\pm 1$ to $N_\bigtriangleup$ 
and $N_\bigtriangledown$ depending on whether
it is traversed clockwise or counter-clockwise).

The argument why the phase for a given path  ${\cal L}$ is the sum of
the phases due to the enclosed elementary triangles is constructed
inductively.  First consider the special case where the closed path
consists of a single non-intersecting
closed loop.  This implies that each vertex on the loop is visited
a single time except for the first vertex which is visited twice,
at the beginning and end of the loop. The phase from this
loop is equal to the sum of phases of elementary triangular
loops into which it can be decomposed.
 Say it is so for all paths enclosing up to $m$ triangles,
and then consider a loop enclosing $m+1$ triangles; we must argue that
the phase accumulated around the loop enclosing $m+1$ triangles is
 equal to the sum of the phases enclosed by two smaller area paths
with $m_1$ and $m_2$ triangles ($m_1+m_2=m+1, m_{1,2}\le m$).

\begin{figure}
\begin{center}
\includegraphics[width=0.85\linewidth]{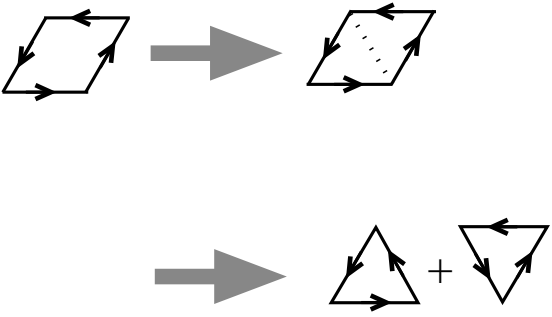}
\caption{Decomposition of a non-crossing loop into
elementary triangles}
\label{fig:loopdec}
\end{center}
\end{figure}

Consider then an arbitrary non-intersecting loop ${\cal L}=({\vec L}_1,{\vec
L}_2,\ldots, {\vec L}_n)$, which we break into two loops by adding an
internal segment that splits the loop into two
loops. Let the segment be the
sequence of vectors ${\vec L'}_1,{\vec L'}_2,\dots,{\vec L'}_p$.
 It is easy to show
that
\begin{equation}
\left[
\zeta_{-\vec L'_1}\;
\zeta_{-\vec L'_2}\;
\cdots
\zeta_{-\vec L'_p}
\right]
\left[
\zeta_{\vec L'_p}
\cdots
\zeta_{\vec L'_2}\;
\zeta_{\vec L'_1}
\right]
=\openone
\; ,
\end{equation}
so that
\begin{widetext}
\begin{equation}
e^{i\Upsilon_{\cal L}}=
\left[
\zeta_{\vec L_n}
\cdots
\zeta_{\vec L_{j+2}}\;
\zeta_{\vec L_{j+1}}\;
\;
\zeta_{-\vec L'_1}\;
\zeta_{-\vec L'_2}\;
\cdots
\zeta_{-\vec L'_p}
\right]
\left[
\zeta_{\vec L'_p}
\cdots
\zeta_{\vec L'_2}\;
\zeta_{\vec L'_1}
\zeta_{\vec L_j}
\cdots
\zeta_{\vec L_2}\;
\zeta_{\vec L_1}
\right]
=e^{i\Upsilon_{{\cal L}_1}}\;e^{i\Upsilon_{{\cal L}_2}}
\; ,
\end{equation}
\end{widetext}
for two loops ${\cal L}_{1,2}$ that encircle areas smaller than the
original loop ${\cal L}$ , {\it i.e.}, $l_{1,2}\le m$ and with
$l_{1}+l_{2}=m+1$, as required by the induction argument.
An example of a decomposition of a non-crossing loop into
elementary triangles is given in Fig.~\ref{fig:loopdec}.

\begin{figure}
\begin{center}
\includegraphics[width=0.65\linewidth]{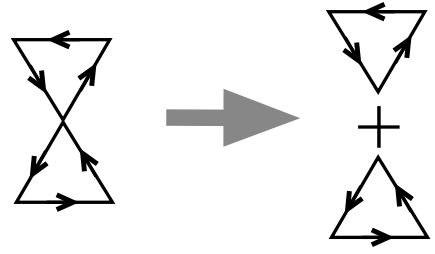}
\caption{Decomposition of a path into
non-crossing loops}
\label{fig:pathdec}
\end{center}
\end{figure}

Finally we must argue that an arbitrary closed path can be
decomposed into a set of non-intersecting loops such that the
total phase of the path is the sum of phases for each
non-intersecting loop. This can be seen as follows.
Consider following an arbitrary oriented path. Consider the first
time that any vertex on the path is visited for a second time.
Label this vertex $V_1$.
The sequence of edges between the first and second visit to $V_1$
defines a non-intersecting closed loop, $L_1$. (This follows since
the first intersection on the path occurs at $V_1$ at the
instant of the second visit.) The phase due to this closed
non-intersecting loop may be calculated unambiguously and
makes an additive contribution to the total phase of
the path. This follows since the product of the Klein factors
around a closed loop is always proportional to the identity matrix.
Now excise the closed loop, $L_1$, from the path. The remaining path
remains closed.  Consider the first time that any vertex
on this excised path is visited twice, at some other
vertex, $V_2$. This defines another closed loop $L_2$ which
we again excise. We continue in this way until we finally
revisit the first point on the path. This final revisitation
defines a final closed loop $L_n$.
An example of the decomposition of a path into
non-crossing loops is given in Fig.~\ref{fig:pathdec}.
Note that when a path (or an
excised path)
makes a U-turn and retraces a link this counts as a closed
loop of length 2 and zero area. The total phase
associated with the path is the sum of phases of each
non-intersecting closed loop.  These phases can each be
expressed as a sum of phases of elementary triangles into
which they can be decomposed. Adding up the net number of
up triangles and down triangles in all the non-intersecting loops
gives us Eq. (\ref{phaseY}) for the total phase.

Let us now turn to the correlation function entering in Eq.~(\ref{eq:Z_Y}):
\begin{eqnarray}
\lefteqn{
\langle
e^{i \vec{L}_{n} \cdot \vec{\Phi}(\tau_n)}
\ldots
e^{i \vec{L}_{2} \cdot \vec{\Phi}(\tau_2)}
e^{i \vec{L}_{1} \cdot \vec{\Phi}(\tau_1)}
\rangle_0 = } \nonumber \\
& \delta_K(\sum_j \vec{L}_j)&
 \exp{\left[ \frac{1}{g}\sum_{j>k} \vec{L}_j \cdot \vec{L}_k
      \ln{|\tau_j - \tau_k|}^2 \right]},
\label{eq:loop}
\end{eqnarray}
where $\delta_K(\vec{L}) = 1$ if $\vec{L}=0$, and 
$\delta_K(\vec{L}) = 0$ otherwise. 
The factor $\delta_K(\sum_j \vec{L}_j)$ enforcing the charge
vectors ${\vec L}_j$ must form a closed path, as we mentioned above,
explicitly follows from of charge conservation in the Coulomb gas
expansion.

Using this expression, in conjunction with the phase factor
we discussed above, we can rewrite the partition function as:
\begin{widetext}
\begin{eqnarray}
Z_Y=\sum_{n}
\;\Gamma^n
\int_{\tau_{n-1}}^\infty \!\!\!\!\!  d\tau_{n}\;
\int_{\tau_{n-2}}^{\tau_{n}} \!\!\!\!\!  d\tau_{n-1}\;
\cdots
\int_{-\infty}^{\tau_{2}} \!\!\!\! d\tau_{1}\;
\sum_{{\cal L}=(\vec L_1,\dots,\vec L_n)}
\delta_K(\sum_j \vec{L}_j)\;\;e^{i\Upsilon_{\cal L}}
\;\;
e^{\frac{1}{g}\sum_{j>k} \vec{L}_j \cdot \vec{L}_k
      \ln{|\tau_j - \tau_k|}^2 }
\;.
\label{eq:Z_Y2}
\end{eqnarray}
\end{widetext}
Thus, the partition function $Z_Y$ is written in a similar form
to the so-called Coulomb gas (particles interacting with the
logarithmic potential) 
in one dimension.
The difference from the standard Coulomb gas is in the
phase factor $e^{i\Upsilon_{\cal L}}$, which is indeed the consequence
of the Fermi statistics of the electron and of the Aharonov-Bohm phase.
We may call this a ``generalized Coulomb gas.''
In fact, below we show that this is indeed identical to the generalized
Coulomb gas which appeared in a quite different problem --
the dissipative Hofstadter model on a triangular lattice.

\subsection{Generalized Coulomb gas from the dissipative Hofstadter model}

Dissipative quantum mechanics (also known as
quantum Brownian motion)
is the problem
of the quantum motion of a single particle subject to 
dissipation (see Ref.~\cite{Leggett-etal} for a review).
In the presence of any periodic potential, single-particle quantum mechanics
implies that the particle never localizes due to the tunneling effect.
However, if the dissipation is sufficiently strong, the
tunneling can be suppressed and the particle may be localized.
When the dissipation is Ohmic, there is a surprising mapping
of the problem to the boundary problem of the free boson field
theory. Here, the bulk of the free boson is related to
the heat bath in the dissipative quantum mechanics, and only the
boundary field is related to the original particle. 
By integrating over the bulk degrees of freedom, the boundary
problem can be reduced to the one-dimensional Coulomb gas.

Dissipative quantum mechanics in the presence of the periodic potential and 
the magnetic field
is known as the dissipative Hofstadter problem.
It is again related to the boundary problem of the free boson,
or the following one-dimensional generalized Coulomb gas.
The ``free'' action for the dissipative Hofstadter model, namely
the effective action in the absence of the potential (but with
the magnetic field and the dissipation), is
\begin{equation}
S_0[\vec{X}] = \frac{1}{2} \int \frac{d\omega}{(2\pi)^2}
\left[
\alpha |\omega| \delta_{\mu \nu} + \beta \omega \epsilon_{\mu \nu} \right]
X^*_{\mu}(\omega) X_{\nu}(\omega),
\label{eq:S_0}\end{equation}
where $\mu,\nu=1,2$ and $\epsilon_{12}=-\epsilon_{21}=1$,
$\epsilon_{11}=\epsilon_{22}=0$. $\alpha$ and $\beta$ are
related to the dissipation and the magnetic field, respectively.
This determines the propagator:
\begin{eqnarray}
\lefteqn{
 D_{\mu \nu}(\tau) = \langle X_{\mu}(\tau) X_{\nu}(0) \rangle = }
\nonumber \\
&&  \!\!\!\!\!\!
- \frac{\alpha}{\alpha^2 + \beta^2}\ln{\tau^2} \;\delta_{\mu \nu}
+ i \pi \frac{\beta}{\alpha^2+\beta^2}\; \sgn{\,\tau} \;\epsilon_{\mu \nu} .
\label{eq:D}\end{eqnarray}
We now introduce a potential term with the periodicity of the
triangular lattice (notice the difference from the ``rectangular'' one
in Ref.~\cite{CF}), as
\begin{equation}
S_V[\vec{X}]
= - V e^{i \delta/3}\int d\tau \sum_{a=1}^3
e^{i \vec{K}_a \cdot \vec{X}} + \mbox{c.c.},
\label{eq:CFpot}
\end{equation}
where $V$ and $\delta$ are chosen to be real.
Expanding the partition function in powers of $V$, one obtains
\begin{widetext}
\begin{eqnarray}
Z_\shbox{DHM}=\sum_{n}
\;V^n
\int_{\tau_{n-1}}^\infty \!\!\!\!\!  d\tau_{n}\;
\int_{\tau_{n-2}}^{\tau_{n}} \!\!\!\!\!  d\tau_{n-1}\;
\cdots
\int_{-\infty}^{\tau_{2}} \!\!\!\! d\tau_{1}\;
\sum_{\{L_j\}}
\left[
\kappa_{\vec L_n}
\cdots
\kappa_{\vec L_2}\;
\kappa_{\vec L_1}
\right]
\;
\langle
e^{i \vec{L}_{n} \cdot \vec{X}(\tau_n)}
\ldots
e^{i \vec{L}_{2} \cdot \vec{X}(\tau_2)}
e^{i \vec{L}_{1} \cdot \vec{X}(\tau_1)}
\rangle_0
\;,
\label{eq:Z_DHM}
\end{eqnarray}
\end{widetext}
where again
${\vec L}_j=q_j{\vec K}_{a_j}$
is one of the six vectors
$\pm \vec{K}_{1,2,3}$
($q_j=\pm 1$ and $a_j=1,2,3$), and the
$\kappa_{\vec L_j}=e^{iq_j\delta/3}$ in the $[\ldots]$ term keep track
of phase factors due to $\delta$. The correlation function is given by
\begin{eqnarray}
\lefteqn{\langle
e^{i \vec{L}_n \cdot \vec{X}(\tau_n)}
\ldots
e^{i \vec{L}_2 \cdot \vec{X}(\tau_2)}
e^{i \vec{L}_1 \cdot \vec{X}(\tau_1)}
\rangle_0 = } \nonumber \\
&\delta_K(\sum_j \vec{L}_j) &  \exp{\Big[  \frac{\alpha}{\alpha^2+\beta^2}
    \sum_{j>k} \vec{L}_j \cdot \vec{L}_k
      \ln{|\tau_j - \tau_k|}^2  } 
\nonumber \\
&&   \!\!\!\!\!\!\!\! \!\!\!\!\! \!\!\!\!\!
-i \pi \frac{\beta}{\alpha^2+\beta^2} \sum_{j>k}\vec{L}_j \times \vec{L}_k
\;\sgn(\tau_j - \tau_k) \Big].
\label{eq:loopCF}
\end{eqnarray}
(Because of the time ordering, notice that $\sgn(\tau_j - \tau_k)=1$
for all terms in the sum.)

Notice that the $\kappa_{{\vec L}_j}$ in the $[\ldots]$ of
Eq.~(\ref{eq:Z_DHM}) commute with one another, in contrast to the
Klein factors $\zeta_{{\vec L}_j}$ in the $[\ldots]$ of
Eq.~(\ref{eq:Z_Y}). However, the $\langle\ \rangle_0$ term in
Eq.~(\ref{eq:loopCF}) contains phase factors not present in
Eq.~(\ref{eq:loop}). While the phases entering the perturbative
expansions for $Z_Y$ and $Z_\shbox{DHM}$
are arising from different sources,
their net effect turns out to be identical, as we show below.

The phases in the expansion of the dissipative Hofstadter model can
also be associated with the closed paths ${\cal L}=({\vec L}_1,{\vec
L}_2,\ldots, {\vec L}_n)$. The contribution from the
$\vec{L}_j\times\vec{L}_k$ terms amounts to a flux through the
oriented area ${\cal A}_{\cal L}$ enclosed by ${\cal L}$.
To see this we again decompose any closed path into a set of
closed non-intersecting loops. The oriented area of such
a non-intersecting loop is
\begin{equation}
{\cal A}_{\cal L}=-\frac{1}{2}\sum_{j>k}\vec{L}_j\times\vec{L}_k
\; .
\end{equation}
Thus, the accumulated phase $\Lambda_{\cal L}$ due to the magnetic
flux is
\begin{eqnarray}
\Lambda_{\cal L}&&=2\pi \frac{\beta}{\alpha^2+\beta^2} \;{\cal A}_{\cal L}
\\
&&= 2\pi \frac{\beta}{\alpha^2+\beta^2}\; A \left[
N_\bigtriangleup({\cal L})+
N_\bigtriangledown({\cal L})
\right]
\;,
\end{eqnarray}
where ${A}=\frac{\sqrt{3}}{4}$
is the area of an elementary triangle,
and $N_\bigtriangleup({\cal L})$, $N_\bigtriangledown({\cal L})$
are the {\it net} numbers of enclosed elementary up and down triangles,
as before.

Now we turn to the phase due to the $[\dots]$ in
Eq.~(\ref{eq:Z_DHM}). The total phase factor for a given closed path ${\cal
L}=({\vec L}_1,{\vec L}_2,\ldots, {\vec L}_n)$,
\begin{equation}
e^{i\;\Xi_{\cal L}}=
\left[
\kappa_{\vec L_n}
\cdots
\kappa_{\vec L_2}\;
\kappa_{\vec L_1}
\right]
\; ,
\end{equation}
is the sum of the phase contributions from all the elementary
triangles enclosed by the path. The argument for this is carried out
by induction, identically to what we have used for the products of the
$\zeta_{\vec{L}_j}$ above.

The phase contributions from ``up'' and ``down'' triangles are again
different. The phase factor for a counterclockwise loop on the
up triangle is determined determined to be
\begin{equation}
e^{i \delta/3}\;e^{i \delta/3}\;e^{i \delta/3}=
e^{i \delta}\label{upDHM}
\end{equation}
while that for the counterclockwise loop on the down triangle is
\begin{equation}
e^{-i \delta/3}\;e^{-i \delta/3}\;e^{-i \delta/3}=
e^{-i \delta}\label{downDHM}
\;.
\end{equation}
Namely, loops on the triangular lattice pick up a phase as if there is
a staggered magnetic flux of $\delta$ and $-\delta$ in each elementary
triangle. The total phase accumulated due to the loop ${\cal L}$ is
thus
\begin{equation}
\Xi_{\cal L}=
N_\bigtriangleup({\cal L})\;\delta-
N_\bigtriangledown({\cal L})\;\delta
\; ,
\end{equation}
where $N_\bigtriangleup({\cal L})$ and $N_\bigtriangledown({\cal L})$
are the net numbers of up and down triangles enclosed by the loop.

The total phase factor due to the loop ${\cal L}$ is the sum
$\tilde\Upsilon_{\cal L}=\Lambda_{\cal L}+\Xi_{\cal L}$:
\begin{equation}
\tilde\Upsilon_{\cal L}=
N_\bigtriangleup({\cal L})\;\tilde\upsilon_\bigtriangleup+
N_\bigtriangledown({\cal L})\;\tilde\upsilon_\bigtriangledown
\; ,
\end{equation}
where
\begin{eqnarray}
\tilde\upsilon_\bigtriangleup&&=2\pi \frac{\beta}{\alpha^2+\beta^2} A+\delta
\label{vup}\\
\tilde\upsilon_\bigtriangledown&&=2\pi \frac{\beta}{\alpha^2+\beta^2} A-\delta
\; .\label{vdown}
\end{eqnarray}

We can now rewrite the partition function for the dissipative
Hofstadter model on the triangular lattice as:
\begin{widetext}
\begin{eqnarray}
Z_\shbox{DHM}=\sum_{n}
\;V^n
\int_{\tau_{n-1}}^\infty \!\!\!\!\!  d\tau_{n}\;
\int_{\tau_{n-2}}^{\tau_{n}} \!\!\!\!\!  d\tau_{n-1}\;
\cdots
\int_{-\infty}^{\tau_{2}} \!\!\!\! d\tau_{1}\;
\sum_{{\cal L}=(\vec L_1,\dots,\vec L_n)}
\delta_K(\sum_j \vec{L}_j)\;\;e^{i\tilde\Upsilon_{\cal L}}
\;\;
e^{\frac{\alpha}{\alpha^2+\beta^2}\sum_{j>k} \vec{L}_j \cdot \vec{L}_k
      \ln{|\tau_j - \tau_k|}^2 }
\;.
\label{eq:Z_DHM2}
\end{eqnarray}
\end{widetext}
Following Callan and Freed~\cite{CF},
we may regard the DHM partition function
as the partition function of a generalized classical 1-dimensional Coulomb
gas.  $\tau_j$ is the position of the j$^{th}$ particle
and $\vec L_j$ is its generalized ``charge'', a vector quantity
taking on six possible values.

\subsection{Mapping of the Y-junction to the dissipative Hofstadter model}
\label{subsec:mapping}
The expression for $Z_\shbox{DHM}$ in Eq.~(\ref{eq:Z_DHM2}) is identical to
that for $Z_{Y}$ in Eq.~(\ref{eq:Z_Y2}),
provided the following conditions are satisfied.
First we need the relation
\begin{equation}
\frac{\alpha}{\alpha^2+\beta^2} = \frac{1}{g},
\label{eq:alpha}
\end{equation}
for the scaling dimension of the leading perturbation to be matched
between the two theories.
Furthermore, in order to match the phase factor, 
we require
$\tilde\Upsilon_{\cal L}\equiv\Upsilon_{\cal L} \mod{2\pi}$, or
equivalently
\begin{eqnarray}
&&\tilde\upsilon_\bigtriangleup \equiv
\upsilon_\bigtriangleup \; \mod{2\pi},
\label{eq:upcond}
\\
&&\tilde\upsilon_\bigtriangledown \equiv
\upsilon_\bigtriangledown \; \mod{2\pi}
\label{eq:downcond}
\;,
\end{eqnarray}
i.e.
\bea
{\sqrt{3}\pi \over 2}{\beta \over \alpha^2+\beta^2}+\delta
&=&\phi \ \ \mod{2\pi} , \nonumber \\
{\sqrt{3}\pi \over 2}{\beta \over \alpha^2+\beta^2}-\delta
&=&\pi -\phi \ \ \mod{2\pi} .
\eea
The above two equations imply:
\bea
\sqrt{3}  \frac{\beta}{\alpha^2 + \beta^2} &=& (2n - 1) \nonumber \\
\delta &=& \phi + \pi (1/2-n).
\label{eq:n}\eea
for an arbitrary integer, $n$. Because the phases, $\phi$ and $\delta$, are
only defined modulo $2\pi$, there is actually an infinite number of
different choices of $\alpha$ and $\beta$ labeled by the integer $n$.
Eqs. (\ref{eq:alpha}) and (\ref{eq:n}) give:
\bea \alpha &=& {3g\over 3+(2n-1)^2g^2}\nonumber \\
\beta &=& {\sqrt{3}(2n-1)g^2\over 3+(2n-1)^2g^2}.\label{eq:ndef}\eea

To understand the ambiguity in the integer $n$, 
let us define the Coulomb gas charge density
\be \vec \rho (\tau )\equiv \sum_{j=1}^N\vec L_j \; \delta (\tau -\tau_j).\ee
The generalized ``energy'', $E$, of an
arbitrary classical gas configuration is determined by the $N$
positions, $\tau_j$ and charges, $L_j$.  This generalized
``energy'' is now a complex quantity and $\exp (-E/T)$
(where $T$ is the classical temperature)  is given by   Eq. (\ref{eq:Z_DHM}).
i.e.
\begin{widetext}
\be -\frac{E}{T}=\sum_{j=1}^n[\ln V +i\delta\, q_j/3] +
 \frac{\alpha}{\alpha^2+\beta^2}
    \sum_{j>k} \vec{L}_j \cdot \vec{L}_k
      \ln |\tau_j - \tau_k|^2
-i \pi \frac{\beta}{\alpha^2+\beta^2} \sum_{j>k}\vec{L}_j \times \vec{L}_k
 \;\sgn(\tau_j - \tau_k),
\ee \end{widetext}
where $q_j$ is defined by $\vec L_j\equiv q_j\vec K_{a_j}$ as before.
Since $\vec L_j\times \vec L_k$ is always $0$ or $\pm \sqrt{3}/2$,
$\exp [-E/T]$ is invariant under a shift:
\be {\pi \sqrt{3}\over 2}{\beta \over \alpha^2+\beta^2}\to
{\pi \sqrt{3}\over 2}{\beta \over \alpha^2+\beta^2}+2\pi .\ee
Eqs. (\ref{vup}), (\ref{vdown}) imply that there is also a symmetry under:
\bea {\pi \sqrt{3}\over 2}{\beta \over \alpha^2+\beta^2}&\to&
{\pi \sqrt{3}\over 2}{\beta \over \alpha^2+\beta^2}+\pi \nonumber \\
\delta &\to & \delta - \pi ,\eea
which corresponds to shifting the integer $n$ in Eq. (\ref{eq:ndef}) by $1$.
Since the energy of any configuration is independent of $n$,
it follows that the Coulomb gas charge density correlation function,
$\langle \rho_\mu (\tau )\rho_\nu (\tau ')\rangle$, is also
$n$-independent.

On the other hand, the correlation function $\langle
X^\mu (\tau )X^\nu (\tau ')\rangle$, {\it does} depend on $n$.
These two correlation functions can be simply related to
each other by adding an external source, $\vec a(\tau )$, to the action,
coupling to the Coulomb gas charge density. That is,
we modify the energy of the Coulomb gas by:
\be -E/T \to -E/T -i\int d\tau\; \vec a(\tau )\cdot \vec \rho (\tau ).\ee
The Coulomb gas correlation function is given by the functional
derivative of the partition function with respect to the source term:
\begin{equation}
  \frac{ \partial^2 \ln{Z}}{\partial a^{\mu}(\omega_n )
\partial a^{\nu}(\omega_n' )}
  = - {1\over (2\pi )^2}\langle \rho_{\mu}(-\omega_n ) \rho_{\nu} (-\omega_n' )\rangle .
\label{eq:Zaa}
\end{equation}
Note that the Coulomb gas energy arises from the functional
integral over $\vec X$:
\be  e^{-E/T}={1\over Z_0}
V^n e^{i(\delta /3)\sum_jq_j}
 \int [d\vec X (\tau )]e^{-S_q(\vec X)}, \ee
where:
\be S_q \equiv S_0(\vec X)-i\int d\tau \vec X(\tau )\cdot \vec \rho (\tau ).
\ee
Here $Z_0$ is the functional integral with $S_q$ replaced by $S_0$,
which is given by Eq. (\ref{eq:S_0}).
Thus, adding a source term to the Coulomb gas corresponds to
a modification of the potential term in the DHM action:
\begin{equation}
S_V[\vec{X}]
\to  - V e^{i \delta/3}\int d\tau \sum_{a=1}^3
e^{i \vec{K}_a \cdot [\vec{X}(\tau )+\vec a(\tau )]} + \mbox{c.c.},
\label{eq:CFpot2}\ee
Alternatively, we may shift $\vec X (\tau )\to \vec X (\tau )-\vec a(\tau)$,
removing the source term from the potential term in the DHM action
but inserting it into the non-interaction term, $S_0$:
\begin{widetext}
\be S_0\to {1\over 2}\int {d\omega_n \over (2\pi )^2}
[X^\mu (\omega_n )-a^\mu ]^*(\omega_n )
\;\left(D^{-1}\right)_{\mu \nu}(\omega_n )\;
[X^\nu (\omega_n )-a^\nu ] (\omega_n ).\ee
\end{widetext}
Here we have introduced the inverse of the DHM propagator, defined
in Eq. (\ref{eq:D}), which is simply the matrix appearing
in $S_0$, in Eq. (\ref{eq:S_0}):
\be \left( D^{-1}\right)_{\mu \nu}\equiv \alpha \delta_{\mu \nu}|\omega |
+\beta \epsilon_{\mu \nu}\omega .\ee
Taking a second derivative with respect to $a^{\mu}$ we see that:
\begin{widetext}
\be
\langle \rho_{\mu}(-\omega ) \rho_{\nu} (-\omega   ' )\rangle
=\delta (\omega   +\omega   ')\left( D\right)^{-1}_{\mu \nu}(-\omega   )
-{1\over (2\pi )^2}\left( D^{-1}\right)_{\mu \sigma}(-\omega   )
\langle X^{\sigma}(-\omega   )X^{\lambda}(-\omega   ')\rangle
\left( D\right)^{-1}_{\lambda \nu}(\omega   '),\label{eq:rrXX}
\ee
or, equivalently:
\begin{equation}
\langle X^{\mu}(-\omega_n )X^{\nu}(-\omega_n ')\rangle
=(2\pi )^2\delta (\omega_n+\omega_n')
D^{\mu \sigma}(-\omega_n )
-(2\pi )^2 D^{\mu \sigma}(-\omega_n )
\langle \rho_{\sigma }(-\omega_n ) \rho_{\lambda} (-\omega_n ' )\rangle
 D^{\lambda \nu}(\omega_n ')\label{eq:XX}
\end{equation}
\end{widetext}
Due to the complicated dependence of $\alpha$ and $\beta$ on $n$,
given by Eq. (\ref{eq:ndef}), and hence the complicated dependence of
$D^{\mu \nu}$ on $n$,  we see that
$\langle X^{\mu}X^{\nu}\rangle$ is $n$-dependent.

However,
what we are interested in here, for obtaining properties of the
Y-junction, is only the Coulomb gas correlation functions,
which are $n$-independent.  Hence we may choose any value of
$n$ that we find convenient. It will sometimes be useful to
deduce the long time behavior of $\langle X^{\mu}X^{\nu}\rangle$
by RG and physical arguments and then use Eq.~(\ref{eq:rrXX})
to deduce the Coulomb gas correlation function. In so doing
we may choose any value of $n$. In general,
the behavior of $\langle X^{\mu}X^{\nu}\rangle$ will
depend strongly on $n$ but this $n$-dependence will cancel
when we compute $\langle \rho_{\mu}\rho_{\nu}\rangle$ using
Eq. (\ref{eq:rrXX}).

Let us now use this mapping between the Y-junction problem and the
dissipative Hofstadter model and consider the case of $g>1$. Then, as
we have already discussed, the electron hopping is a relevant
perturbation in the disconnected limit. In the DHM, for any choice of
$n$, $V$ is a relevant perturbation for $g>1$. We would like to find
the infrared stable fixed point reached in the low energy limit. A
simple guess is that it occurs at $V\to \infty$.  The stability of the
$V\to \infty$ fixed point can be determined using the instanton
method~\cite{CF}.  Namely, in the strong potential limit, the
$\vec{X}$ field is pinned at one of the minima of the potential
(\ref{eq:CFpot}).  The leading perturbation to this limit is given by
a tunneling process between the neighboring minima, represented by an
``instanton.''  This is a classical solution, $\vec X_{\shbox{cl}}(\tau )$
which interpolates between two adjacent minima of the potential
as $\tau$ goes from $-\infty$ to $\infty$. The nature of
these solutions is discussed in Callan and Fried and, in more detail,
in Ref.~\cite{Freed-Harvey}. To determine the scaling dimension
of the tunneling operator we must consider the classical action
of a dilute multi-instanton configuration which tunnels successively
from minimum to minimum with each tunneling event well separated
in time from the rest. This instanton gas is dilute because,
in the limit of large $V$, the action of an instanton goes to
infinity, so the density of instantons goes to zero.
In the dilute limit, we can approximate
such a multi-instanton by a sum of the form:
\be \vec X(\tau ) \approx \vec X_0 +
\sum_{j=1}^n\vec X_{j}(\tau -\tau_j),\label{eq:inst}\ee
where
\bea \vec X_{j}(\tau )  &\to & 0,\ \  (\tau \to -\infty)\nonumber \\
&\to & \vec M_{j},\ \  (\tau \to \infty ).\eea
Here the $j^{\shbox{th}}$ tunneling event takes place at time $\tau_j$
 and involves the particle tunneling between two
adjacent minima of $V(\vec X)$, displaced by $\vec M_{j}$.  In
the dilute limit each function $\vec X_{j}(\tau )$ is a
single instanton solution.  In the long time low energy limit
of interest to us, we may regard the single instanton solutions
as having the form:
\be \vec X_{a_j}(\tau ) \approx \vec M_{j} \,f(\tau ),\ee
for a real function, $f(\tau )$ which interpolates between
$0$ and $1$.
The Fourier transform of the multi-instanton
solution can then be approximated as:
\be \vec X(\omega )\approx \sum_j \vec M_{j}\,e^{i\omega \tau_j}
f(\omega ).\ee
The low frequency behavior of $f(\omega )$, which
follows from the long time asymptotic behavior of the interpolating 
function $f(\tau)$, is given by:
\be f(\omega ) \to {1\over i\omega}
\label{eq:fas}.\ee
The important interaction term
in the instanton action comes simply from $S_0$ and is
obtained by substituting Eq. (\ref{eq:inst}) into $S_0$ in
Eq. (\ref{eq:S_0}).  This gives the
instanton interaction term in the action:
\begin{widetext}
\be S_{\shbox{int}}\approx
\sum_{i>j}M_{i,\mu}M_{j,\nu}
\int {d\omega \over (2\pi )^2}\;e^{i\omega (\tau_j-\tau_i)}\;
\left[ \alpha |\omega |\delta_{\mu \nu}+\beta \epsilon_{\mu \nu}\omega\right]|f(\omega )|^2.
\ee \end{widetext}
In the dilute limit, where the  time separations
are large, we may use the asymptotic form of $f(\omega )$ in
Eq. (\ref{eq:fas}).
In this we recover the Coulomb gas energy:
\begin{widetext}
\be S_{\shbox{int}}\approx -{\alpha \over (2\pi )^2}
\sum_{i>j}\vec M_i\cdot \vec M_j \ln
|\tau_i-\tau_j|^2 +{i\pi \beta \over (2\pi )^2}
\sum_{i>j}\vec M_i\times \vec M_j\;
\hbox{sgn} (\tau_i-\tau_j).\ee \end{widetext}
We see that this expansion is equivalent to the weak $V$ expansion
and therefore that the tunneling events are relevant only if
\be|\vec M_i|^2\alpha^2/(2\pi )^2  <1.  \ee

Thus we see that the separation of the minima of the potential,
$|\vec M_i|$ is a crucial quantity in determining the stability
of the localized fixed point. This separation is determined by
a simple classical analysis of the potential, $V(\vec X)$
in Eq. (\ref{eq:CFpot}). The stationary points of $V(\vec X)$
occur at the points:\be
{\vec X} = {4\pi \over 3}\sum_{i=1}^2 n_i\;\vec K_i, \ee
independent of $\delta$ and forming
a triangular lattice.  However, the energies (values of $V$)
 at these stationary
points depend on $\delta$.  For general values of $\delta$, these
energies are different at the 3 sub-lattices into which the
triangular lattice can be partitioned:
\bea V(\vec 0) &=& -6V \cos (\delta /3)\nonumber \\
V\left( {4\pi}\;\vec K_1/3\right)
 &=& -6V\cos [(2\pi -\delta )/3]\nonumber \\
V\left( -{4\pi}\;\vec K_2/3\right)&=& -6V\cos [(2\pi +\delta )/3].\eea
The sub-lattice containing $\vec 0$, is a triangular lattice
with basis vectors, $2\pi \vec R_i$ ($i=1,2$) with
\be \vec R_i\equiv (2/\sqrt{3})\;\vec K_i\times \hat z.
\label{eq:Rdef}\ee
Of course, the other sub-lattices are simply displaced from this
one by the fixed vectors, $(4\pi /3)\vec K_1$ and $(4\pi /3)\vec K_2$.
Therefore, the vectors $\vec M_i$ occurring in the instanton
solutions are $\pm 2\pi \vec R_1$, $\pm \vec 2\pi R_2$
and $\pm 2\pi (\vec R_1+\vec R_2)$ with length $4\pi /\sqrt{3}$.
As we vary $\delta$ the relative energies on the three sub-lattices
change.
For $\delta =0$, the potential minima occur on the $\vec X=0$
sublattice.  As we increase $\delta$ from zero,
the minima remains on the $\vec X=0$ sublattice until we
reach $\delta =\pi$. For $\pi < \delta < 2 \pi$, the minima
are rather on the $\vec X=(4\pi /3)\vec K_1$ sub-lattice.
Right at $\delta =\pi$, there are degenerate minima on both $\vec X=0$ and
$\vec X=(4\pi /3)\vec K_1$ sub-lattices. These two sub-lattices
define a honeycomb lattice with lattice spacing $4\pi /3$,
smaller by a factor of $1/\sqrt{3}$ compared to
that of the triangular lattices.

Thus, from the dilute instanton analysis, we conclude that, for
general $\delta$,
the localized phase is stable when
\be \Delta\equiv {\alpha |2\pi R_i|^2\over (2\pi )^2}={4\alpha \over 3}
= {4g\over 3+(2n-1)^2g^2}>1.\ee
While this does not prove rigorously that
the DHM is in a localized phase for small $V$ when
this inequality is satisfied, it is
reasonable to believe that the system is in a localized
phase for large enough $V$, in this region of $\alpha$.

For $n\neq 0$ or $1$, the localized phase is never stable,
for any value of $g$.  This follows since the maximum
value of $\Delta$, with respect to $g$, for fixed $n$
is $2/[\sqrt{3}|2n-1|]$, which is $<1$ for $n\neq 0$ or $1$.
On the other hand, for $n=0$ or $1$, $\Delta <1$, for
$1<g<3$. Thus is it reasonable to assume that
the behavior of the Y-junction, at least for large $\Gamma$,
corresponds to the localized phase of the DHM in
the $n=0$, or $1$ representation. This result appears
at first peculiar because we might just as well have
chosen a different value of $n$, in which case
we would conclude that the DHM is {\it not} in the
localized phase. For $g>1$, the DHM is not in the
freely diffusing phase either, since $V$ is relevant.
It therefore must be in some intermediate phase which
is neither freely diffusing nor localized. Remarkably,
Eq. (\ref{eq:XX}) allows us to determine $\langle XX \rangle$
in these intermediate phases by determining $\langle \rho
\rho \rangle$ using $n=0$ or $1$~\cite{CF}.

However, an ambiguity still remains because both $n=0$ and $1$
give stable localized fixed points.  However, the localized
fixed points for these two values of $n$ predict different
$\langle \rho \rho \rangle$.  Consistency requires that,
while both of these fixed points is stable at large $V$,
the RG flow goes all the way to the localized fixed point,
starting from a small $V$, for only one of these two
values of $n$. For the other value of $n$ the RG
flow should go to an intermediate $V$ fixed point, which
must give the same $\langle \rho \rho \rangle$.

The particular cases $\phi = \pm \pi /2$, which maximally break
time-reversal, are instructive in understanding the resolution of this
ambiguity. From Eq. (\ref{eq:n}), for $\phi=\pi/2$, the choice $n=1$ ($\beta$
positive) gives $\delta=0$, and the potential minima forms a triangular
lattice as usual. However the other choice $n=0$ ($\beta$ negative) gives
$\delta=\pi$, for which the potential minima form a honeycomb lattice with
spacing $4\pi /3$, making the strong potential limit unstable for all $g$.
Similarly, for $\phi = - \pi/2$, $n=0$ ($\beta$ negative) is the unique
choice that gives a stable fixed point. This suggests that these localized
fixed points $\chi_{\pm}$ reflect the breaking of time reversal symmetry due
to the magnetic flux $\phi$. Indeed, the conductance at these fixed points
exhibits a chiral behavior breaking the time reversal invariance, as we show
below.

Let us next consider the case when $\phi$ is very close to, but not
exactly equal to, $\pi /2$. If one chooses $n=1$, $\delta$ will be
close to zero, and the potential minima forms a triangular lattice
with a large separation between the true minima and secondary minima.
However, if one chooses  $n=0$ instead of $n=1$, the minima of
the potential $V(\vec X)$ will form a triangular lattice with spacing
$4\pi /\sqrt{3}$, but there are local minima with only slightly larger
energy such that the distance between a true minimum and the nearest
local minima is $4\pi /3$. We could then envision another type of
approximate instanton solution where the particle tunnels from a true
minimum to a local minimum where it lingers for a long but finite time
but must eventually tunnel to a true minimum. The typical time that
the particle spends at a minimum is determined by the instanton
fugacity, and grows exponentially as $V\to \infty$. Thus this other
type of instanton might never be important for sufficiently large $V$.
However, for intermediate values of $V$, these instantons could play
an important role, effectively preventing the RG flow to go to
infinite $V$ for the $n=0$ choice. This argument suggests that the
choice of $n=1$ should represent the stable fixed point around a
vicinity of $\phi=\pi/2$. Of course, similar statements hold regarding
the case $\phi$ very close to $-\pi /2$, for which $n=0$ should
represent the stable fixed point.

We are thus led to conjecture that the localized fixed point for $n=1$ ($n=0$)
corresponds to the stable fixed point of the Y-junction, for $1<g<3$,
even for small bare $V$, for a range of $\phi$ around $\pi /2$ ($-\pi /2$).
While none of our arguments have proven rigorously that the
RG flow really goes all the way from $V=0$ to $V=\infty$ (i.e. to the
localized phase) for any values of $g$ and $\phi$, this
seems like the most ``economical'' assumption since otherwise there
would necessarily be intermediate $V$ fixed points for all values of $n$.
On the other hand, for $\phi =0$ or $\pi$, the $n=0$ and $n=1$ fixed points
appear equally stable. Since, as we show in the next sub-section,
the conductance exhibits broken time-reversal symmetry for these localized
fixed points, we expect that neither of them is stable at $\phi =0$. There
must be some other stable fixed point, corresponding to intermediate $V$,
in this case. We refer to this fixed point as M (for mysterious).

It is convenient to now think about the RG flow in the space of the
physical parameters of the Y-junction, $\Gamma$ and $\phi$. We have
proposed (unique) stable fixed points for $\phi$ near $\pi /2$ ($\chi_+$)
or $-\pi /2$ ($\chi_-$) and another fixed point, M, stable for $\phi =0$.
While the $\chi_{\pm}$ fixed points are stable against a small change in
the flux, we do not know whether the M fixed point has such a stability.
The simplest assumption is that it is does not. In this case the RG flows
might go to the
$\chi_{\pm}$ fixed points for any non-zero $\phi$, starting from
arbitrarily small $V$. Alternatively, it is possible that
the M fixed point is stable against adding a small flux.  In that
case, there would have to be additional unstable fixed points
defining the boundaries between the basins of attraction of the
M, $\chi_+$ and $\chi_-$ fixed points. So again, an ``economy''
principle, suggests the simple picture with only 3 stable fixed points,
for $1<g<3$, as sketched in Fig.~\ref{fig:Chiflow}.

\subsection{Calculation of the conductance tensor}


Let us calculate the conductance tensor $G_{jk}$
as defined in Eq.~(\ref{eq:def-of-conduc-tensor}) at
the chiral fixed points $\chi_{\pm}$.
There are two convenient ways of introducing the voltages drops. They
can either occur right at the junction, or else can be spread over finite
regions of the wires which are long compared
to microscopic length scales like $k_F^{-1}$
but could still be short compared to the length of the wires
and other long length scales such as an inelastic scattering length.
Both lead to the same results. The former
approach has the advantage that it allows the conductance tensor to be
expressed in terms of the Coulomb gas correlation function.  The
latter approach is useful when the BCFT formalism is used and is the
only method at our disposal in the case of the Dirichlet fixed
point. We discuss each in turn, one in this section and the other in
the next. For convenience, we set the electron charge, $e$, to $-1$. (In
our convention the actual electron charge $e<0$.) 
We also set $\hbar =1$.

We first consider the case where the voltage drop occurs right at the
junction. We then modify the tunneling Hamiltonian by time-dependent
phases, $A_j(t)$:
\be H_T\to -\Gamma e^{i\phi /3}\sum_j\psi^\dagger_j\psi_{j-1}e^{i(A_j-A_{j-1})}
+h.c.\label{HTA}\ee
Note that, in the case where $A_j(t)=-V_jt$, that this
corresponds to the replacement:
\be \psi^\dagger_j(t)\to e^{-iV_jt}\psi^\dagger_j(t).\ee
This corresponds to a potential $V_j$ on wire $j$, so we see that:
\be dA_j/dt = -V_j.\ee
It is convenient to define $I_j$ as
the current on wire $i$, headed towards the junction:
\be I_j= -dN_j/dt=i[N_j,H].
\ee
Here $N_j$ is the total number of particles on wire $j$.  Since
only the tunneling terms destroy conservation of particle
number on each wire separately, we readily obtain:
\begin{widetext}
\be I_j=-i\Gamma\left[ e^{i\phi /3}\,e^{i(A_j-A_{j-1})}\,
\psi^\dagger_j\psi_{j-1}
+e^{-i\phi /3}\,e^{i(A_j-A_{j+1})}\,
\psi^\dagger_j\psi_{j+1}\right] +h.c.\ee
Expanding to first order in the $A_j$'s gives:
\be  \langle I_j \rangle = \langle I_j^0 \rangle 
-[(A_j-A_{j-1})+(A_j-A_{j+1})]\;
E_T/3.\label{Ij}\ee
Here $E_T$ is the tunneling energy:
\be E_T=-\Gamma e^{i\phi /3}\sum_j 
\langle \psi_j^\dagger \psi_{j-1} \rangle +c.c.
=-3\Gamma e^{i\phi /3} \langle \psi_j^\dagger \psi_{j-1} \rangle +c.c.\ee
where the $Z_3$ symmetry was used in the last step. $I_j^0$ is
the current operator to zero$^{\shbox{th}}$ order in the $A_k$:
\be I_j^0\equiv -i\Gamma \left[e^{i\phi /3}\psi^\dagger_j\psi_{j-1}
+e^{-i\phi /3}\psi^\dagger_j\psi_{j+1}\right] + h.c.\ee
$ \langle I_j^0 \rangle $ is calculated to first order in $A_j(t)$ using:
\be
H\approx H_0+H'=H_0+\sum_j I_j^0\;A_j.
\ee
\be  \langle I_j^0(t) \rangle =-i\int_{-\infty}^t dt'\; \langle [I_j^0(t),H'(t')] \rangle 
=-i\sum_k\int_{-\infty}^tdt'\; \langle [I_j^0(t),I_k^0(t')] \rangle \;A_k(t')
=\sum_k \int_{-\infty}^\infty dt'\;K_{jk}^{\shbox{ret}}(t-t')\,A_k(t').\ee
\end{widetext}
Here $K_{ij}^{\shbox{ret}}$ is the retarded Green's function
of the $I_j^0$ operators. Fourier transforming,
\be A_j(\omega )\equiv \int dt \;e^{i\omega t}\;A_j(t),\ee
and including the ``diamagnetic term'' in Eq. (\ref{Ij}), we obtain
the Kubo formula in the form:
\be  \langle I_j \rangle (\omega )=\sum_k G_{jk}(\omega )\,V_k(\omega ).\,\ee
where the conductance matrix is given by:
\be G_{jk}(\omega )={1\over i\omega }\left[K_{jk}^{\shbox{ret}}
(\omega )-C_{jk}\right],
\label{eq:asp-cond}
\ee
and the $\omega$-independent symmetric matrix, $C_{jk}$ is given by:
\bea C_{jj}&=&2E_T/3,\nonumber \\
C_{jk}&=& -E_T/3\ \  (j\neq k).\eea
Note that 
\be 
\sum_jC_{jk}=\sum_kC_{jk}=0
\;,
\ee
which combined with Eqs.~(\ref{eq:conduc-aspen}) and (\ref{eq:asp-cond})
implies
\be \sum_jK_{jk}^{\shbox{ret}}
=\sum_kK_{jk}^{\shbox{ret}}=0.\ee
The retarded Green's function may be obtained by analytic continuation
from the Matsubara Green's function:
\be K^M_{jk}(\omega_n)\equiv -\int d\tau \;e^{i\omega_n\tau}\;
\langle {\cal T}\, I^0_j(\tau )\,I^0_k(0) \rangle .\ee
Here $\omega_n=2\pi nT$ for integer $n$,
 at temperature $T$, $\tau$ is imaginary
time and ${\cal T}$ represents time ordering. The retarded Green's
function is obtained, as usual, by:
\be K^{\shbox{ret}}_{jk}(\omega)=\lim_{\delta \to 0^+}K^M_{jk}(\delta
-i\omega).\ee
The conductivity can be conveniently derived from the imaginary
time path integral, with a vector potential, $A_j(\tau )$, defined
as in Eq. (\ref{HTA}) which
depends on imaginary time. The imaginary time partition function,
$Z_I[A_j(\tau )]$, is obtained from the path integral of
$\exp [-S_0-\int d\tau \;H_T(\tau )]$, where $S_0$ is the imaginary
time action
for the three independent leads. We see that:
\be {\delta \ln Z_I\over \delta A_j(\tau )}=- \langle I_j(\tau ) \rangle \ee
and, at $A_j\to 0$,
\be {\delta^2 \ln Z_I\over \delta A_j(\tau )\,\delta A_k(\tau ')}
=-K^{M}_{jk}(\tau -\tau ')+\delta (\tau -\tau ')C_{jk}.\ee
Fourier transforming:
\be {\delta^2 \ln Z_I\over \delta A_j(\omega_n) \,\delta A_k(\omega_n  ')}
={\delta (\omega_n +\omega_n ')\over 2\pi}[-K^M_{jk}(\omega_n)+C_{jk}].\ee
The conductivity can be obtained from this by analytic continuation.
Explicitly, writing:
\be {\delta^2 \ln Z_I\over \delta A_j(\omega_n ) \,\delta A_k(\omega_n  ')}
={1\over (2\pi )^2}\delta (\omega_n +\omega_n ')Z_{jk}(\omega_n),\ee
the dc conductance is given by:
\be
G_{jk}=\lim_{\omega_n\to 0}{1\over 2\pi
\omega_n}Z_{jk}(\omega_n).\label{Zcond}\ee
Here the limit $\omega_n\to 0$ must be taken by first continuing
$\omega_n$ into the complex plane and then taking $\omega_n\to 0$
along the imaginary axis with a small positive real part.
This version of the Kubo formula is convenient because this second
derivative of $\ln Z_I$ is proportional to
the Coulomb gas density correlation function in the DHM. To make
this connection it is convenient
 to rewrite the three vector potentials, $A_j(\tau )$
in terms of a pair of vector potentials, $a^\mu (\tau )$ such that:
\be A_j-A_k=\sum_l\epsilon_{jkl}\;\vec K_l\cdot \vec a.\label{a}\ee
Here $\epsilon_{jkl}$ is the antisymmetric unit tensor
($\epsilon_{123}=1$ etc.) and the Greek indices run over $1$ and $2$ only.
Then the tunneling term is modified to:
\be H_T\to i\Gamma e^{i\phi /3}\sum_{j=1}^3\eta_ie^{i\vec K_j\cdot
(\vec \Phi +\vec a)}+ h.c.\ee
The corresponding modification of the DHM action is the shift:
\be \vec X\to \vec X +\vec a.\ee
This is precisely the source term coupled to the Coulomb gas
density discussed above, which appeared in Eq. (\ref{eq:CFpot2}).

We now see that the Y-junction conductivity is directly related to the Coulomb
gas correlation function. To make the connection we use:
\be \sum_jK^{\mu}_jK^{\mu}_j=(3/2)\delta^{\mu \nu}\ee
to invert (\ref{a}):
\be \vec a=\sum_{j,k,l}\vec K_j\epsilon_{jkl}(A_k-A_l)/3.\label{aA}\ee
This implies:
\be {\partial a^\mu \over \partial A_k}={2\over 3}\sum_{j,l}K^\mu_j
\epsilon_{jkl}={2\over 3}\sum_jK^\mu_j\epsilon_{jk}.\ee
Here we use an  anti-symmetric $3\times 3$  tensor, $a_{jk}$, defined by:
\be \epsilon_{ij}\equiv \sum_k\epsilon_{ijk}\ee
with $\epsilon_{12}=\epsilon_{23}=\epsilon_{31}=1$.
We may thus write:
\begin{widetext}
\begin{equation}
  \frac{ \partial^2 \ln{Z}}{\partial A_j(\omega_n ) \,\partial A_k(\omega_n')}
  = \frac{4}{9} \sum_{m,n,\mu,\nu}
  \epsilon_{jm} \epsilon_{kn} K_m^{\mu} K_n^{\nu}
 \; \frac{ \partial^2 \ln{Z}}{\partial a^{\mu}(\omega_n) \,\partial a^{\nu}
(\omega_n')}.
\label{eq:ZAA}
\end{equation}
The derivative on the right hand side of Eq. (\ref{eq:ZAA}) is proportional
to the Coulomb gas correlation function. $\langle \rho_\mu (\omega_n)
\rho_\nu (\omega_n')\rangle$ as we observed in Eq. (\ref{eq:Zaa}).

There are two simple limits of the DHM.  One is the case
where the potential, $V$, is irrelevant. Then the correlators
of the $X$-fields are determined by $S_0$:
\be  \langle X^{\sigma}(-\omega_n )X^{\lambda}(-\omega_n ') \rangle 
=(2\pi )^2 \delta (\omega_n +\omega_n')
\;(D^{-1})^{\sigma \lambda}(-\omega_n ),\ee
\end{widetext}
and so the Coulomb gas correlation function vanishes. This corresponds
to the limit of disconnected wires and zero conductance.
The other simple limit  the corresponds to the potential, $V$,
renormalizing to infinity and pinning the $X$ co-ordinates at one
of its minima. The Coulomb gas correlation function is then
given by the first term only in Eq. (\ref{eq:rrXX}).
Thus,
\be Z_{jk}(\omega_n)=
 -\frac{4}{9} \sum_{m,n,\mu,\nu}
  \epsilon_{jm} \epsilon_{kn} K_m^{\mu} K_n^{\nu}\;
D_{\mu \nu}(-\omega_n)
.\ee
Using
 this gives:
\be Z_{jk}(\omega_n)={2\over 3}(3\delta_{jk}-1)\,\alpha \,|\omega_n|
+{2\over \sqrt{3}}\beta \,\omega_n \,\epsilon_{jk}.\ee
Thus, from Eq. (\ref{Zcond}), the dc conductivity is given by:
\be G_{jk}={e^2\over h}\left[{2\over 3}(3\delta_{jk}-1)\,\alpha +
{2\over \sqrt{3}}\,\beta\,
\epsilon_{jk}\right] .\ee
We have restored the factors of $e^2$ and $\hbar$ which we had previously set
equal to one. The values of $\alpha,\beta$ for the chiral fixed points are
obtained from Eq.~(\ref{eq:ndef}) with $n=1$ and $n=0$, and lead to the
conductance tensor
\begin{equation}
  G_{jk}^{\pm} =\frac{e^2}{h}\;
{2g\over (3+g^2)} [(3 \delta_{jk}  -1)
         \pm g \,\epsilon_{jk}],
\label{eq:GDGA}
\end{equation}
with $\pm$ corresponding to the fixed points $\chi_{\pm}$. It is interesting
to check that the same result can be obtained by applying electric fields
uniformly to the entire wire $j$, which we do in the next section, where we
address the Y-junction problem using BCFT.


\section{Approaches based on boundary conditions}
\label{sec:approaches}
\newcommand{\bomega}{\bar{\omega}}


After studying the junctions of three quantum wires using the mapping
into the dissipative Hofstadter model in the previous section,
we now turn to a study based on boundary conditions satisfied by the
bosonic fields describing the degrees of freedom of the wires.
The boundary condition of the boson fields represents the
physics at the junction.
In particular, conformally invariant boundary conditions
corresponds to RG fixed points of the junction problem.
Among them, stable fixed points are of special importance,
as they govern the
physical properties of the junction
at low bias voltages and low temperatures.
From these boundary conditions, one can determine the
conductance tensor of the junctions as well as the scaling of
perturbations to the fixed point, {\it i.e}, the perturbing
operators and the physical processes that they correspond to.
The study of conformally invariant boundary conditions will
allow us to identify other fixed points in addition to
the ones discussed in section~\ref{sec:DHM}.

We carry out the analysis based on boundary conditions in two
ways. The first approach is based on the formalism of boundary
conformal field theory (BCFT), which is a systematic treatment of
conformally invariant boundary conditions for a given (bulk) CFT. The
second way is an approach that we introduce in this paper, which we
refer to as the method of delayed evaluation of boundary conditions
(DEBC).

A theory with free $c$-component bosons is a conformal field theory
with central charge $c$. The classification of all possible
conformally invariant boundary conditions, even for free bosons, is a
rather difficult and rich problem especially for $c \geq 2$.
For $c \geq 2$, it is possible to have a ``magnetic flux''
at the boundary, as was
discussed in Refs.~\cite{CF} in the context of Dissipative
Hofstadter Model. Nontrivial boundary conditions, which have no
simple description in terms of the original boson field, have been
discussed in Refs.~\cite{YiKane,QBM} for $c=2$ free boson.  As the
junction of 3 wires is mapped to the boundary problem of $c=2$ free
bosons, we can expect these features to be relevant in the present
problem.  App.~\ref{sec:BCFT}, \ref{sec:refboson} and 
\ref{sec:BC_cond} review aspects of BCFT, BCFT 
for  free bosons and the calculation of the 
conductance using BCFT, respectively. 

In the bosonization approach, the Fermi statistics of the electrons
may be taken care of by the Klein factors. On the other hand, in the
BCFT, conformally invariant boundary conditions, in particular the
nontrivial ones, are formulated in terms of {\em boundary states.}
Boundary states belong to the Hilbert space of the theory with the
periodic boundary condition in the ``space'' direction.  The
implication of the Fermi statistics of electrons on the boundary state
formulation is not obvious.

Considering the Fermi statistics of the electrons, we show in 
section~\ref{sec:twist} that the
structure of the Hilbert space is ``twisted'' compared to that of the
standard free boson.  The twisted structure affects the possible
boundary states, the scaling dimensions of the boundary operators, and
stability.  We show that the derived boundary states are indeed
consistent with electrons with Fermi statistics.  Thus the twisted
structure encodes the effect of the Fermi statistics.

The formulation of the problem of tunneling between multiple quantum
wires within the DEBC approach, presented in section~\ref{sec:DEBC}, 
provides another angle of attack, which
gives results consistent  with the BCFT approach. The DEBC method
allows us to give a simple physical interpretation of the operators
that become relevant and enforce the correct boundary conditions, as
well as the operators that correspond to perturbations around the
fixed point boundary condition. 

The DEBC method consists of doubling the number of degrees of freedom
in the problem by keeping both the bosonic fields $\varphi_i$, $i=1,2,3$
in each lead as well as its conjugate momentum $\theta_i$. Notice that
such construction, in terms of both sets of bosonic fields $\varphi_i$
and $\theta_i$, is the usual one when describing systems on the
infinite line, but is unusual if we want to describe the system on the
half line $x>0$, appropriate for describing the quantum wires that
meet at a junction. There must be a boundary condition imposed at
$x=0$ which cuts in half the number of degrees of freedom, when
describing the system on the half-line. It is only with this imposed
boundary condition that the operators above describe the electronic
degrees of freedom on the the half-line. The essence of the method is
that we delay the choice of boundary condition (hence the name
``delayed evaluation'') until after we write the operators
corresponding to the physical tunneling of electrons (and pairs or
multiplets), working with twice the physical degrees of freedom. With
the tunneling operators in hand, we can then study the correct
boundary condition that must be imposed on $\varphi_i$ and $\theta_i$.

In Sec.~\ref{sec:fpts} we use both methods to discuss various 
fixed points of the Y-junction.

\section{Fermi statistics and the twisted structure in bosonization}
\label{sec:twist}
\begin{widetext}

The BCFT of free bosons can  be applied to the system of
interacting electrons via bosonization.
However, owing to the Fermi statistics of the electron, 
there are various differences compared to the standard
compactified boson field theory, reviewed in App.~\ref{sec:refboson}.
In this section we formulate the effect of the Fermi statistics
in terms of a twisted structure in the Hilbert space of the
free boson theory.
This approach was first developed in Ref.~\cite{WongAffleck}.
In this paper, we further propose a simplified treatment of the twisted
structure for the junction problem, and apply it to the
junction of three wires.

\subsection{Bosonization of a single wire}

Let us start from the bosonization of a single wire,
carefully taking the effect of the Fermi statistics into account.

\subsubsection{Derivation of the twisted structure}

Since the Matsubara formalism implies 
{\it anti-periodic} boundary conditions in imaginary time, 
anti-periodic boundary conditions in the space direction will also occur after
a modular transformation. 
These must be imposed independently on left and right-movers:
\be \psi_{L/R}(t,x )=-\psi_{L/R}(t,x+\beta ).\ee
It will be important to determine the corresponding b.c.'s on 
the bosons, using the bosonization formulas reviewed in Sec. IV. 
Let us first consider the case, $g=1$, corresponding to free fermions. 
The b.c.'s are then:
\be \exp [i\sqrt{2}\varphi_{L/R}(t,\beta )]=-\exp [i\sqrt{2}\varphi_{L/R}(t,0 )].\label{bcLR}\ee
Because $[\varphi_{L/R},\varphi_{L/R}]\neq 0$ it is not so straightforward to 
deduce the boundary conditions on the boson fields from Eq. (\ref{bcLR}). One 
way of proceeding is to determine the b.c.'s on the bosons which reproduce 
the finite-size spectrum implied by anti-periodic b.c.'s on the fermions. 
The allowed wave-vectors for single-particle fermion states are:
\be k= {2\pi (n+1/2)\over \beta},\ \  (n \in Z).\label{kferm}\ee
Recognizing that $k$ is measured from $\pm k_F$ for
right and left-movers respectively, 
we see that $n$ may be positive or negative in Eq. (\ref{kferm})
for both left and 
right movers.
The negative $n$ states correspond to hole states for right-movers 
and electron states for left-movers.  It is straightforward to calculate 
the energy of an arbitrary multi-particle fermionic state with this linearized 
dispersion relation. Consider the energy of an arbitrary state of
the right-movers.
The lowest energy state of charge $Q_R \in Z$, is:
\be E(Q_R)={2\pi \over \beta}
\sum_{n=0}^{|Q_R|-1}(n+1/2)={2\pi \over \beta}{{Q_R}^2\over 2}.
\label{fssF}\ee
Here $Q_R$ may be positive or negative, corresponding
to creating particles or holes. 
An arbitrary multi particle-hole state can be constructed,
on top of the charge $Q_R$ 
groundstate, by first raising the lowest $n_m$ fermions by $m$ levels, then 
raising the next $n_{m-1}$ by $m-1$ levels, etc., finally raising $n_1$
fermions by $1$ level. 
Thus the energy of an arbitrary excited state of the right-movers
may be written:
\be E={2\pi \over \beta}\left[{Q_R^2\over 2}+\sum_{m=1}^\infty
mn_m\right],
\label{fssR}\ee
where $Q_R$ is an arbitrary integer and the $n_m$'s are arbitrary
non-negative integers. 

Precisely this spectrum arises from right-moving bosons.
We assume that the right and left moving boson field,
$\phi_{R/L}$ are periodic variables:
\be
\varphi_{R/L}(t,x+\beta )\leftrightarrow
\varphi_{R/L}(t,x)+\sqrt{2}\pi n,\ \  (n\in Z).\label{pbcR}\ee
This is reasonable since all physical operators, local in fermion
fields, are single-valued with this identification.  
We then find that the anti-periodic b.c.'s on fermions arise from
{\it periodic} b.c.'s on bosons. 
To see this, consider the mode expansion for periodic right-movers:
\be \varphi_R (x_-)=\hat \varphi_0^R +{\pi \over \beta}\hat{P}_R x_-
+{1\over \sqrt{2}}\sum_{n=1}^\infty {1\over \sqrt{n}}\left[ 
a_n^R\exp [-inx_- { 2\pi \over \beta}]+ \mbox{h.c.}\right].
\label{modeR}\ee
Here $\hat{P}_R$, the momentum variable conjugate to $\hat \varphi_0^R$, 
has eigenvalues $\sqrt{2}$ times an integer.
consistent with the assumed periodic b.c.'s. Substituting 
this mode expansion into the right-moving Hamiltonian:
\be H_R={1\over 2\pi}\int_0^\beta (\partial_-\varphi_R)^2 \; dx, \ee
gives precisely the spectrum of Eq. (\ref{fssR}), upon the identification, 
$\hat{P}_R \leftrightarrow \sqrt{2} Q_R$.
Thus we see that {\it anti-periodic}
b.c.'s on fermions turn into {\it periodic} b.c.'s on
the {\em chiral} boson field.

Alternatively, it is also possible to derive the
periodic b.c.'s on right-moving bosons 
from anti-periodic b.c.'s on right-moving fermions from Eq. (\ref{bcLR}), 
taking into account the boson commutation relations. The equal-time 
commutator follows from 
Eq. (\ref{modeR}):
\be [\varphi_R(x),\varphi_R(0)]=(i\pi /2)\epsilon_P(x),\label{phirC}\ee
where $\epsilon_P(x)$ is the function, anti-symmetric in $x$, 
which is constant everywhere except at $x=n\beta$, for $n\in Z$, 
where it jumps by 2: 
\be \epsilon_P(x)=(2n+1),\ \  [n\beta <x <(n+1)\beta ].\label{epPdef}\ee
We multiply both sides of (the ``right version'')
of Eq. (\ref{bcLR}) on the right by 
$\exp{[-i\sqrt{2}\varphi_R(\delta )]}$ where $\delta$
is a small positive quantity, giving:
\be \exp [i\sqrt{2}\varphi_R(\beta )]\exp [-i\sqrt{2}\varphi_R(\delta )]
=-\exp [i\sqrt{2}\varphi_R(0)]\exp [-i\sqrt{2}\varphi_R(\delta)].
\label{prod}
\ee
This can be rewritten:
\be \exp \{ i\sqrt{2}[\varphi_R(\beta )-\varphi_R(\delta )]\}\exp \{[\varphi_R(\beta ),\varphi_R(\delta )]\}
=-\exp \{ i\sqrt{2}[\varphi_R(0)-\varphi_R(\delta )]\}\exp \{[\varphi_R(0 ),\varphi_R(\delta )]\}\label{comm}
\ee
Since $0<\beta -\delta <\beta$ and $-\beta <0-\delta<0$,
using Eq. (\ref{phirC}) 
this becomes:
\be i\exp \{ i\sqrt{2}[\varphi_R(\beta )-\varphi_R(\delta )]\}=i
\exp \{ i\sqrt{2}[\varphi_R(0)-\varphi_R(\delta )]\}\ee
We may finally let $\delta \to 0^+$, giving:
\be \exp \{ i\sqrt{2}[\varphi_R(\beta )-\varphi_R(0 )]\}=1,\label{PER}\ee
and hence periodic b.c.'s:
\be \varphi_R(\beta )=\varphi_R(0) \pmod{\sqrt{2}\pi n}.\ee
Note that if we had instead chosen the infinitesimal $\delta <0$, so that 
$\beta <\beta -\delta <2\beta$ and $0<0-\delta <\beta$, the 
two commutators in Eq. (\ref{comm}) would still differ
by $\pi$ from Eq. (\ref{epPdef}) 
so that we would still obtain Eq. (\ref{PER}). 

\subsubsection{Finite-size spectrum and bulk operators}

The left-moving fermions with anti-periodic b.c.'s have an identical
spectrum to the right-movers 
which again is given by periodic b.c.'s on the left-moving bosons. Adding 
left and right contributions, we obtain the mode expansions:
\bea
\varphi (t,x)&=&\hat \varphi_0+
{\sqrt{2}\pi \over \beta}[(\hat Q_L+\hat Q_R)t+(\hat Q_L-\hat Q_R)x]
\nonumber \\ &&
+{1\over \sqrt{2}}\sum_{n=1}^\infty {1\over \sqrt{n}}\left[ 
a_n^R\exp{(-inx_- {2\pi  \over \beta} )}+
a_n^L\exp{(-inx_+ {2\pi  \over \beta} )}+ \mbox{h.c.}\right],
\nonumber \\
\theta (t,x)&=&\hat \theta_0+
{\sqrt{2}\pi \over \beta}[(\hat Q_L-\hat Q_R)t+(\hat Q_L+\hat Q_R)x]
\nonumber \\ &&
+{1\over \sqrt{2}}\sum_{n=1}^\infty {1\over \sqrt{n}}\left[ 
a_n^R\exp{(-inx_- {2\pi \over \beta} )}-
a_n^L\exp{(-inx_+ {2\pi \over \beta} )}+ \mbox{h.c.}\right].
\label{phitheta}\eea
Here \bea 
\hat \varphi_0&\equiv &\hat \varphi^R_0+\hat \varphi_0^L\nonumber \\ 
\hat \theta_0&\equiv& \hat \varphi^R_0-\hat \varphi_0^L.\eea
While it is convenient to define a new set of integer quantum numbers:
\bea Q&\equiv& Q_L+Q_R\nonumber \\
\tilde Q&\equiv & Q_L-Q_R,\label{mndef}\eea
it is important to realize that they do not run independently over the 
integers, but rather obey the ``gluing condition'' \cite{WongAffleck}:
\be Q=\tilde Q \pmod{2}
\label{eq:glue1}
\ee
Thus, the periodic identifications of the fields $\varphi (t,x)$ and 
$\theta (t,x)$ do not simply correspond to them being angular
variables. They are of the form:
\bea \varphi &\sim& \varphi + \tilde Q\sqrt{2}\pi, \nonumber \\
\theta &\sim& \theta + Q\sqrt{2}\pi,
\label{twist}\eea
but rather with the gluing condition~(\ref{eq:glue1}).
We refer to this as a ``twisted structure''. 
It simply reflects the fact that the bosons arise from 
bosonizing fermions. 
It can be verified that the finite size energies (with anti-periodic 
b.c.'s on the fermions) are in 
one-to-one correspondance with the (bulk) operator content.
For example, the right-moving oscillator vacuum
with quantum number $Q_R<0$ corresponds to 
the lowest dimension (right-moving) operator with charge $Q_R$.
Taking 
into account Fermi statistics, this operator is:
\be \prod_{j=0}^{Q_R-1}\left[ \left(\partial_{x_-}\right)^j \psi_R\right],\ee
with the (right-moving) conformal dimension
\be
 \sum_{i=0}^{Q_R-1} \big( \frac{1}{2} + j \big) = \frac{{Q_R}^2}{2} .
\ee
Comparing with eq.~(\ref{fssF}),
we see that the usual relationship between the finite size energy and 
scaling dimension, $\Delta$, is obeyed: $E=(2\pi /\beta )\Delta$.
The corresponding bosonized operator is $\exp [iQ_R\sqrt{2}\varphi_R]$.


We have discussed extensively the case $g=1$, corresponding to free fermions. 
However, these results can be seen to carry over to the case of interacting 
fermions, with general values of $g$.  In general, the mode expansion 
for $\varphi$ is:
\bea \varphi (t,x)&=&\hat \varphi_0+{\sqrt{2}\pi \over \beta}
\left[{1\over g}\hat Qt+{\hat {\tilde Q}}x
\right]
\nonumber \\ &&
+{1\over \sqrt{2g}}\sum_{n=1}^\infty {1\over \sqrt{n}}\left[ 
a_n^R\exp{(-inx_- {2\pi \over \beta} )} +
a_n^L\exp{(-inx_+ {2\pi \over \beta} )} + \mbox{h.c.}
\right]\label{modeg}
\eea
and the corresponding spectrum is:
\be E={2\pi \over \beta}\left[{Q^2\over 4g}+{g\tilde Q^2
\over 4}+\sum_{m=1}^\infty m(n_m^L+n_m^R)\right ],\label{fssFB}\ee
again with the gluing condition~(\ref{eq:glue1}).
The corresponding {\em bulk} primary operators are:
\be \exp [i(Q\varphi + \tilde Q \theta )/\sqrt{2}],\label{Fops}\ee
with dimensions:
\be \Delta = {Q^2\over 4g}+{g\tilde Q^2\over 4}.
\ee

When we form boundary states from the Hilbert space with the
anti-periodic b.c.'s on 
the fermions, the condition $[{\cal H}_R-{\cal H}_L]|A \rangle=0$
will simplify this twisted structure, as we will see. 

\subsubsection{Boundary states and boundary operators}

Let us now consider the boundary operators, finite size spectrum 
and boundary states.
We consider BC's on the fermions either of the free type:
\be \psi_L(t,0)=\psi_R(t,0),\ee
corresponding to $\theta (0)=\theta_0$ (a constant),
i.e. N b.c.'s on $\varphi$ 
or else of Andreev type:
\be \psi_L(t,0)=\psi_R^\dagger (t,0)\ee
corresponding to D BC's on $\varphi$.
The N BC turns the operators of Eq. (\ref{Fops}) into:
\be \exp{[\frac{i}{\sqrt{2}}(Q\varphi (t,0) + \tilde Q \theta (t,0))]}
\to \exp{[\frac{i}{\sqrt{2}}(Q\varphi (t,0)+  \tilde Q \theta_0)]}
\propto \exp [iQ\sqrt{2} \varphi_L(t,0)],\ee
of dimension:
\be
\Delta_N=\frac{Q^2}{2g}.
\label{eq:single-DeltaN}
\ee
We find that the value of $\tilde Q$ is now immaterial
to the scaling dimension. 
Consequently, we can allow $Q$ to take any integer value, even or odd,
despite the gluing condition~(\ref{eq:glue1}).
Also note that, 
in the free fermion case, $g=1$, we obtain $\Delta_N=Q^2/2$, corresponding 
to the operator spectrum of a single left-mover, which we can obtain 
by regarding the right movers as the continuation of the left-movers to 
the negative $x$-axis. 

As reviewed in~\ref{sec:approaches},
the spectrum of the {\em boundary} operators
is related to the FSS on the finite-width strip with the same boundary
condition imposed on the both ends.
To find the corresponding FSS,
let us consider the strip $0<x<l$, and impose the N b.c. on $\varphi$
at the both ends $x=0,l$.
We notice that the N b.c. (on $\varphi$)
corresponds to the D b.c. on the dual field $\theta$.
The mode expansion
consistent with the D b.c.'s on $\theta$
\bea \theta (0)&=0&   \pmod{\sqrt{2}\pi} \nonumber \\
\theta (l)&=0& \pmod{\sqrt{2}\pi}
\eea
reads:
\be \theta = {\sqrt{2}\pi \over l}\hat Qx
+ {1\over \sqrt{2}g}\sum_{n=1}^\infty 
2\sin{n\pi x \over l}
[a_n^L\exp{(-int \frac{\pi}{l})}+\mbox{h.c.}].\ee
This implies the Hamiltonian:
\be H={\pi \over l}\left[{(\hat Q)^2 \over 2g}+\sum_{n=1}^\infty na_n^{L\dagger}a_n^L\right].
\label{fssFN}
\ee
This is indeed consistent with the scaling dimension of the
boundary primary operator in Eq.~(\ref{eq:single-DeltaN}).

On the other hand, as discussed in Sec.~\ref{sec:approaches},
the boundary condition generally can be described in terms of
the corresponding boundary state upon modular transformation.
Naturally, the N boundary state for the present case
is similar to the N boundary state of a standard free boson.
However, here the gluing condition~(\ref{eq:glue1}) is important
in determining the boundary state.
Since the N boundary state (on $\varphi$) is Dirichlet boundary state
on $\theta$, the winding of $\theta$ along the boundary is zero.
Consequently, we have the quantum number $Q=0$.
Because of the gluing condition~(\ref{eq:glue1}), the quantum number
$\tilde{Q}$ is now restricted to {\em even} integers.
The N boundary state is:
\be
|N(\theta_0)\rangle = g^{1/4}
\exp \left[ \sum_{n=1}^\infty 
a_n^{L\dagger}a_n^{R\dagger} \right] \sum_{\tilde Q}^{'}
e^{-i\sqrt{2}\tilde Q\theta_0} |(\tilde Q,0)\rangle .
\label{NBSF}
\ee
Here the $'$ on the summation indicates that $\tilde Q$ is restricted to even 
integers.
Upon performing the modular transformation 
we obtain:
\be \langle N(\theta_0)|\exp [-lH^P_\beta ]|N(\theta_0)\rangle 
={1\over \eta (q)}\sum_{m=-\infty}^\infty
\exp{- \frac{\pi \beta}{l}\frac{m^2}{2g}}.
\ee
Here $m$ is summed over {\it all} integers.
This corresponds to the finite size spectrum of Eq. (\ref{fssFN}),
demonstrating the consistency.

We can repeat the similar analysis for the D b.c. on $\varphi$.
The primary boundary operators have 
dimensions:
\be \Delta_D={g(\tilde Q)^2\over 2},\ee
where $\tilde{Q}$ can be even or odd, and the mode expansion becomes:
\be
\varphi = {\sqrt{2}\pi \over l}\hat{\tilde Q}x +
{1\over \sqrt{2}g}\sum_{n=1}^\infty 
2\sin{n\pi x \over l} [a_n^L \exp{(-int \frac{\pi}{l})} +\mbox{h.c.}],
\ee
where $\tilde Q$ can be even or odd.  
This implies the Hamiltonian:
\be
H={\pi \over l}\left[{g \hat{\tilde{Q}}^2 \over 2}+
\sum_{n=1}^\infty na_n^{L\dagger}a_n^L\right].
\label{fssFD}
\ee
The D boundary state is:
\be |D(\varphi_0)\rangle = g^{-1/4}
\exp \left[ -\sum_{n=1}^\infty 
a_n^{L\dagger}a_n^{R\dagger} \right] \sum_{Q}^{'} \exp [-i\sqrt{2}Q\varphi_0]
|(0,Q)\rangle ,\label{DBSF}\ee
where the sum is restricted to even $Q$ since $\tilde Q$ is now set to zero.
We now obtain:
\be \langle D(\varphi_0)|\exp [-lH^P_\beta ]|D(\varphi_0)\rangle 
={1\over \eta (q)}\sum_{m=-\infty}^\infty
\exp{[-\frac{\pi \beta}{l}\frac{gm^2}{2}}],
\ee
consistent with the FSS of Eq.~(\ref{fssFD}).

\subsection{Junction of two wires}
\label{sec:twist2wires}

Now let us consider the implications of the twisted structure
for the junction of two wires.

\subsubsection{Structure of the boundary states}

We introduce 2 fermion fields $\psi_{j,L/R}$ ($j=1,2$), corresponding
to each wire. 
In order to apply the boundary state formalism,
we define them on a circle of circumference $\beta$ with the
antiperiodic boundary condition.
Bosonizing each fermion fields,
we introduce 2 boson fields $\varphi_j$ and their duals, $\theta_j$.
They independently obey the twisted periodic identification
of Eq.~(\ref{twist}):
\begin{eqnarray}
\varphi_j &\sim& \varphi_j + \tilde Q_i\sqrt{2}\pi \nonumber \\
\theta_j &\sim& \theta_j + Q_j \sqrt{2}\pi \nonumber \\
\tilde{Q}_j &=&  Q_j  \pmod{2}
\label{twist2V}
\end{eqnarray}
It is convenient to define linear combinations:
\begin{eqnarray}
\Phi_0=\frac{\varphi_1+\varphi_2}{\sqrt 2}
\quad
&{\rm and}&
\quad
\Phi=\frac{\varphi_1-\varphi_2}{\sqrt 2}
\label{eq:rot2phi}
\\
\Theta_0=\frac{\theta_1+\theta_2}{\sqrt 2}
\quad
&{\rm and}&
\quad
\Theta=\frac{\theta_1-\theta_2}{\sqrt 2}
\; .
\label{eq:rot2theta}
\end{eqnarray}
As already discussed in Sec.~\ref{sec:approaches},
as long as the charge is conserved at the junction,
$\Phi_0$ always obeys the Neumann boundary condition.
The Hamiltonian will thus decouple
into separate terms for $\Phi$ and $\Phi_0$, 
with $\Phi_0$ decoupling from the interactions.
Namely, only $\Phi$ and $\Theta$ will appear in the boundary interactions.

However, the periodic identification conditions (referred to as 
``gluing conditions'' in \cite{WongAffleck}) now link these 2 fields, 
as well as linking the $\Phi$ and $\Theta$ fields.
Explicitly, defining:
\bea
Q_0 &\equiv& Q_1+Q_2\nonumber \\
Q   &\equiv & Q_1-Q_2,\label{Q_0,Qdef}\eea
corresponding to $\Phi_0$ and $\Phi$, and 
similarly defining $\tilde{Q}_0$ and $\tilde{Q}$, we see that these obey
the restrictions, which are equivalent to
$Q_j =\tilde{Q}_j \pmod{2}$:
\bea Q_0=\tilde Q_0=Q&=&\tilde Q \pmod{2} \nonumber \\
Q_0+Q+\tilde Q_0+\tilde Q &=& 0 \pmod{4}. \label{twist2}\eea
the mode expansions for the fields $\Phi_0$ and $\Phi$ are:
\bea
\Phi_0 (t,x)&=&\hat \Phi^0_0+{\pi \over \beta}
\left[{1\over g}Q_0t+\tilde Q_0x \right] +
{1\over \sqrt{2g}}\sum_{n=1}^\infty {1\over \sqrt{n}}\left[ 
a_n^{R0}\exp{(-inx_- \frac{2\pi}{\beta} )} +
a_n^{L0}\exp{(-inx_+ \frac{2\pi}{\beta} )} + \mbox{h.c.} \right]
\nonumber \\
\Phi (t,x)&=&\hat \Phi_0+
{\pi \over \beta}\left[{1\over g}Qt+\tilde Qx \right]+
{1\over \sqrt{2g}}\sum_{n=1}^\infty {1\over \sqrt{n}}\left[ 
a_n^{R} \exp{(-inx_- \frac{2\pi}{\beta} )} +
a_n^{L} \exp{(-inx_+ \frac{2\pi}{\beta} )} + \mbox{h.c.}
\right],
\label{mode2}\eea
and similarly for $\Theta_0$, $\Theta$.
The $Q$, $Q_0$, $\tilde Q$ and $\tilde Q_0$ quantum 
numbers obey the gluing conditions of Eq. (\ref{twist2}). 

The FSS is given by two copies of Eq. (\ref{fssFB}).
This can be written as:
\be E={2\pi \over \beta}\left[ {Q_0^2+Q^2\over 8g}+{g(\tilde Q_0^2+\tilde Q^2)
\over 8}+\sum_{m=1}^\infty m(n_m^{L0}+n_m^{R0}+n_m^L+n_m^R)\right],
\label{fssFB2}\ee
where, again the gluing conditions of Eq. (\ref{twist2}) are obeyed. 

Analogous to what was found in Sec.~\ref{sec:approaches}
in the 3-wire case, $\Phi_0$ always obeys N BC's at $x=0$.
This is the ``bare'' BC corresponding to decoupled leads, 
and it is not changed by the interactions since 
$\Phi_0$ does not appear in them. On the other hand, 
$\Phi$ can obey N or D BC's corresponding to  
perfect reflection or perfect transmission through 
the junction, respectively.  

It is straightforward to determine directly the full 
boundary operator spectrum with either set of b.c.'s. 
(Henceforth we will label these two sets of b.c.'s as simply 
N or D.)  In the N case we simply get 2 copies of 
the operator spectrum of a single wire with 
N b.c.'s, since $\varphi_j$ ($j=1,2$) independently 
obey N b.c.'s in this case.  As usual, this 
spectrum of operator dimensions is in one-to-one 
correspondance with the FSS with N b.c.'s at both 
ends of a strip of length $l$.
This is, as in Eq. (\ref{fssFN}):
\be E_N={\pi \over l}
\left[{Q_1^2\over 2g}+{Q_2^2\over 2g}+
\sum_{m=1}^\infty m(n_m^1+n_m^2)\right] 
,\ee
where $Q_1$ and $Q_2$ range independently over all integers. 
Equivalently, we can write this in terms of the charge quantum
numbers $Q_0$ and $Q$ defined in Eq. (\ref{Q_0,Qdef}) and the new
oscillator operators, which are
associated with the fields $\Phi_0$ and $\Phi$:
\be
E_N={\pi \over l}
\left[{Q_0^2\over 4g}+{Q^2\over 4g}+\sum_{m=1}^\infty m(n_m^0+n_m)\right] ,
\label{fssN2}\ee
where now:
\be Q=Q_0 \pmod{2}
\ee
The N boundary state is just a product of
the single fermion N boundary states.
Taking 
the case where both phases $\Theta_0$ and $\Theta_0^0$ are set to zero, this 
can be written as:
\be |N\rangle = g^{1/2}
\exp{\left[ \sum_{n=1}^\infty 
\left( a_n^{L0\dagger}a_n^{R0\dagger}+a_n^{L\dagger}a_n^{R\dagger}
\right)
\right]} 
\sum_{\tilde Q_0,\tilde Q}^{'} 
|(\tilde Q_0,0)\rangle \otimes |(\tilde Q,0)\rangle .\label{NBSF2}
\ee
Here, as follows from Eq.~(\ref{twist2}),
with $Q=Q_0=0$, $\tilde Q_0$ and $\tilde Q$ are even 
and $\tilde Q+\tilde Q_0=0 \pmod{4}$.

The FSS with the D b.c.'s at both ends of the strip is equivalent
to the FSS of a single 
wire with periodic BC's on an interval of length $2l$, as follows 
from an unfolding transformation \cite{WongAffleck}. This is given by
Eq. (\ref{fssFB}) with $\beta$ replaced by $2l$.  We may also 
replace the $n_m^L$, $n_m^R$ quantum numbers by a new basis,
$n_m^0$, and $n_m$, and identify the total charge $Q$
with $Q_0$ and $\tilde Q$ with the same quantity related
to the linear combination $\varphi$ in Eq. (\ref{eq:rot2phi}). 
\be
E_D={\pi \over l}\left[{Q_0^2\over 4g}+{g\tilde Q^2
\over 4}+\sum_{m=1}^\infty m(n_m^0+n_m)\right ],\label{fssD2}\ee
with
\be Q_0=\tilde Q \pmod{2} .\ee
The corresponding boundary state is:
\be |D\rangle = \exp \left[ \sum_{n=1}^\infty 
\left( a_n^{L0\dagger}a_n^{R0\dagger}-a_n^{L\dagger}a_n^{R\dagger} \right) \right] 
\sum_{\tilde Q_0, Q}^{'} 
|(\tilde Q_0,0)\rangle \otimes |(0,Q)\rangle ,
\label{DBSF2}\ee
where the sum is restricted by the gluing conditions of Eq. (\ref{twist2}), 
in the case $Q_0=\tilde Q=0$, to 
$\tilde Q_0$ and $Q$ even with $\tilde Q_0+Q=0$ [mod $4$].

In fact, we are generally only interested in the $Q_0=0$ boundary operators 
since these are the only ones which can appear in the Hamiltonian.
Operators with $Q_0 \neq 0$ corresponds to a vertex operator involving
$\Theta_0$ and thus are forbidden by charge conservation.
Setting $Q_0=0$ in Eq. (\ref{fssN2}) 
or (\ref{fssD2}) we see that $Q$ or $\tilde Q$ respectively must be even. 
The dimensions of the primary operators can thus be written as:
\bea \Delta_N&=& \frac{n^2}{g} \nonumber \\
\Delta_D &=& gm^2,\label{specper}\eea
with $n$ and $m$ arbitrary integers. 
These are precisely the same as the boundary scaling dimensions of 
primary fields for a single {\it periodic boson} with N or D BC's [
Eq. (\ref{Delta_N}) or (\ref{Delta_D}).]
These reproduce the known results on the junction of two
wires.~\cite{Kane,Furusaki,WongAffleck}

We note that, however, in the presence of a resonant
level at the junction, the charge-nonconserving boundary operators
with $Q_0 \neq 0$ should be also considered.
Suppose for example that there is a single resonant level, exactly
at the Fermi energy, at the junction.
One may imagine the two wires connected to a quantum dot. 
Let $d$ and $d^\dagger$ be the annihilation/creation operator of
an electron in that level.
Then the hopping of an electron between a wire and the dot
can be written as $\psi_j d^{\dagger} + \mbox{h.c.}$.
As $d$ and $d^{\dagger}$ have zero scaling dimension,
the scaling dimension of the hopping operator is determined
by the electron annihilation operator $\psi_j$ at the
end of the wire.
Therefore, in the presence of the resonant level,
even if the charge is overall conserved,
we have to consider the ``charge-nonconserving'' boundary operators.
Indeed, the effect of the resonant tunneling at the junction
has been studied in Refs.~\cite{Kane,Furusaki} for the case of two
wires and in Ref.~\cite{Nayak} for three wires.

Nevertheless, in this paper, we focus on the problem without the resonant
levels. We discuss only the charge-conserving boundary operators,
for which we develop a simplified treatment in the following.
It should be kept in mind, however, that it does not cover the
resonant case.

\subsubsection{Projection to an effective single component boson}

While we have succeeded in correctly reproducing known results on
the junction of two wires, the gluing conditions obviously
complicates the calculation.
On the other hand, if we restrict the analysis to
the charge-conserving boundary operators,
there is a simpler method to find the dimensions of such boundary
operators.
This will be generalized in the next sub-section 
to the 3-wire case.

Note that in any boundary state which is N for $\Phi_0$, 
only $Q_0=0$ states occur.
The restrictions of Eq. (\ref{twist2}) reduce to:
\bea \tilde Q_0=Q=\tilde Q&=&0 \pmod{2} \label{eq:twistQmod2}
\\
Q+\tilde Q+\tilde Q_0 &=& 0 \pmod{4}. \label{twistQ0}\eea
Defining:
\bea \tilde Q_0&\equiv& 2\tilde n_0\nonumber \\
Q&=&2n\nonumber \\
\tilde Q &=& 2\tilde n,\eea
these integers obey the condition:
\be n+\tilde n = \tilde n_0 \pmod{2}. \ee
A general boundary state of this type can be written:
\be
|B \rangle = |N^0_0 \rangle \otimes
\sum_{\tilde m,m \in Z }|\psi_{4\tilde m,4m} \rangle
+ |N^1_0 \rangle \otimes
\sum_{\tilde m,m \in Z } |\psi_{4\tilde m+2,4m+2}\rangle.
\label{bsent}
\ee
Here $|\psi_{\tilde Q,Q}\rangle$
are states in the Hilbert space of $\Phi$, 
proportional to $|(\tilde Q,Q)\rangle$.  Actually
only $|\psi_{2\tilde n,0}\rangle$ is non-zero
in the N state and only $|\psi_{0,2n}\rangle$  
is non-zero
in the D state but we since we wish to consider both cases at once 
we use this more general notation. 
$|N^0_0\rangle$ and $|N^1_0\rangle$ are N states in the Hilbert 
space of $\Phi_0$, given explicitly by:
\bea
|N^0_0\rangle &=& (2g)^{1/4}
\exp{\left[ \sum_{n=1}^\infty a_n^{L0\dagger}a_n^{R0\dagger}\right]}
\sum_{\tilde m_0\in Z} |(4\tilde m_0,0)\rangle \nonumber \\
|N^1_0 \rangle &=& (2g)^{1/4}
\exp{\left[ \sum_{n=1}^\infty a_n^{L0\dagger}a_n^{R0\dagger}\right]}
\sum_{\tilde m_0\in Z}|(4\tilde m_0+2,0) \rangle .
\eea
Here we have let $\tilde n_0=2\tilde m_0$ in $|N^0_0 \rangle$ and 
$\tilde n_0=2\tilde m_0+1$ in $|N^1_0 \rangle$.
The diagonal partition function is given by:
\be Z_{BB}=Z_{N^0_0}\sum_{\tilde m,m}Z_{4\tilde m,4m}+Z_{N^1_0}
\sum_{\tilde m,m \in Z }Z_{4\tilde m+2,4m+2}.\ee
Here $Z_{N^i_0}$ is a diagonal partition function in the $\Phi_0$ 
Hilbert space and the second factors are diagonal partition functions 
in the $\Phi$ Hilbert space:
\be Z_{2\tilde n,2n}\equiv \langle \psi_{2\tilde n,2n}|
\exp [-lH^P_\beta ]|\psi_{2\tilde n,2n}\rangle .\ee
The former are given explicitly by:
\bea
Z_{N^0_0} &=&
{\sqrt{2g} \over \eta (\tilde q)}\sum_{\tilde m_0 \in Z}
\tilde q^{g\tilde m_0^2}
={1\over \eta (q)}\sum_{Q_0 \in Z }q^{{Q_0}^2/(4g)}\nonumber \\
Z_{N^1_0}&=&
{\sqrt{2g} \over \eta (\tilde q)}
\sum_{\tilde m_0 \in Z }\tilde q^{g (2 \tilde{m}_0+1)^2/4}
={1\over \eta (q)}\sum_{Q_0 \in Z}e^{i\pi Q_0}q^{{Q_0}^2/(4g)}.\eea

The complete spectrum of boundary operators can be read off
from the partition function $Z_{BB}$.
However, the scaling dimension coming from $Z_{N^0_0}$ and
$Z_{N^1_0}$ corresponds to a boundary operator involving the
center-of-mass field $\Phi_0$.
In particular, those includes a nontrivial dimensions ${Q_0}^2/(4g)$
comes from the vertex operator $e^{i Q_0 \Phi_0/2}$ which
changes the total charge.
Thus they do not appear if we consider only the perturbations
which respects the charge conservation at the junction.
The other operators with integer dimensions corresponding to
the expansion of $1 / \eta(q)$ comes from derivatives
of $\Phi_0$. They are all irrelevant or marginal.

Thus, in order to identify potentially relevant operators,
we can consider the boundary operators which involves
the dynamical field $\Phi$ only.
For this purpose, we may replace both $Z_{N^0_0}$ and $Z_{N^1_0}$ by
\be
  Z_{N^0_0},Z_{N^1_0}\to 1
\ee
The resultant effective partition function is
\be 
   Z_{BB} \to \sum_{\tilde n,n}Z_{2\tilde n,2n},
\label{eq:ZBBred}
\ee
from which we can read off the spectrum of boundary operators
involving $\Phi$ only.

This ``reduced partition function''
can be obtained by simply ignoring the $\Phi_0$ sector 
and dropping the constraint of Eq. (\ref{twistQ0}), thus allowing 
$n$ ($=Q/2$) and $\tilde n$ ($=\tilde Q/2$) to be arbitrary integers.
Namely, Eq.~(\ref{eq:ZBBred}) is identical to the
diagonal partition function arising
from a ``reduced'' boundary state 
entirely in the $\Phi$ sector of the form:
\be |B \rangle_R \propto \sum_{\tilde n,n}|\psi_{2\tilde n,2n} \rangle,
\label{bsunent}
\ee
in which the second constraint of Eq. (\ref{twistQ0}) has been simply ignored. 

Thus, to obtain the dimensions of boundary operators in $\Phi$ sector,
we may simply forget about the second constraint of Eq. (\ref{twistQ0})
and hence replace the entangled boundary state of Eq. (\ref{bsent}) by
the simpler reduced one, written only in the $\Phi$ sector, of 
Eq. (\ref{bsunent}).  This corresponds to a boundary state for 
a single boson obeying standard periodic b.c.'s.  In particular 
it leads to the primary operators spectra of Eq. (\ref{specper}).
Therefore, as long as we restrict ourselves to $Q_0=0$ operators,
we may drop the $\Phi_0$ boson entirely and ignore the
twisted structure that results from Fermi statistics.
This simplicity appears to be related to the 
fact that the Klein factors are trivial for a 2 wire junction. 

In contrast to the above,
we will see that a similar reduction leaves a nontrivial effect for
the junction of three wires.

\subsection{Three wires}

Let us finally turn to the model of interest here: a 3-wire junction.  
As in Subsection~\ref{sec:twist2wires}
on the 2-wire junction, we define 3 fermion fields
on a circle of circumference $\beta$.
Upon bosonizing, we introduce 
3 fields and their duals, $\varphi_j$ and $\theta_j$, which obey the 
fermionic type (twisted) periodic identification of Eq. (\ref{twist2V}) where 
now $j=1, 2$ or $3$. We then introduce the 3 linear combinations of 
fields $\Phi_j$ as in Eq.~(\ref{eq:newbasisphi}). We then combine $\Phi_1$ 
and $\Phi_2$ into a 2-component vector field, $\vec \Phi$ and 
similarly for $\vec \Theta$. We now consider the mode expansions 
for the fields $\vec \Phi$, $\Phi_0$, $\vec \Theta$ and $\vec \Theta_0$.
We introduce charge and dual charge quantum numbers $Q_j$ and $\tilde Q_j$ 
for each of the original boson, $\varphi_j$,
as in Eq. (\ref{modeg}), (\ref{mndef}).
We define the total charge and dual charge:
\bea Q_0&\equiv& Q_1+Q_2+Q_3, \nonumber \\
\tilde Q_0 &\equiv& \tilde Q_1+\tilde Q_2+\tilde Q_3.\eea
We are only interested in b.c.'s which are N for $\Phi_0$, 
so that the corresponding b. states are constructed from 
$Q_0=0$ states. It is convenient to parameterize the vectors 
of integers, $Q_i$ satisfying this condition by:
\be
  (Q_1,Q_2,Q_3) \equiv m_1(0,1,-1)+m_2(-1,0,1).
\ee
Clearly $m_1=Q_2$, $m_2=-Q_1$ and letting the $m_i$ run 
over all integers uniquely parameterizes all $Q_i$ satisfying 
$Q_0=0$. On the other hand, $\tilde Q_0$ need not be zero 
and it is convenient to parameterize the $\tilde Q_i$ 
in a different way:
\be
 (\tilde Q_1,\tilde Q_2,\tilde Q_3)=n_0(1,1,1)-n_1(0,1,1)-n_2(1,0,1).
\ee 
This can be inverted to give:
\be
 (n_0,n_1,n_2)=(\tilde Q_1+\tilde Q_2-\tilde Q_3,-\tilde Q_3+\tilde Q_1,
                 -\tilde Q_3+\tilde Q_2).
\ee
Thus any set of integer dual charges, $\tilde Q_j$, can be uniquely expressed 
in terms of the integers, $n_j$. The conditions $Q_j=\tilde Q_j \pmod{2}$
imply:
\bea
  n_0&=&2p_0\ \  (\hbox{i.e. $n_0$ is even})\nonumber \\
  n_j&=&m_j \pmod{2}.
\label{twist3}
\eea
In this parameterization, the total dual charge is:
\be \tilde{Q}_0=2(3p_0-n_1-n_2).\ee
This implies that $\tilde Q_0$ is even and:
\be \tilde Q_0/2 = -n_1-n_2 \pmod{3}. \ee
Thus we see that a general boundary state which is N
with respect to $\Phi_0$ will be of the form:
\be
|B \rangle = \sum_{k=-1,0,1}|N^k_0 \rangle \otimes
\sum_{n_1+n_2=-j \pmod{3}}' |\psi(n_1,n_2,m_1,m_2) \rangle.
\ee
Here the $'$ over the summation sign indicates that,
in addition to the explicitly 
noted restriction, the conditions:
\be
  n_k=m_k \pmod{2}, \ \  j=1,2
\label{nmres}
\ee 
must also be imposed. 
The states $|\psi(n_1,n_2,m_1,m_2)\rangle$
are in the sector of the 2-component 
field $\vec \Phi$, with charges and dual charges labelled by the 
integers $m_j$ and $n_j$.  The states $|N^k_0 \rangle$,
for $k=-1,0$ or $1$, 
are in the $\Phi_0$ sector and are given explicitly by:
\be |N^k_0 \rangle \equiv
\left(\frac{3g}{2} \right)^{1/4}
\exp{\left[ \sum_{n=1}^\infty a_n^{L0\dagger}a_n^{R0\dagger}\right]}
\sum_{\tilde r_0}|(2(3\tilde r_0+k),0)\rangle.
\ee
The corresponding ``partial Neumann'' partition functions
in the $\Phi_0$ sector 
may be readily calculated. 
\be
Z_{N^k_0}= \sqrt{3g}
{1\over \eta (\tilde q)}\sum_{\tilde r_0 \in Z} \tilde q^{g(3\tilde r_0+k)^2/6}
={1\over \eta (q)}\sum_{Q_0 \in Z }e^{-2\pi i Q_0 k /3}q^{{Q_0}^2/(6g)} .
\ee
As in Subsection~\ref{sec:twist2wires},
we are only interested in the
boundary operators in the $\vec{\Phi}$ sector,
and replace $Z_{N^k_0}$ by unity.

We may equivalently get the dimensions of 
all such primary boundary operators by using a reduced boundary state, which 
exists in the reduced Hilbert space of the 2 component boson field,
$\vec \Phi$.  
That is:
\be
  |B \rangle \to \sum'|\psi(n_1,n_2,m_1,m_2)\rangle.
\label{eq:redbs3w}
\ee 
The restrictions of Eq.~(\ref{nmres}) remain, but the other conditions, 
that entangle the $\vec \Phi$ sector with the $\Phi_0$ sector disappear 
upon restricting to boundary states with $Q_0=0$ and boundary 
operators with $Q_0=0$. 

The windings $\Delta \vec{\Phi} \equiv \vec{\Phi}(\beta) - \vec{\Phi}(0)$ and
$\Delta \vec{\Theta} \equiv \vec{\Theta}(\beta) - \vec{\Theta}(0)$ along the
boundary follow from the integers $(n_1,n_2,m_1,m_2)$. $\Delta \vec{\Phi}$
and $ \Delta \vec{\Theta}$ can be obtained from the windings of the three
pairs of boson fields, $\Delta \varphi_i=\tilde Q_i\sqrt{2}\pi$ and $\Delta
\theta_i=Q_i\sqrt{2}\pi$, using the rotations in
Eqs.(\ref{eq:newbasisphi},\ref{eq:newbasistheta}) and the vectors ${\vec
K}_j$ and ${\vec R}_j$ defined, respectively, in Eqs.(\ref{eq:Kvec}) and
(\ref{eq:Rdef}). One obtains
\begin{eqnarray}
\vec{\Phi}(\beta) - \vec{\Phi}(0) &=&
2 \pi (n_1 \frac{\vec{R}_1}{2} + n_2 \frac{\vec{R}_2}{2}) \\
 \vec{\Theta}(\beta) - \vec{\Theta}(0) &=&
  2 \pi (m_1 \vec{K}_1 + m_2 \vec{K}_2).
\end{eqnarray}
Thus the
``reduced boundary state''~(\ref{eq:redbs3w})
would arise from a 2-component boson field with the mode expansion:
\be
\vec{\Phi} (t,x) =
{2\pi\over \beta}\left[{1\over g}\sum_{j=1,2} m_j \vec{K}_j t
+{1 \over 2}\sum_{j=1,2} n_j \vec{R}_j x\right] 
+{1\over \sqrt{2g}}\sum_{n=1}^\infty {1\over \sqrt{n}}\left[ 
\vec a_n^L\exp [-inx_+ {2\pi \over \beta}]+
\vec a_n^R\exp [-inx_- {2\pi \over \beta} ]+ h.c.
\right],
\label{modered}
\ee
where the integers $n_j$ and $m_j$ 
are restricted by Eq. (\ref{nmres}). 
The corresponding energies are:
\be E={2\pi \over \beta}\left[ {1\over 2g}\left( \sum_j m_i\vec K_i\right)^2
+{g\over 2}\left({\sum_j n_j\vec R_i\over 2}\right)^2+\ldots \right],
\label{specred}\ee
where the $\ldots$ represents the oscillator mode terms. Of 
course the restriction of Eq. (\ref{nmres}) must be applied. 

Let us compare the mode expansion~(\ref{modered})
to the corresponding Eq.~(\ref{eq:modec})
for the standard compactified boson, with
$\Lambda$ being the triangular lattice spanned by $\vec{R}_1$
and $\vec{R}_2$.
While they are similar, we find that they are in fact different
and cannot be identical with any choice of the compactification
lattice $\Lambda$ for the standard free boson.
Namely, the fermionic nature of 
the problem has modified the mode expansion, via the 
factor of 1/2 multiplying the second term inside
the first square bracket in Eq. (\ref{modered}) and the restriction 
of Eq. (\ref{nmres}) on the quantum numbers $n_i$ and $m_i$.
In contrast 
to the two wire case, the fermionic nature of the problem 
survives and makes important changes to the boundary states 
and the corresponding ``reduced mode expansion''
of Eq. (\ref{modered}).
This  may be related to the fact that the Klein factors 
cannot be eliminated for the 3-wire junction whereas 
they can be eliminated in the 2-wire case. 

\end{widetext}

\section{The method of delayed evaluation of boundary condition (DEBC)}
\label{sec:DEBC}

In this section we introduce the DEBC method. The method consists of doubling
the number of degrees of freedom in the problem by keeping both the bosonic
fields $\phi_i$ and its conjugate momentum $\theta_i$, for each semi-infinite
lead $i$. The redundancy is eliminated by imposing the boundary condition at
$x=0$, which cuts in half the number of degrees of freedom, down to the
correct number for the semi-infinite wires. The basic idea behind the method
is that we delay the choice of boundary condition until after we write the
operators corresponding to the (multi)particle tunneling processes at the
junction. The correct (or stable) boundary condition that must be imposed on
$\phi_i$ and $\theta_i$ is determined {\it a posteriori}, taking into account
the scaling dimensions of the tunneling operators given the different choices
of boundary conditions.

Explicitly, we write the electron operators, both right and left movers (or
incoming to and outgoing from the junction at $x=0$), for each of the
semi-infinite leads, labelled by $i$, defined in the half line $x>0$:
\begin{equation}
\Psi^{L,R}_i\propto e^{i\frac{1}{\sqrt{2}}(\varphi_i \pm \theta_i)}
\ .
\label{eq:RLdef}
\end{equation}
Klein factors can be absorbed into the zero mode terms of the mode
expansion of the $\varphi_i$ and $\theta_i$ bosons. For example, the
fermionic commutations between different leads can be taken care of by
imposing that $[\varphi^{(0)}_i,\varphi^{(0)}_j]=i2\pi\alpha_{ij}$,
with an anti-symmetric $\alpha_{ij}=-\alpha_{ji}$ and unit entries
$\alpha_{ij}=\pm 1$. (A different represention of Klein factors using
zero modes is introduced in Appendix \ref{sec:fqh} when we discuss
fractional quantum Hall junctions.) The $\Theta_i$ fields can 
be taken to commute for $i\neq j$. 

Notice that the above construction, in terms of both sets of bosonic fields
$\varphi_i$ and $\theta_i$, is the usual one when describing systems on the
infinite line, and thus we are working with twice the number of degrees of
freedom for the semi-infinite line. What is missing is the boundary condition
at the origin $x=0$ (the choice of which we are delaying) which contraints
$\varphi_i$ and $\theta_i$ and thus cuts in half the number of degrees of
freedom for systems on the half-line.

The subsequent step is to construct all tunneling operators between the leads
using the electron operators in Eq. (\ref{eq:RLdef}), keeping both $\varphi$
and $\theta$ fields, and imposing later the appropriate boundary condition.

For concreteness, let us present below the DEBC method through
an example, using
the simpler problem of the two-wire junction for this purpose.
In the next section, we
apply the method to the problem of interest in this paper, the three-wire
junction.


For the two-wire problem, the single particle tunneling terms at
$x=0$ can be expressed in terms of the $\Psi^{R,L}_j$ as:
\begin{eqnarray}
T^{RL}_{21}&&= {\Psi^{R}_2}^\dagger\;\Psi^{L}_1\big|_{x=0}
\propto e^{i\frac{\varphi_1-\varphi_2}{\sqrt{2}}}
\;e^{i\frac{\theta_1+\theta_2}{\sqrt{2}}}
\label{term-tun12}
\\
T^{LL}_{21}&&= {\Psi^{L}_2}^\dagger\;\Psi^{L}_1\big|_{x=0}
\propto e^{i\frac{\varphi_1-\varphi_2}{\sqrt{2}}}
\;e^{i\frac{\theta_1-\theta_2}{\sqrt{2}}}
\label{term-tuntilde12}
\\
T^{RL}_{11}&&= {\Psi^{R}_1}^\dagger\;\Psi^{L}_1\big|_{x=0}
\propto 
e^{i\sqrt{2}\;\theta_1}
\label{term-back}
\; .
\end{eqnarray}
The fields $\varphi_i$ and $\theta_i$ above are all evaluated at the
boundary $x=0$. Other terms can be obtained by simply taking the
Hermitian conjugate or by exchanging $RL$ in these three combinations
above. Physically, these terms correspond to tunneling between the
leads [terms (\ref{term-tun12}) and (\ref{term-tuntilde12})], and
backscattering [term (\ref{term-back})].

Written in terms of the rotated basis
Eqs.~(\ref{eq:rot2phi},\ref{eq:rot2theta}), we have
\begin{eqnarray}
T^{RL}_{21}&&
\propto e^{i\Phi}
\;e^{i\Theta_0}
\propto 
e^{i\Phi}
\label{Bterm-tun12}
\\
T^{LL}_{21}&&
\propto e^{i\Phi}
\;e^{i\Theta}
\label{Bterm-tuntilde12}
\\
T^{RL}_{11}&&
\propto 
e^{i(\Theta + \Theta_0)}
\propto 
e^{i\Theta}
\label{Bterm-back}
\; ,
\end{eqnarray}
where we have used the fact that charge conservation requires that
$\Phi_0$ must obey the Neumann boundary condition, or equivalently,
$\Theta_0$ must obey the Dirichlet boundary condition
($\Theta_0$ is pinned at the boundary).

The boundary conditions on the other pair of bosons, $\Phi$ and
$\Theta$, can be expressed in terms of chiral fields as 
\be \phi_R={\cal
R}\phi_L+C,\ee
 with $\Phi=\frac{1}{\sqrt{g}}(\phi_R+\phi_L)$ and
$\Theta=\sqrt{g}\;(\phi_R-\phi_L)$, where ${\cal R}$ is a
rotation and $C$ is a constant.
 Since we just have a single pair of ${RL}$ fields, the only
possibilities are ${\cal R}=\pm 1$. We will soon see that in the
problem of three leads there will be two pairs of ${RL}$ boson fields,
and more general rotations ${\cal R}_\delta$ are allowed.

The scaling dimension of a general vertex operator
\begin{equation}
V=e^{i a  \sqrt{g}\;\Phi +i  b\frac{1}{\sqrt{g}}\;\Theta}=
e^{i(a+ b) \phi_R
+ i(a- b)  \phi_L}
\; ,
\end{equation}
is given by
\begin{equation}
\label{eq:dim1D}
\Delta_V={1\over 4}
|{\cal R}(a+b)+(a-b)|^2
\; .
\end{equation}
For $N$ boundary conditions, $\Theta$ is pinned, ${\cal R}=1$ and
$\Delta_V=a^2$. For $D$ boundary conditions, $\Phi$ is pinned instead,
${\cal R}=-1$ and $\Delta_V=b^2$. These results for $\Delta_V$ could
have been easily guessed, but going through the exercise with the
rotation ${\cal R}$ is useful as a warm up for the three lead case.

We now analyze the scaling dimensions of the three sets of operators
(\ref{Bterm-tun12})-(\ref{Bterm-back}) for different boundary
conditions on $\Phi$.

\vspace{3mm}
$\bullet$ {\it Neumann boundary condition} -- In this case
\begin{equation}
\Delta_{T^{RL}_{21}}=g^{-1}
\quad , \quad
\Delta_{T^{LL}_{21}}=g^{-1}
\quad , \quad
\Delta_{T^{RL}_{11}}=0.
\end{equation}
The operator $T^{RL}_{11}$ is proportional to the identity (dimension
zero), which is consistent with the fact that it is this
backscattering operator that is responsible for pinning the $\Theta$
field for $N$ boundary condition (on $\Phi$). This is the perfect
backscattering fixed point, stable as long as the other two operators
${T^{RL}_{21}}$ and ${T^{LL}_{21}}$ are irrelevant, which is the case
for $g<1$.

\vspace{3mm}
$\bullet$ {\it Dirichlet boundary condition} -- In this case
\begin{equation}
\Delta_{T^{RL}_{21}}=0
\quad , \quad
\Delta_{T^{LL}_{21}}=g
\quad , \quad
\Delta_{T^{RL}_{11}}=g.
\end{equation}
Now it is the operator $T^{RL}_{21}$ which is proportional to the
identity (dimension zero), and this transmission operator is
responsible for pinning the $\phi$ field for $D$ boundary
condition. This is the perfect transmission fixed point, stable as
long as the other two operators ${T^{LL}_{21}}$ and ${T^{RL}_{11}}$
are irrelevant, which is the case for $g>1$.

It is clear from this approach that the two-wire junction can be
understood in terms of the boson fields $\Phi$ and $\Theta$, which
are dual to each other.
They can be regarded as subject to the standard compactification
$\Phi \sim \Phi + 2 \pi$ and $\Theta \sim \Theta + 2\pi$,
as can be seen from
Eqs.~(\ref{term-tun12},\ref{term-tuntilde12},\ref{term-back}).
Also, the DEBC approach reproduces rather simply the results for a junction
between two wires as in the Kane-Fisher
model~\cite{Kane}. We now apply it to the three-wire junction.


\section{Fixed points for the junction of three wires}
\label{sec:fpts}
Now let us discuss the RG fixed points in the junction of three wires.
As we have discussed, each fixed point corresponds to a conformally
invariant boundary condition of the boson field theory.

In the DEBC approach, we need the $x=0$ tunneling operators between the
three wires.
They are easily expressible in terms of the vectors $\vec K_j$
defined in Eq.~(\ref{eq:Kvec}), with the two-component field $\vec{\Phi}$
defined in Eqs.~(\ref{eq:newbasisphi},\ref{eq:newbasistheta}). 

The tunneling operators involving
the $RL$ combinations can be split into three groups:

$\bullet$ $+$ cycle:
\begin{eqnarray}
T^{RL}_{21}={\Psi^{R}_2}^\dagger\Psi^{L}_1\big|_{0}
\propto e^{i{\vec K_3}\cdot\vec\Phi}\;
e^{i\frac{1}{\sqrt{3}}
{\hat z}\times{\vec K_3}\cdot\vec\Theta}\;e^{i\sqrt{\frac{2}{3}}\Theta_0}
&&
\\
T^{RL}_{32}={\Psi^{R}_3}^\dagger\Psi^{L}_2\big|_{0}
\propto e^{i{\vec K_1}\cdot\vec\Phi}\;
e^{i\frac{1}{\sqrt{3}}
{\hat z}\times{\vec K_1}\cdot\vec\Theta}\;e^{i\sqrt{\frac{2}{3}}\Theta_0}
&&
\\
T^{RL}_{13}={\Psi^{R}_1}^\dagger\Psi^{L}_3\big|_{0}
\propto e^{i{\vec K_2}\cdot\vec\Phi}\;
e^{i\frac{1}{\sqrt{3}}
{\hat z}\times{\vec K_2}\cdot\vec\Theta}\;e^{i\sqrt{\frac{2}{3}}\Theta_0}
&&
\end{eqnarray}

$\bullet$ $-$ cycle:
\begin{eqnarray}
T^{RL}_{12}={\Psi^{R}_1}^\dagger\Psi^{L}_2\big|_{0}
\propto e^{-i{\vec K_3}\cdot\vec\Phi}\;
e^{i\frac{1}{\sqrt{3}}
{\hat z}\times{\vec K_3}\cdot\vec\Theta}\;e^{i\sqrt{\frac{2}{3}}\Theta_0}
&&
\\
T^{RL}_{23}={\Psi^{R}_2}^\dagger\Psi^{L}_3\big|_{0}
\propto e^{-i{\vec K_1}\cdot\vec\Phi}\;
e^{i\frac{1}{\sqrt{3}}
{\hat z}\times{\vec K_1}\cdot\vec\Theta}\;e^{i\sqrt{\frac{2}{3}}\Theta_0}
&&
\\
T^{RL}_{31}={\Psi^{R}_3}^\dagger\Psi^{L}_1\big|_{0}
\propto e^{-i{\vec K_2}\cdot\vec\Phi}\;
e^{i\frac{1}{\sqrt{3}}
{\hat z}\times{\vec K_2}\cdot\vec\Theta}\;e^{i\sqrt{\frac{2}{3}}\Theta_0}
&&
\end{eqnarray}

$\bullet$ backscattering:
\begin{eqnarray}
T^{RL}_{11}&=&{\Psi^{R}_1}^\dagger\Psi^{L}_1\big|_{0}
\propto e^{-i\frac{2}{\sqrt{3}}
{\hat z\times\vec K_1}
\cdot\vec\Theta}\;e^{i\sqrt{\frac{2}{3}}\Theta_0}
\\
T^{RL}_{22}&=&{\Psi^{R}_2}^\dagger\Psi^{L}_2\big|_{0}
\propto e^{-i\frac{2}{\sqrt{3}}
{\hat z\times\vec K_2}
\cdot\vec\Theta}\;e^{i\sqrt{\frac{2}{3}}\Theta_0}
\\
T^{RL}_{33}&=&{\Psi^{R}_3}^\dagger\Psi^{L}_3\big|_{0}
\propto e^{-i\frac{2}{\sqrt{3}}
{\hat z\times\vec K_3}
\cdot\vec\Theta}\;e^{i\sqrt{\frac{2}{3}}\Theta_0}
\end{eqnarray}
The $LL$ and $RR$ combinations lead to another group:

$\bullet$ $LL-RR$:
\begin{eqnarray}
T^{LL}_{21}&=&{\Psi^{L}_2}^\dagger\Psi^{L}_1\big|_{0}
\propto e^{i{\vec K_3}\cdot\vec\Phi}\;
e^{+i{\vec K_3}\cdot\vec\Theta}
\\
T^{LL}_{32}&=&{\Psi^{L}_3}^\dagger\Psi^{L}_2\big|_{0}
\propto e^{i{\vec K_1}\cdot\vec\Phi}\;
e^{+i{\vec K_1}\cdot\vec\Theta}
\\
T^{LL}_{13}&=&{\Psi^{L}_1}^\dagger\Psi^{L}_3\big|_{0}
\propto e^{i{\vec K_2}\cdot\vec\Phi}\;
e^{+i{\vec K_2}\cdot\vec\Theta}
\\
T^{RR}_{21}&=&{\Psi^{R}_2}^\dagger\Psi^{R}_1\big|_{0}
\propto e^{i{\vec K_3}\cdot\vec\Phi}\;
e^{-i{\vec K_3}\cdot\vec\Theta}
\\
T^{RR}_{32}&=&{\Psi^{R}_3}^\dagger\Psi^{R}_2\big|_{0}
\propto e^{i{\vec K_1}\cdot\vec\Phi}\;
e^{-i{\vec K_1}\cdot\vec\Theta}
\\
T^{RR}_{13}&=&{\Psi^{R}_1}^\dagger\Psi^{R}_3\big|_{0}
\propto e^{i{\vec K_2}\cdot\vec\Phi}\;
e^{-i{\vec K_2}\cdot\vec\Theta}
\end{eqnarray}
To these sets of operators we must add their Hermitian conjugates.

In the following, we consider several conformally invariant
boundary conditions which can be expressed as
\be
\vec{\phi}_R={\cal R}_{\xi}\;\vec{\phi}_L\label{rotphi}
+\vec C, \ee
 where
\begin{equation}
{\cal R}_\xi =\left(
\begin{array}{ll}\cos\xi&-\sin\xi\\ \sin\xi&\cos\xi \end{array}
\right),
\label{R_xi}
\end{equation}
\begin{eqnarray}
 \vec{\Phi}&=&\frac{1}{\sqrt{g}}\;(\vec \phi_R+\vec \phi_L) \\
 \vec{\Theta} &=& \sqrt{g}\;(\vec \phi_L-\vec \phi_R)
\end{eqnarray}
and $\vec C$ represents a constant vector. 
$\Theta_0$ is pinned by charge conservation, and
can be ignored in all operators above.

Consider a general operator
\begin{equation}
V=e^{i\vec a \sqrt{g}\;\vec \Phi +i \vec b\frac{1}{\sqrt{g}}\;\vec\Theta}=
e^{i(\vec a - \vec b)\cdot \vec \phi_R
+ i(\vec a + \vec b)\cdot \vec \phi_L}
\; .
\end{equation}
For the boundary condition~(\ref{rotphi}), its
boundary scaling dimension is:
\begin{eqnarray}
\label{eq:dim}
\Delta_V
&&={1\over 4}\;
|{\cal R}^{-1}_\xi(\vec a - \vec b)+(\vec a + \vec b)|^2
\\
&&={1\over 4}\;
|{\cal R}_\xi(\vec a+\vec b)+(\vec a-\vec b)|^2
\\
&&={1\over 2}\;\big[
|\vec a|^2+|\vec b|^2
+({\cal R}_\xi\vec a)\cdot\vec a
-({\cal R}_\xi\vec b)\cdot\vec b
\nonumber
\\
&&\:\:\:\:\:\:\:\:
-({\cal R}_\xi\vec a)\cdot\vec b
+({\cal R}_\xi\vec b)\cdot\vec a
\big]
\nonumber
\end{eqnarray}

With the scaling dimension of the tunneling operators in hand, we can now
analyze the stability of the fixed points that are characterized by a choice
of boundary condition.
We will carry out this analysis in the following
section, where we present, discuss, and parallel the results obtained using
both BCFT taking into account the twisted structure of the Hilbert space and
the DEBC approach.

\subsection{Disconnected fixed point --
      Neumann boundary condition on $\vec\Phi$}

The choice $\xi=0$ in Eq.~(\ref{rotphi}) 
pins $\vec\Theta=\vec C$ (back-reflection condition).
This amounts to the Dirichlet boundary condition on
$\vec{\Theta}$, equivalently the Neumann boundary condition on
$\vec{\Phi}$.

Let us calculate the scaling dimensions for the $\pm$
cycle operators. Since ${\cal R}_0={\bf 1}$, $\Delta_V=|\vec
a|^2$. Take $T^{RL}_{21}$ as an example, so $\vec a=\frac{1}{\sqrt
g}\vec K_3$ and $\vec b=\sqrt{\frac{g}{3}}\hat z\times\vec K_3$ enter
in Eq.~(\ref{eq:dim}):
\begin{equation} 
\Delta_{T^{RL}_{21}}=\frac{1}{g}\;|\vec K_3|^2=\frac{1}{g}
\end{equation}
The result is the same for all  6 $\pm$ cycle operators, as well as for
the 6 $LL-RR$ operators. Upon fixing the boundary condition, many of these
operators are redundant; the number of independent operators can be easily
accounted for, and we find that there are only 3 independent operators plus
their 3 Hermitian conjugates. All these 6 independent operators are
irrelevant for $g<1$, and so the backscattering fixed point is stable for
this range of $g$.

The case just discussed above is the simplest boundary conditions for
the problem of the junction of three wires. It corresponds to
disconnected wires. Indeed, Neumann boundary condition on $\vec{\Phi}$
(or Dirichlet on $\vec{\Theta}$) is equivalent to Neumann boundary
condition on all $\varphi_{1,2,3}$ (or Dirichlet on $\theta_{1,2,3}$).

Let us now connect this to the standard BCFT approach, and construct
the boundary state and the cylinder amplitude explicitly. As it is
reviewed in App.~\ref{sec:refboson} for the standard free boson,
the Neumann boundary state is given by a superposition of Ishibashi
states.
Since the boundary condition for $\vec{\Theta}$ is Dirichlet,
the boundary state must have zero winding $\Delta \vec{\Theta}=0$,
namely $m_1=m_2=0$.
Let we define the bosonic Ishibashi state
\begin{equation}
| (n_1,n_2,0,0) \rangle\rangle
\equiv
\exp{\left[ \sum_n \vec{a}^{L\dagger}_n \cdot \vec{a}^{R\dagger}_n \right]}
| (n_1, n_2, 0,0) \rangle,
\end{equation}
where $| (n_1,n_2,0,0)\rangle$ is the
oscillator vacuum with the quantum number $(n_1,n_2,m_1=0,m_2=0)$.

The most stable Neumann-type boundary state would be given by the
linear superposition over all the possible Ishibashi states.
The important consequence of the twisted structure is that,
because of the condition~(\ref{nmres}), 
$n_1$ and $n_2$ cannot run over all integers when $m_1=m_2=0$.
Instead, they can only take even integers values.
Thus the most stable N boundary is given by
\begin{equation}
| N \rangle =
g_N \sum_{n'_1,n'_2 \in \mathbf{Z}} | (2 n'_1, 2 n'_2,0,0 ) \rangle \rangle,
\end{equation}
where the prefactor $g_N$ represents
the ground-state degeneracy~\cite{gtheorem} of the N boundary state.
(It is {\em not} the Tomonaga-Luttinger parameter $g$
introduced in Eq.~(\ref{eq:S0}).)

The calculation of the cylinder amplitude with
the Neumann boundary state at the both ends then follows,
exactly as in the case of the standard boson.
The diagonal cylinder amplitude is given by
\begin{eqnarray}
Z_{NN}(\tilde{q}) &=& \langle N | e^{-l H^P_{\beta}} | N \rangle
\nonumber \\
&=& \frac{{g_N}^2}{\eta(\tilde{q})^2}
\sum_{n'_1, n'_2 \in \mathbf{Z}}
\tilde{q}^{ \frac{g}{4} (n'_1 \vec{R}_1 + n'_2 \vec{R}_2)^2 }.
\end{eqnarray}
This can be identified with the
corresponding partition function~(\ref{eq:NBCc}) 
of the standard compactified free boson,
with $\Lambda$ being the triangular lattice of lattice constant
$2/\sqrt{3}$, spanned by $\vec{R}_1$ and $\vec{R}_2$.
Thus we find
\begin{equation}
g_N = \left( \frac{g^2}{3}\right)^{1/4},
\end{equation}
because $V_0(\Lambda) = 2/\sqrt{3}$.
The modular transform of Eq.~(\ref{eq:ZNNc})
should also be given by Eq.~(\ref{eq:ZNNc-o}), namely
\begin{equation}
 Z_{NN}(q) = \frac{1}{\eta(q)^2} \sum_{l_1,l_2 \in \mathbf{Z}}
  \tilde{q}^{\frac{1}{g}( l_1 \vec{K}_1 + l_2 \vec{K}_2)^2}.
\label{eq:Y-ZNN}
\end{equation}
From Eq.~(\ref{eq:Y-ZNN}) we can immediately read off the
scaling dimensions of the boundary operators.
The leading ones have the smallest scaling dimension $1/g$,
and correspond to the $\pm$ cycle operators discussed above
based on DEBC.
The BCFT analysis confirms that they are indeed the leading
operators and the N fixed point is stable for $g<1$ and
is unstable for $g>1$.

For the Neumann boundary condition on $\vec{\Phi}$, the
conductance tensor vanishes: $G_{jk}=0$.
This is most easily seen by
observing the Neumann boundary condition $\vec{\Phi}$ is nothing but
the Neumann boundary condition on the original fields $\varphi_{1,2,3}$.
Alternatively, we find that $G_{jk}$ vanishes
by an explicit calculation of Eq.~(\ref{eq:Y-cond}).

\subsection{Andreev reflection fixed point --
Dirichlet boundary condition on $\vec\Phi$}

The other simple boundary condition for the free boson is the
Dirichlet boundary condition on $\vec\Phi$, which is dual to 
Neumann. In fact, the Dirichlet boundary condition on $\vec{\Phi}$ is
the Neumann boundary condition on $\vec{\Theta}$. The choice of angle
$\xi=\pi$ pins $\vec\Phi=\vec C$. Plugging ${\cal R}_\pi=-{\bf 1}$ into
Eq.~(\ref{eq:dim}), we obtain $\Delta_V=|\vec b|^2$. Using this, we
find, for example, that

\begin{eqnarray} 
\Delta_{T^{RL}_{33}}&=&\frac{4g}{3}\;|\hat z\times \vec K_3|^2=\frac{4g}{3}
\label{eq:DP-TRL33}
\\
\Delta_{T^{RL}_{21}}&=&\frac{g}{3}\;|\hat z\times \vec K_3|^2=\frac{g}{3}
\label{eq:DP-TRL21}
\\
\Delta_{T^{LL}_{21}}&=&g\;|\vec K_3|^2=g
\label{eq:DP-TLL21}
\ .
\end{eqnarray}
There are 3 independent operators with dimension $4g/3$, 3 with dimension
$g/3$, and 3 with dimension $g$, plus their respective Hermitian conjugates.
The leading irrelevant terms are the $\pm$ cycle tunneling operators. This
fixed point is stable for $g>3$.

Within the BCFT approach, in the standard compactified boson case, the
scaling dimensions of the boundary operators in the Neumann and
Dirichlet boundary states are given by the reciprocal lattices as
discussed in App.~\ref{sec:refboson}.  If the present problem of
the junction were mapped to the standard compactified free bosons, the
Dirichlet boundary state would have the leading boundary operator with
the scaling dimension $4g/3$, which is irrelevant for $g>3/4$.  Thus,
the Dirichlet boundary condition would be stable for $3/4 < g$.  In
particular, both the Neumann and Dirichlet boundary conditions are
then stable for $3/4 < g <1$. This is discussed by Yi and
Kane~\cite{YiKane} in the context of the quantum Brownian motion on a
triangular lattice.  There is a duality $g \leftrightarrow 4/(3g)$
which exchanges the Neumann and Dirichlet boundary conditions.


These considerations for the standard compactified boson
exactly matches the potential argument presented
in Sec.~\ref{sec:potential}.
As discussed there, it would not apply to the 
present problem in which electrons hop between different wires.
In Sec.~\ref{sec:potential}, the fermi statistics of electrons
were taken of care by the Klein factors, to derive sensible results.
In the present context, the difference from the standard boundary
problem of free boson field theory is given by the twisted structure
of the Hilbert space discussed in Section~\ref{sec:twist}.
This affects the stability of the fixed point in an important manner.
As it will be shown below,
the most stable Dirichlet-type boundary state turns out to
be the strong pair-hopping limit, D$_\shbox{P}$,
discussed in Sec.~\ref{sec:potential}.

The most stable Dirichlet-type boundary state is
given by a superposition of all the Ishibashi states 
allowed under $\Delta \vec{\Theta}=0$, namely $n_1=n_2=0$.
These Ishibashi states are given as
\begin{equation}
| (0,0,m_1,m_2) \rangle\rangle
\equiv
\exp{\left[ - \sum_n \vec{a}^{L\dagger}_n \cdot \vec{a}^{R\dagger}_n \right]}
| (0, 0, m_1,m_2) \rangle,
\end{equation}

As in the construction of the N boundary state,
because of the parity condition~(\ref{nmres}),
$m_1$ and $m_2$ take
only even integer values.
Thus we have
\begin{equation}
| D_P \rangle =
g_{D_P} \sum_{m'_1,m'_2 \in \mathbf{Z}}
| (0,0,2m'_1,2m'_2) \rangle \rangle .
\end{equation}
The corresponding diagonal cylinder amplitude is
\begin{eqnarray}
&&Z_{D_P D_P}(\tilde{q}) \langle D_P | e^{-l H^P_{\beta}} | D_P \rangle
\nonumber \\
&=& \frac{{g_{D_P}}^2}{\eta(\tilde{q})^2}
\sum_{m'_1, m'_2 \in \mathbf{Z}}
\tilde{q}^{ \frac{1}{4g} (2 m'_1 \vec{K}_1 + 2 m'_2 \vec{K}_2)^2} .
\end{eqnarray}

Modular transforming using Eq.~(\ref{eq:LatticeMod}), now
with $\Lambda$ being the triangular lattice with the lattice
constant $2$, we obtain
\begin{equation}
 Z_{D_P D_P}(q) = \frac{1}{\eta(q)^2} \sum_{l_1,l_2 \in \mathbf{Z}}
  q^{\,g\left( \frac{1}{2} l_1 \vec{R}_1 + \frac{1}{2} l_2 \vec{R}_2\right)^2}.
\label{eq:Y-ZDD}
\end{equation}
Now the dimension of the leading boundary operator is given by
\begin{equation}
\frac{g}{4} |\vec{R}_j|^2 = \frac{g}{3} .
\end{equation}
Furthermore, the dimensions of the
second-leading and third-leading operators are related to
the distance between next-nearest and next-next-nearest neighbors
on the triangular lattice and given by $g$ and $4g/3$, respectively.
These exactly agree with the scaling dimensions of tunneling
operators~(\ref{eq:DP-TRL21}),(\ref{eq:DP-TLL21})
and~(\ref{eq:DP-TRL33}) obtained from the DEBC method.
Furthermore, these scaling dimensions coincide with those
that can be obtained from the instanton argument for the
\DP fixed point, in which $\vec{\Phi}$ is pinned to
the triangular lattice with the lattice constant $2\pi/\sqrt{3}$.

The present analysis confirms that $g/3$ is the scaling dimension
of the leading boundary operator,
and thus the \DP fixed point is stable for $g>3$ and unstable
for $g<3$.
This is quite different from what would obtain for the stability
of the Dirichlet boundary state if the problem were mapped
to the standard compactified boson.
The difference is due to the twisted structure, which originates
from the fermionic nature of the electron.

The present result can also be understood from the duality.
Eq.~(\ref{modered}) shows that the twisted boson theory has
the duality
\begin{eqnarray}
 \vec{\Phi} &\rightarrow& \vec{\Theta} \\
 g &\leftrightarrow& \frac{3}{g},
\label{eq:gduality}
\end{eqnarray}
instead of $g \leftrightarrow 4/(3g)$ for the standard boson.
The duality~(\ref{eq:gduality}) was found in Ref.~\cite{Nayak}
by a different argument.
Our analysis shows that the duality exists as a structure
of the effective Hilbert space, beyond a particular
choice of the perturbation.

Let us now discuss the conductance.
For the Dirichlet boundary condition,
a similar calculation to that in App.~\ref{sec:BC_cond} gives
\begin{equation}
\lim_{\omega \rightarrow 0_+}
\frac{1}{\pi  \omega L} \int_0^L dx
\int_{-\infty}^{\infty} \!\!\!d\tau\;e^{i\omega\tau}
\;
\langle J^{\mu}(y,\tau) J^{\nu}(x,0) \rangle
= 2 g \delta^{\mu \nu}
\end{equation}
We thus find, from Eq.~(\ref{eq:Y-cond}),
\begin{equation}
 G_{kl} = 2 g \frac{e^2}{h} \sum_{\mu=1,2} v_{k\mu} v_{l\mu}
\end{equation}
The summation $\sum_{\mu=1,2}$ can be interpreted as an inner product
of the {\em two-dimensional} vectors $\vec{v}_k$ and $\vec{v}_l$.
From eq.~(\ref{eq:vtr}), we find
\begin{equation}
\vec{v}_k \cdot \vec{v}_l = (\delta_{kl} - \frac{1}{3} )
\end{equation}
Therefore we recover the results in Ref.~\cite{Nayak} as
\begin{equation}
G_{kl}  = \left\{
 \begin{array}{lc}
  \frac{4g}{3} \frac{e^2}{h} & (k=l) \\
  - \frac{2g}{3} \frac{e^2}{h} & (k\neq l)
   \end{array}
\right .
\end{equation}
This derivation and the result on the conductance equally apply to
\DP and \DN fixed points;
they are indistinguishable in terms of the conductance.
As pointed out in
Ref.~\cite{Nayak}, the single-terminal conductance in this case is
larger than the single ideal wire.  This exhibits an enhanced
conductance due to the Andreev reflection.

\subsection{Chiral fixed points $\chi_\pm$}

\subsubsection{Chiral fixed points from DEBC}

In the DHM representation discussed in Section~\ref{sec:DHM}, the
quantal phases due to the Fermi statistics and the magnetic flux
piercing through the junction was accounted by the ``boundary magnetic
field''.  In the representation with $n=0,1$ the chiral fixed points
$\chi_{\pm}$ correspond to the localized phase of the DHM, which
amounts to the Dirichlet boundary condition on the field $\vec{X}$.
We emphasize that the field $\vec{X}$ is not identical to the fields
$\vec{\Theta}$ and $\vec{\Phi}$ we have been using.  The $\chi_{\pm}$
fixed point corresponds to
  neither Dirichlet nor Neumann boundary conditions
in terms of $\vec{\Theta}$ and $\vec{\Phi}$.
As we will see in the following, the
field $\vec{X}$ in fact corresponds to a chiral rotation of
$\vec{\Theta}$ or $\vec{\Phi}$.

Indeed, by inspecting the operators in the $\pm$ cycles, we would find
that if these coupling constants dominate the low-energy physics (flow to
strong coupling), then they set boundary conditions that are mixed in
the $\vec\Phi$ and $\vec\Theta$ fields. 

For example, pinning the phases in the $+$~cycle corresponds to the
boundary condition 
\begin{equation}
\chi_+:\;\;{\vec K_i}\cdot\vec\Phi+
\sqrt{\frac{1}{3}}\;(\hat z\times \vec K_i)\cdot\vec\Theta=
\vec C.
\end{equation} 
And pinning this combination corresponds to the choice of angle
$\xi$ such that 
\be \tan\frac{\xi}{2}=\frac{\sqrt{3}}{g}.\label{xichi}\ee

With this choice of boundary condition or angle $\xi$, the scaling
dimension, for example, of a backscattering term such as $T^{RL}_{33}$
is obtained with $\vec a=0$ and $\vec b=-2\sqrt{\frac{g}{3}}\hat
z\times\vec K_3$ in Eq.~(\ref{eq:dim}). Using that $|\hat z \times
\vec v|^2=|\vec v|^2$ and $[{\cal R}_\xi(\hat z\vec v)]\cdot (\hat
z\times\vec v)= ({\cal R}_\xi\vec v) \cdot \vec v=|\vec
v|^2\;\cos\xi$ for any vector $\vec v$:
\begin{eqnarray}
\Delta_{T^{RL}_{33}}
&&=2\;\frac{g}{3}\left[|\hat z\times \vec K_3|^2
-[{\cal R}_\xi(\hat z\times \vec K_3)]\cdot(\hat z\times \vec K_3)\right]
\nonumber\\
&&=2\;\frac{g}{3}\;|\vec K_3|^2\;[1-\cos\xi]
=\frac{4g}{3+g^2}
\end{eqnarray}
after some simple algebra.

Similarly, the scaling dimensions of the $-$~cycle terms (such as
$T^{RL}_{12}$) as well as those of the $LL-RR$ terms (such as $T^{LL}_{12}$)
can be shown to be also $\frac{4g}{3+g^2}$. Again, these operators are not
independent after the choice of boundary condition; in total, there are 3
independent operators and their 3 Hermitian conjugates.
This agrees with the dimension of the leading operator for
the fixed point $\chi_+$,
from the instanton analysis in Sec.~\ref{sec:DHM}.

Now, had we chosen to pin the phases in the $-$~cycle set of
operators, or select the boundary condition
\begin{equation}
\chi_-:\;\;{\vec
K_i}\cdot\vec\Phi-
\sqrt{\frac{1}{3}}\;(\hat z\times \vec K_i)\cdot\vec\Theta=\vec C, 
\end{equation} 
we would have obtained an analogous case, in which
$\tan\frac{\xi}{2}=-\frac{\sqrt{3}}{g}$. The dimensions of the
backscattering and $+$ cycle terms, as well as of the $LL-RR$ terms,
are again $\frac{4g}{3+g^2}$. This fixed point is stable for
$1<g<3$.
Which of the $\chi_\pm$ is selected depends on the
time-reversal symmetry breaking flux.

\subsubsection{Relation to the DHM representation} 
\label{subsubsec:rel_DHM}
Let us connect the present analysis
to the ones based on
the dissipative Hofstadter model, more explicitly.

First we formulate the DHM (for $V=0$)
as a boundary problem of a $c=2$ free boson field theory.
This can be done in two different ways.
One of them is to begin 
with the (1+1) dimensional action:
\be
S=\int_{-\infty}^\infty d\tau \int_0^\infty dx \left[
{\alpha \over 4\pi}(\partial_\mu \vec X)^2+{\beta \over 4\pi}
\epsilon^{\mu \nu}\epsilon_{ab}\partial_\mu X_a\partial_\nu X_b\right].
\label{S2d}
\ee
We do not impose any b.c. on $\vec X$ at $x=0$, instead letting 
the field be dynamical on the boundary.
By integrating out the bulk part of the free boson fields, $\vec X$, 
this is reduced to the (non-interacting part of the)
one-dimensional action of Eq.~(\ref{eq:S_0})
introduced in Ref.~\cite{CF}.
To see this explicitly, we may 
decompose the fields $\vec X(\tau ,x)$ inside the path integral 
into a sum of classical and quantum terms:
\be \vec X(\tau ,x)=\vec X^{\shbox{cl}}(\tau ,x)+\vec X(\tau ,x)'.\ee
Here $\vec X^{\shbox{cl}}$ obeys the classical equations of motion:
\be \partial^2\vec X^{\shbox{cl}}=0.\ee
[Note that the term proportional to $\beta$ in the 
action of Eq. (\ref{S2d}) makes no contribution to the Euler-Lagrange 
equations because it is a total derivative.] $\vec X^{\shbox{cl}}$ 
is chosen to obey the boundary condition:
\be \vec X^{\shbox{cl}}(\tau ,0)=\vec X_b(\tau ).\label{Xclbc}\ee
On the other hand, the quantum part, $\vec X'(\tau ,x)$ obeys 
a D b.c., $\vec X'(\tau ,0)=0$. With this choice of b.c. on $\vec X'$, 
using the fact that $\vec X^{\shbox{cl}}$ obeys the Euler-Lagrange 
equations, we see that:
\be S(\vec X^{\shbox{cl}}+\vec X')=S(\vec X^{\shbox{cl}})+S(\vec X'),\ee
i.e. the cross-term vanishes. Furthermore, in calculating 
Green's functions of $X(\tau ,0)$, $\vec X'$ doesn't appear 
due to its D b.c. Therefore, the calculation of these boundary 
Green's functions reduces to a functional integral over 
the boundary fields, $\vec X_b(\tau )$ with a classical action:
\be S_b[\vec X_b]=S(\vec X^{\shbox{cl}})\ee
where $S$ is the two-dimensional action of Eq. (\ref{S2d}) and 
the dependence on the boundary fields, $\vec X_b(\tau )$ is 
induced through the b.c. on $\vec X^{\shbox{cl}}$ of Eq. (\ref{Xclbc}). 
Fourier expanding $\vec X^\tau$:
\be \vec X(\tau ) = \sum_n\vec X_n e^{i\omega_n\tau},\ee
gives the classical solution:
\be
\vec X^{\shbox{cl}}(\tau ,x)=\sum_n\vec X_ne^{i\omega_n\tau
-|\omega_n|x},
\ee
and the boundary action~(\ref{eq:S_0}).

The alternative approach is to take the bulk action
with no magnetic field term:
\be S=\int_{-\infty}^\infty d\tau \int_0^\infty dx \left[
{\alpha \over 4\pi}(\partial_\mu \vec X)^2\right].
\label{S2d'}\ee
and impose a chiral boundary condition:
\begin{equation}
 \vec{X}_R = {\cal R}_{\xi '} \vec{X}_L
\label{bcch}
\end{equation}
for some rotation angle $\xi '$. 

We wish to show that correlation functions at $x=0$
for Eq.~(\ref{S2d'}) with~(\ref{bcch}) are the same,
for an appropriate value of $\xi'$,
as those obtained from the boundary action 
of Eq.~(\ref{eq:S_0}).
This can be shown using a limiting procedure where we 
first calculate correlation functions for fields $\vec X(\tau_i ,x_i)$
and then take the limit where $x_i \to 0^+$. Since we are considering 
the non-interacting theory at this point,
it is sufficient to show 
this for the 2-point function. 
Using:
\be 
\langle X^\mu_L(\tau +ix) X^\nu_L(\tau '+ix') \rangle
={-1\over 2\alpha}\ln [(\tau -\tau ')+i(x-x')]
\ee
and the fact that the boundary condition of Eq. (\ref{bcch}) implies:
\be \vec X_R (\tau ,x)= {\cal R}_{\xi '} \vec{X}_L (\tau ,-x),\ee
for $x>0$, we obtain:
\begin{widetext}
\bea 
&& \langle X^a(\tau ,x)X^b(0,x') \rangle=
\langle [X^a_L(\tau + ix)+{\cal R}_{\xi '}^{ac}X_L^c(\tau - ix)]
[X^b_L(ix')+{\cal R}_{\xi '}^{bd}X^d_L(-ix')] \rangle \nonumber \\
&=& -{1\over 2\alpha}
\bigl\{\delta^{ab}\ln [\tau +i(x-x')]+
\delta^{ab}\ln [\tau -i(x-x')]
+{\cal R}_{\xi '}^{ab}\ln [\tau +i(x+x')]+
{\cal R}_{\xi '}^{ba} \ln [\tau  - i(x+x')]\bigr\}.
\eea
Using the explicit form of ${\cal R}_{\xi '}$ given in Eq. (\ref{R_xi}) this
can be written:
\be
 \langle X^a(\tau ,x)X^b(0,x') \rangle = 
-{1\over 2\alpha}\bigl\{ [1+\cos \xi ' ]\delta^{ab}\ln [\tau^2+(x-x')^2]
-\epsilon^{ab}\sin{\xi'}
\bigl( \ln{[\tau - i(x+x')]} - \ln{[\tau +i(x+x')]} \bigr)
\bigr\} .
\ee
Now letting $x$, $x'\to 0^+$, we obtain:
\be
\langle X^a(\tau ,+0)X^b(0,+0) \rangle \to 
-{1+\cos \xi ' \over 2\alpha }\delta^{ab}\ln \tau^2 
+i\pi {\sin \xi ' \over 2\alpha}
\epsilon^{ab}[\hbox{sgn} (\tau )-1].
\ee
\end{widetext}
Choosing:
\begin{equation}
 \tan{\frac{\xi'}{2}} = \frac{\beta}{\alpha} ,
\label{eq:xi'}
\end{equation}
this reduces to the correlation function of the DHM model, Eq.~(\ref{eq:D}),
apart from a constant term:
\be
\delta \langle X^a(\tau ,+0)X^b(0,+0) \rangle =
-i\pi {\beta \over \alpha^2+\beta^2}\epsilon^{ab}.
\ee
This constant term has no effect on the perturbative expansion
discussed in Sec.~\ref{sec:DHM}
and used to connect the DHM with the Y-junction model.
This follows since the 
extra contribution to Eq.~(\ref{eq:loopCF})
is proportional to
\be
e^{\sum_{j=1}^n\sum_{k=1}^n\vec L_j\times \vec L_k} = 1.
\ee
Thus it is shown~\cite{CKMY}
that imposing the chiral b.c. of Eq. (\ref{bcch}) on 
2-component bosons with the standard action of Eq. (\ref{S2d'}) with 
the rotation angle, $\xi '$ of Eq. (\ref{eq:xi'}) gives the 
DHM model.

In fact, it is possible to derive a general connection between the 
fields $\vec X(\tau ,x)$ introduced here and the original fields 
$\vec \Phi (\tau ,x)$ introduced in Sec.~\ref{sec:set-up}.
Naively, by comparing  the actions, we might expect that:
\be
  \vec X=\sqrt{g/\alpha}\vec \Phi \ \ ?
\ee
However, this does not take into account the b.c. correctly. 
Recall that the disconnected wire fixed point corresponded to N b.c. 
on $\vec \Phi$,
\be
\vec \Phi_L=\vec \Phi_R.
\ee
However, the above discussion establishes that 
this N. $V=0$,
fixed point corresponds to
the chiral b.c. of Eq. (\ref{bcch}) on $\vec X$. Thus 
we tentatively identify:
\bea
\vec{X}_R&=& \sqrt{g/\alpha} {\cal R}_{\xi'} \vec{\Phi}_R\nonumber \\
\vec X_L&=& \sqrt{g/\alpha} \vec \Phi_L.\label{XPhi}
\eea
To check the consistency of this identification, consider the 
chiral b.c. on $\vec \Phi$, given by Eq. (\ref{rotphi}) 
with the rotation angle of Eq. (\ref{xichi}).  As we argued 
in Sec.~\ref{sec:DHM},
this fixed point corresponds to the localized 
phase of the DHM which in turn corresponds to a simple 
D b.c. on $\vec X$:
\be \vec X_L=-\vec X_R.\label{XD}
\ee
Consistency of Eq. (\ref{XPhi}) [with $\xi '$ given 
by Eq. (\ref{eq:xi'})], which maps $\vec X$ into $\Phi$,
with the b.c. on $\vec \Phi$ of Eq. (\ref{rotphi}), (\ref{xichi})
and the b.c. on $\vec X$ of Eq. (\ref{XD}) requires:
\be
{\cal R}_{\xi '}=-{\cal R}^{-1}_{\xi},
\ee
or
\be \xi '=\pi -\xi .\ee
This implies $\tan (\xi ' /2)=\cot (\xi /2)$ and hence 
is consistent with
Eqs.~(\ref{eq:xi'}) and (\ref{xichi}) for 
$\xi '$ and $\xi$ together with the value of $\beta /\alpha$ 
determined in Eq.~(\ref{eq:n}) for the case $n=1$, corresponding 
to the stable chiral fixed point. 

\subsubsection{Boundary states and the generalized chiral fixed points}

We have now established that the chiral fixed points $\chi_{\pm}$
corresponds to the chiral boundary condition~(\ref{rotphi})
on the field $\vec{\Phi}$.
Let us construct the corresponding boundary state explicitly,
taking the twisted structure
discussed in Sec.~\ref{sec:twist} into account.

The winding number along the boundary can be written as
\begin{eqnarray}
 \Delta \vec{\phi}_L &=&
\frac{1}{2}\left( \sqrt{g} \Delta \vec{\Phi} +
\frac{1}{\sqrt{g}} \Delta \vec{\Theta} \right) \\
 \Delta \vec{\phi}_R &=&
\frac{1}{2}\left( \sqrt{g} \Delta \vec{\Phi} - 
\frac{1}{\sqrt{g}} \Delta \vec{\Theta} \right) .
\end{eqnarray}
For the chiral boundary condition~(\ref{rotphi}), we have
\begin{equation}
 \sqrt{g} \Delta \vec{\Phi} - \frac{1}{\sqrt{g}} \Delta \vec{\Theta}
= {\cal R}_\xi ( \sqrt{g} \Delta \vec{\Phi}
+ \frac{1}{\sqrt{g}} \Delta \vec{\Theta} )
\end{equation}
For $\xi \neq 0$, $\pi$, it is clear that we need both $\Delta \vec{\Phi}$
and $\Delta \vec{\Theta}$ nonvanishing.
Moreover, ${\cal R}_\xi$ is a rotation matrix and thus preserves
the length of the vector. This requires $\Delta \vec{\Phi}$ and
$\Delta \vec{\Theta}$ to be mutually orthogonal.

Thus the only oscillator vacua which can appear in the chiral
boundary state is
\begin{equation}
 | (r l_1, r l_2, s l_1, s l_2 ) \rangle,
\label{eq:rsvacuum}
\end{equation}
for arbitrary integers $l_1$ and $l_2$.
Here $(r,s)$ is a fixed set of integers.
In order to satisfy the parity condition~(\ref{nmres}) for arbitrary
integers $l_{1,2}$,
\begin{equation}
 r = s \pmod{2},
\label{eq:rsparity}
\end{equation}
is required.
The chiral rotation angle is fixed by $(r,s)$ as
\begin{equation}
 \tan{\frac{\xi}{2}} = \frac{s}{r} \frac{\sqrt{3}}{g} .
\label{eq:xirs}
\end{equation}
That is, the chiral rotation angle is ``quantized''.

For each vacuum~(\ref{eq:rsvacuum}), we can construct
a chiral bosonic Ishibashi state
\begin{equation} 
| (r l_1, r l_2, s l_1, s l_2 ) \rangle \rangle
  = \exp{\big(
   \sum_{n=1}^{\infty}
\vec{a}^{R\dagger}_n \cdot {\cal R}_{\xi} \vec{a}^{L\dagger}_n
   \big)}
| (r l_1, r l_2, s l_1, s l_2 ) \rangle,
\label{eq:rsIshibashi}
\end{equation}
which is conformally invariant.

A single Ishibashi state does not satisfy Cardy's consistency condition.
Similarly to the case of the standard Neumann/Dirichlet boundary
state, the summation over all integers $l_{1,2}$ could give
a consistent boundary state.
Namely,
\begin{equation}
|\chi \rangle =
g_{\chi} \sum_{l_1,l_2 \in \mathbf{Z}}
| (r l_1, r l_2, s l_1, s l_2 ) \rangle \rangle,
\label{eq:chistate}
\end{equation}
where $g_{\chi}$ is again the ground-state degeneracy~\cite{gtheorem}
of the chiral boundary states.

The diagonal cylinder amplitude is given by
\begin{equation}
 Z_{\chi \chi} =
\frac{{g_{\chi}}^2}{\eta(\tilde{q})^2}
\sum_{l_1,l_2 \in \mathbf{Z}}
  \tilde{q}^{\frac{1}{4}
\left[  l_1 ( \sqrt{g} r \frac{\vec{R}_1}{2} + \frac{s}{\sqrt{g}} \vec{K}_1)
     + l_2  ( \sqrt{g} r \frac{\vec{R}_2}{2} + \frac{s}{\sqrt{g}} \vec{K}_2)
\right]^2} .
\end{equation}
This can be written as
\begin{equation}
 Z_{\chi \chi} =
\frac{{g_{\chi}}^2}{\eta(\tilde{q})^2}
\sum_{\vec{V} \in \Lambda^*_\chi} q^{\frac{1}{4}\vec{V}^2},
\end{equation}
where
$\Lambda^*_\chi$ is a triangular lattice with lattice constant
\begin{equation}
 \sqrt{\frac{r^2g}{3}+\frac{s^2}{g}} .
\end{equation} 
Thus, choosing
\begin{equation}
 g_{\chi} = \sqrt{ \frac{\sqrt{3}}{4} \left(\frac{r^2g}{3}+\frac{s^2}{g} \right)},
\label{eq:gchi}
\end{equation}
the same amplitude in the open string channel is given by
\begin{equation}
 Z_{\chi \chi} =
\frac{1}{\eta(q)^2}
\sum_{\vec{W} \in \Lambda_\chi} q^{\vec{W}^2},
\end{equation}
because of eq.~(\ref{eq:LatticeMod-2}).
The lattice $\Lambda_\chi$ is dual to $\Lambda^*_\chi$, and
is the triangular lattice
with the lattice constant
\begin{equation}
 \sqrt{\frac{4g}{r^2g^2 + 3 s^2}} .
\end{equation}
This formula encodes, as usual in the BCFT approach,
the complete list of scaling dimensions of possible boundary
perturbations to the chiral fixed point.
Namely, the scaling dimensions of all the boundary perturbations
are given by $\vec{W}^2+\hbox{integer}$.
In particular, the scaling dimension of the leading perturbation
corresponding to the minimal length basis vector
of $\Lambda_\chi$, is given by
\begin{equation}
 \Delta_{\chi} = \frac{4g}{3 r^2 + s^2 g^2} .
\label{eq:Deltachi}
\end{equation}

From this spectrum of boundary operators, and
the ground-state degeneracy~(\ref{eq:gchi}),
we see that the boundary state with larger $r$  and $s$ are unstable.
While any given chiral rotation angle $\xi$
may be approximated with arbitrary precision
by taking large $r$ and $s$, such a choice gives a highly
unstable fixed point.

As a special series of chiral boundary states,
let us take $r=1$.
Then, due to the parity condition~(\ref{eq:rsparity}),
$s$ must be odd: $s= 2 n -1$ for an integer $n$.
Now the boundary states are characterized by the chiral
rotation angle
\begin{equation}
  \tan{\frac{\xi}{2}} = (2n-1) \frac{g}{\sqrt{3}} .
\end{equation}
This agrees with the chiral rotation angle for the
chiral fixed points obtained
from the mapping to the DHM.
Indeed, the scaling dimension of the leading perturbation
\begin{equation}
 \Delta_{\chi} = \frac{4g}{3 + (2n-1)^2 g^2}
\end{equation}
also agrees exactly with the instanton analysis on the
chiral fixed points in the DHM approach.
In particular, $n=1,0$ corresponds to the most stable
ones which we denoted as $\chi_{\pm}$.

We have thus succeeded in explicitly constructing
the boundary state for the chiral fixed points.
In general, a chiral fixed point is characterized by
two non-zero integers $(r,s)$ with $r=s \pmod{2}$.
It includes the one-parameter series $r=1,s=2n-1$
which was found in Sec.~\ref{sec:DHM}.
However, except for $(r,s) = (1,\pm 1)$, the chiral
fixed point is always unstable for any given value of $g$.

The only chiral fixed points that can be stable
are $(r,s)=(1,\pm 1)$, that is $\chi_{\pm}$.
The BCFT gives us the complete spectrum of the boundary operator,
and thus more convincing evidence of the stability.
In particular, for $1 < g <3$,
$\chi_{\pm}$ fixed points should be stable against
any boundary perturbation, including the asymmetry and
the change of the flux $\phi$, but excluding
the ``resonant'' type perturbation as discussed in
Sec.~\ref{sec:twist2wires}.

\subsubsection{Conductance}

Let us calculate the conductance at the chiral fixed points
$\chi_{\pm}$ from the above analysis on the boundary condition.
Using $J^\mu_R = {\cal R} J^\mu_L$, we find
\bea
&&\lim_{\omega \rightarrow 0_+}
\frac{1}{\pi  \omega L} \int_0^L dx
\int_{-\infty}^{\infty} \!\!\!d\tau\;e^{i\omega\tau}
\;
\langle J^{\mu}(y,\tau) J^{\nu}(x,0) \rangle
\nonumber \\ &&
=  g
\left[ (1 - \cos{\xi})\delta^{\mu\nu} + \sin{\xi} \epsilon_{\mu\nu}
\right] .
\eea
Thus, from Eq.~(\ref{eq:Y-cond}) we have
\begin{eqnarray}
G_{jk} &=& g \frac{e^2}{h} \left[
(1 - \cos{\xi}) \vec{v}_j \cdot \vec{v} _k +
\sin{\xi}  \vec{v}_j \times \vec{v}_k
\right]
\\
&=&
g \frac{e^2}{h} 
\left[ (1 - \cos{\xi}) (\delta_{jk} - \frac{1}{3})
+ \frac{\sin{\xi}}{\sqrt{3}} \epsilon_{jk} 
\right] .
\end{eqnarray}
For the chiral rotation angle $\xi$ in Eq.~(\ref{eq:xirs}),
this reduces to the Z$_3$ symmetric form~(\ref{eq:Z3tensor}), with
\begin{eqnarray}
G_S &=& \frac{e^2}{h} \frac{4 g s^2}{g^2 r^2 + 3 s^2}, \\
G_A &=& \frac{e^2}{h} \frac{4 g^2 r s}{g^2 r^2 + 3 s^2}.
\end{eqnarray}
In particular, for the most stable chiral fixed points $\chi_{\pm}$,
where $r=1$ and $s=\pm 1$, we recover
the same result~(\ref{eq:GDGA}) derived from the mapping to
the DHM.

\subsection{Asymmetric fixed point}
\label{sec:asymmetric-FP}

In this paper, we have mostly discussed the junction with $Z_3$
symmetry among the wires.
However, if the junction of three wires could be realized,
it will inevitably contain some asymmetry between the wires.
Therefore, it is important to study the effect of $Z_3$
asymmetry.

Here we restrict our study to the asymmetry at the junction, assuming
the three quantum wires are identical in the bulk.  Although there
would also be some asymmetry in the bulk in reality, it is outside the
scope of the present paper.

There is a natural fixed point corresponding to the limit of the
strong anisotropy: two of the wires are strongly coupled to form a
single wire with perfect transmission at the junction, and the other
is decoupled.  Without losing generality, let us assume that the wires
1 and 2 are strongly coupled and the wire 3 is decoupled.  Let us
discuss the stability of this fixed point, which we call D$_\shbox{A}$.

\subsubsection{Heuristic analysis}

As perturbations to this asymmetric fixed point, we consider the
following three operators:
\begin{description}
\item[electron hopping between wires 3 and 1\& 2]
  $\psi^{\dagger}_1 \psi_3 +\mbox{h.c.}$
\item[electron pair hopping between wires 3 and 1\& 2]
  $\psi^{\dagger}_1 \psi^{\dagger}_1 \psi_3 \psi_3 +\mbox{h.c.}$
\item[electron backscattering in the wire 1\& 2]
  $\psi^{\dagger}_1 \psi_2 + \mbox{h.c.}$
\end{description}
It is natural to expect that the leading perturbation with the
smallest scaling dimension is given by one of these three.  Indeed, we
confirm this expectation later with the BCFT analysis.

In the first two cases, the scaling dimension of the operator is given
by the sum of the scaling dimensions of the operators on wire 3 and
wire 1\& 2.  The disconnected wire 3 is subject to the Neumann boundary
condition on $\varphi_3$ (Dirichlet on $\theta_3$) at the junction.
Thus, from eq.~(\ref{bosend}),
the electron annihilation operator
$\psi_3$ corresponds to the boundary operator $e^{i
\varphi_3/\sqrt{2}}$ at the Dirichlet boundary condition on
$\theta_3$.  Its scaling dimension turns out to be $1/(2g)$.

On the other hand, for the wire 1\& 2, the scaling dimension of the
electron creation operator $\psi^{\dagger}_1$ is given by the bulk
scaling dimension of $\psi^{\dagger}_{1L}$ or $\psi^{\dagger}_{1R}$.
It turns out to be $(g+g^{-1})/4$.  This scaling dimension determines
the Local Density Of States for the electron tunneling into the
TLL~\cite{FisherGlazman}.

As a result, the scaling dimension of the electron hopping operator
$\psi^{\dagger}_1 \psi_3 +\mbox{h.c.}$ is
\begin{equation}
 \frac{1}{2g} + \frac{g + g^{-1}}{4} = \frac{3+ g^2}{4g}.
\end{equation}
This, remarkably, is exactly the inverse of
the scaling dimension~(\ref{eq:Deltachi})
of the leading perturbation at $\chi_{\pm}$ fixed points.
As a consequence, it is relevant for $1<g<3$ and irrelevant
for $g< 1 $ or $3<g$.

Next we consider the electron pair hopping between the wire 3 and 1\& 2. 
The pair annihilation operator in the wire 3 $\psi_3 \psi_3$ is given
by the boundary operator $e^{i \sqrt{2} \varphi_3/\sqrt{g}}$ at the
Dirichlet boundary condition on $\theta_3$, with the scaling dimension
$2/g$.  The leading operator in the pair creation operator
$\psi^{\dagger}_1 \psi^{\dagger}_1$ is given by
\begin{equation}
 \psi^{\dagger}_{1L} \psi^{\dagger}_{1R} \sim
e^{i \sqrt{2} \varphi_2}.
\end{equation}
Although this has the identical form to the above $\psi_3 \psi_3$,
its scaling dimension is different since it must be evaluated
as a bulk operator.
The scaling dimension is given by
\begin{equation}
 2 \times \frac{1}{4} \times \frac{2}{g} = \frac{1}{g} .
\end{equation}
Thus, the scaling dimension of the pair hopping operator is
\begin{equation}
 \frac{2}{g} + \frac{1}{g} = \frac{3}{g}.
\end{equation}
Finally, the scaling dimension of the backscattering operator in
the wire 1\& 2 is known to be $g$ from Ref.~\cite{Kane}.

\subsubsection{BCFT analysis}

Now let us confirm the boundary operator content with the BCFT.
The D$_\shbox{A}$ fixed point corresponds to  Dirichlet boundary
condition on $\Phi_1 = (\varphi_1 - \varphi_2)/\sqrt{2}$,
and Neumann boundary condition on $\Phi_2$.

Thus, the allowed winding numbers in the oscillator vacua is
\begin{eqnarray}
\Delta \vec{\Phi} &=& - 2 \pi n \frac{\vec{R}_3}{2} =
2 \pi n \frac{\vec{R}_1 + \vec{R}_2}{2}, \\
\Delta \vec{\Theta} &=& - 2 \pi m \vec{K}_3 =
2 \pi m (\vec{K}_1 + \vec{K}_2 ),
\end{eqnarray}
where $n,m$ are integers satisfying
\begin{equation}
 n = m \pmod{2},
\end{equation}
from Eq.~(\ref{nmres}).  The unit vectors $\vec K_i$ 
and $\vec R_i$ are defined in Eq. (\ref{Kdef}) and (\ref{eq:Rdef}) 
respectively. 

For each permitted oscillator vacuum
$| (n,n,m,m) \rangle$,
we can construct the bosonic Ishibashi state
\begin{equation} 
| (n , n , m , m ) \rangle \rangle
  = \exp{\big(
   \sum_{k=1}^{\infty}
\vec{a}^{R\dagger}_k \cdot {\cal R}_{\zeta} \vec{a}^{L\dagger}_k
   \big)}
| (n , n , m , m ) \rangle,
\end{equation}
where the rotation angle $\zeta$ is given by
\begin{equation}
  \tan{\frac{\zeta}{2}}=\frac{m}{n}\frac{\sqrt{3}}{g},
\label{eq:zetamn}
\end{equation}
similarly to~(\ref{xichi}).

The boundary state for the D$_\shbox{A}$ fixed point is given by
the superposition of these Ishibashi states over all
possible $n,m$:
\begin{equation}
 | D_A \rangle = g_{D_A} \sum_{n=m \pmod{2}} | (n,n,m,m) \rangle \rangle .
\end{equation}
Although the construction is similar to that of
the chiral boundary state~(\ref{eq:chistate}),
$|D_A\rangle$ is {\em not} a chiral boundary state.
Since the chiral rotation angle $\zeta$ determined by Eq.~(\ref{eq:zetamn})
depends on each Ishibashi state, the boundary state as a whole
cannot be simply related to a chiral boundary condition as
in~(\ref{rotphi}).
Nevertheless, it can be a legitimate conformally invariant
boundary state.  

The diagonal cylinder amplitude in the closed string channel is given by
\begin{eqnarray}
Z_{D_A D_A} &=&
\frac{{g_{D_A}}^2}{\eta(\tilde{q})^2}
 \sum_{n=m \pmod{2}}
\tilde{q}^{\frac{1}{4}
( \sqrt{g} n \frac{\vec{R}_3}{2} + \frac{m}{\sqrt{g}} \vec{K}_3 )^2}
\nonumber \\
&=&
\frac{{g_{D_A}}^2}{\eta(\tilde{q})^2}
\sum_{\vec{U} \in \Lambda^*_A}
\tilde{q}^{\frac{1}{4} \vec{U}^2} ,
\end{eqnarray}
where $\Lambda^*_A$ is the two-dimensional
Bravais lattice spanned by
$(1/\sqrt{g}, \pm \sqrt{g/3})$.
Using Eq.~(\ref{eq:LatticeMod-2}), we fix
\begin{equation} 
g_{D_A} = \left(\frac{1}{3}\right)^{1/4},
\end{equation}
to obtain
\begin{equation}
Z_{D_A D_A} = \frac{1}{\eta(q)^2}
\sum_{\vec{W}\in \Lambda_A} q^{\vec{W}^2},
\end{equation}
where $\Lambda_A$ is the Bravais lattice spanned
by $(\sqrt{g}/2, \pm \sqrt{3/(4g)})$ and is dual of $\Lambda_A$.

The scaling dimensions of the boundary operators can be
read off from the amplitude, as $\vec{W}^2$ for
$\vec{W} \in \Lambda_A$.
These include
\begin{eqnarray}
 (\frac{\sqrt{g}}{2}, \sqrt{\frac{3}{4g}})^2 &=& \frac{3+g^2}{4g}, \\
 (2 \frac{\sqrt{g}}{2}, 0 )^2 &=&  g, \\
 ( 0, 2 \sqrt{\frac{3}{4g}})^2 = \frac{3}{g},
\end{eqnarray}
which exactly match the three operators considered
in the heuristic analysis.

\subsubsection{Restoration of the Z$_3$ symmetry}

We have shown that the
asymmetric fixed point D$_\shbox{A}$ is always unstable except
for $g=1,3$.  At $g=1$, the fixed point belongs to the continuous
manifold of fixed points described by free fermion $S$-matrices.  Thus
it is natural to have marginal operators.  Presumably it belongs to
the similar manifold corresponding to the
$S$-matrices of the free dual fermion, at $g=3$.  For $g<1$, the
backscattering in the wire 2\& 3 is relevant.  It is most natural to
assume that the system flows into the $N$ fixed point, in which all
the three wires are disconnected.

For $1<g<3$, the electron hopping is relevant.  In the time-reversal
non-invariant case $\phi \neq 0,\pi$, it is most natural to assume
that the system flows into the $\chi_{\pm}$ fixed point depending on
the sign of the flux $\phi$. When $\phi=0,\pi$, the system is
time-reversal invariant and thus cannot flow into $\chi_{\pm}$.  We
expect the infrared fixed point to be the nontrivial M(ysterious)
fixed point, which is stable for $1<g<3$ in the time-reversal
invariant models.

For $3<g$, the electron-pair hopping is dominant.  In the
time-reversal non-invariant case $\phi \neq 0,\pi$, it is most natural
to assume that the system flows into the \DP fixed point, which can
be regarded as the limit of strong pair tunneling.  In the
time-reversal invariant cases $\phi=0,\pi$, the situation is
subtle.
As discussed in Sec.~\ref{sec:potential}, the simple strong pair hopping
limit in the time-reversal invariant case suggests that the potential
minima forms a finer lattice.  This fixed point would correspond to
\DN boundary state and thus is unstable for $g<9$.  If this
argument applies to any time-reversal symmetric case, for $3<g<9$ we
expect the system to flow into a nontrivial fixed point.

In any case, our analyses imply that
the system recovers the $Z_3$ symmetry, even if
the microscopic model of the junction is not $Z_3$ symmetric.
This is remarkable compared to many other quantum impurity problems,
in which a high symmetry is required to
reach a nontrivial fixed point.
This feature is favorable in observing the intriguing
behavior of the chiral fixed points $\chi_{\pm}$ in future experiments.




\section{Constraints on conductance from energy conservation}
\label{sec:Econservation}


In this section we will show how energy conservation constraints the
conductance tensor. We will also show that the three sets of boundary
conditions on the bosonic fields studied above, namely Neumann, chiral, and
Dirichlet, correspond to a situation where all the energy exchanged in the
scattering processes off the junction is between the zero modes. Drops or
increases of voltages across the junction are associated with the zero modes,
or current carrying states. The non-zero or oscillator modes are neutral, and
can carry energy but no electric current. Therefore, scattering
without exciting the oscillator modes is the most efficient way of
transmitting current without additional dissipation.

It is simpler to discuss the energy exchanges in the unfolded formalism. Let
$\phi^R_i(t,x)$, for $i=1,2,3$, be the bosonic fields in each wire.
$\phi^R_i(t,x<0)$ correspond to the incoming fields into the junction at
$x=0$, and $\phi^R_i(t,x>0)$ are the outgoing fields after scattering. 
$\phi_R(x)$ obeys periodic boundary conditions, 
\be \phi_R(x+2l)=\phi_R(x)+2\pi n.\ee
Away
from the impurity, the dynamics is governed by 
\begin{equation}
{\cal L}= \sum_{i=1}^3 -\frac{g}{2\pi}\partial_x\phi^R_i
\left(\partial_t+\partial_x\right)\phi^R_i
\; .
\label{eq:scattering-lagrangian1}
\end{equation}
The mode expansion of the fields $\phi^R_i(t,x)$, in a circle of length $2l$,
reads:
\begin{eqnarray}
\label{eq:mode-for-scattering}
\phi^R_i(t,x)=&&
{\hat \phi}^{R}_{0i}+\frac{\pi}{lg}{\hat Q}_i^R\;(t-x)
\\
&&+\frac{1}{\sqrt {2g}}\sum_{n=1}^\infty\frac{1}{\sqrt{n}}
\left[a_{n,i}\;e^{-in\frac{\pi}{l}(t-x)}+h.c.\right]
\; .
\nonumber
\end{eqnarray}
and the Hamiltonian can be written as
\begin{equation}
H=\sum_{i=1}^3 \frac{\pi }{l}\;
\left[ \frac{1}{g}
\left({{\hat Q}^R_i}\right)^2+\sum_{n=1}^\infty n\;a^\dagger_{n,i}a_{n,i}\right]
\; ,
\label{eq:hamil-mode-expansion}
\end{equation}
where the operator ${\hat Q}^R_i$ takes integer
eigenvalues $Q^R_i$. (Notice that the Hamiltonian and the mode
expansion for each of these three chiral fields are exactly those in
Eqs.~(\ref{fssR}) and (\ref{modeR}), respectively, but with $\beta$
replaced by $l$ and $g\neq 1$.)

The scattering picture is as follows. If an incident pulse of some
size $w$ is prepared and imparted into the scatterer at the origin
$x=0$, it will propagate freely according
Eq.~(\ref{eq:scattering-lagrangian1}) as long as the pulse does not
overlap with the origin, {\it i.e.}, its amplitude is vanishingly
small at $x=0$. In this case, the energy of this incoming pulse is
obtained from the Hamiltonian Eq.~(\ref{eq:hamil-mode-expansion}), and
it is independent of the boundary interaction at the origin, which is
what determines the scattering at the junction. Past the junction,
once the outgoing pulse has reached a vanishing overlap with the
origin, its energy is again obtained from the Hamiltonian
Eq.~(\ref{eq:hamil-mode-expansion}). 

Now, in the case of no tunneling between wires, the dynamics described
in Eq.~(\ref{eq:scattering-lagrangian1}) is exact even at $x=0$,
because the no tunneling regime corresponds to free propagation
through the junction; this is the case of perfect backscattering, or
N BC in the unfolded formalism [in which case the Hamiltonian in the
circle of size $2l$ in Eq.~(\ref{eq:hamil-mode-expansion}) should
coincide with that for a finite segment of size $l$ in
Eq.~(\ref{EN1})]. We argue that, once tunneling is turned on, the
Lagrangian should change due to the tunneling terms at $x=0$, but that
away from $x=0$ one can still use the free Lagrangian
Eq.~(\ref{eq:scattering-lagrangian1}) as long as the scattered pulse
goes far enough from $x=0$ so as to have vanishing overlap with the
origin. Finally, we should consider the situation where the scattered
pulse does not go all around the circle and scatters again; such
scattering picture is possible in the limit of $l\to \infty$, so that
scattered charges do not return to take the place of our prepared
initial state.

Let us prepare an incoming state $|{\rm IN}\rangle$ consisting of only
zero modes being occupied; their occupation is dictated by the
voltages $V_i$ applied to the leads. After scattering off the impurity
at $x=0$, the state becomes $|{\rm OUT}\rangle$, which in general has
excitations created in both the zero modes and the oscillator modes
($n>0$). The outgoing currents are solely determined by the
excitations of the zero modes.

Let $I^{\rm in}_i$ and $I^{\rm out}_i$, for $i=1,2,3$, be incoming and
outgoing currents in each of the three leads, respectively. The incoming
currents are related to the voltages on the three leads through
\begin{equation}
I^{\rm in}_i=g \frac{e^2}{h} V_i=\sum_j g\frac{e^2}{h}\,\delta_{ij}\;V_j
\; ,
\label{eq:cond-in}
\end{equation}
whereas the outgoing currents are obtained from the junction conductances
$G_{ij}$ as in Eq.~(\ref{eq:def-of-conduc-tensor}) through
\begin{equation}
I^{\rm in}_i-I^{\rm out}_i=I_i=\sum_j G_{ij} \;V_j
\; ,
\label{eq:cond-out}
\end{equation}
where $I_i$ is the {\it net} current flowing into the junction from
lead $i$.  

The $Q^{R}_i$ quantum number is the total charge of the state, which
in the unfolded formalism is proportional to the total current $I^{\rm
in}_i$ or $I^{\rm out}_i$ for the $|{\rm IN}\rangle$ or $|{\rm
OUT}\rangle$ states, respectively. (The currents are given by the
charges multiplied by the velocity, which we set to unity, divided by
a length $l$). Define
\begin{equation}
\Delta E_0=\sum_{i=1}^3 {\pi  l\over g}\;
\left[ \left({I^{\rm out}_i}\right)^2 
- \left({I^{\rm in}_i}\right)^2\right]
\; .
\label{eq:delta-Edef}
\end{equation}
This is the difference between the energy carried by the zero modes of the
two quantum states $|{\rm IN}\rangle$ and $|{\rm OUT}\rangle$. The incoming
state has no excited oscillator modes, but the outgoing state may have them.
Energy is conserved in the scattering process, so it follows [see
Eq.~(\ref{eq:hamil-mode-expansion})] that
\begin{equation}
\Delta E_0\le 0
\; ,
\end{equation}
with the equality holding when no oscillator modes are excited in $|{\rm
  OUT}\rangle$.

Using the expressions for the conductances
Eqs.~(\ref{eq:cond-in},\ref{eq:cond-out}), and 
substituting in Eq.~(\ref{eq:delta-Edef}),
we obtain
\begin{eqnarray}
\frac{g\Delta E}{l\pi} &=&
\sum_{ijk}\;G_{ij}G_{ik}\;V_jV_k-2\sum_{ij} g\frac{e^2}{h}\, G_{ij} V_i V_j
\label{eq:deltaE}
\\
&=&-\frac{1}{2}
\sum_{ijk}\left(G_{ij}G_{ik}-{2g\frac{e^2}{h}}\,G_{ij}\,\delta_{ik}\right)\;(V_j-V_k)^2
\; ,
\nonumber
\end{eqnarray}
where we have used that $\sum_j G_{ij}=0$, which follows from current
conservation, in order to express the second line 
in  Eq.~(\ref{eq:deltaE}) in
terms of voltage differences.

Now, the most general conductance tensor that satisfies the ${\bf Z}_3$
symmetry of the three lead problem is given by Eq.~(\ref{eq:Z3tensor}).
 Recall that a non-zero
$G_A$ is allowed if time-reversal symmetry is absent. 
Defining dimensionless parameters
\be
G\equiv G_S\,\frac{h}{e^2}
\quad
{\rm and}
\quad
\Delta\equiv \frac{1}{\sqrt{3}}\;G_A\,\frac{h}{e^2}
\ee
 for convenience, and inserting the expression for $G_{ij}$ in
Eq.~(\ref{eq:Z3tensor}) into the second line of Eq.~(\ref{eq:deltaE}),
we obtain after some elementary manipulations that
\begin{equation}
\frac{g\Delta E}{l\pi} =
\frac{3}{8}
\frac{e^4}{h^2}
\left[{G^2}-\frac{4g}{3}\;G+\Delta^2\right]\;
\sum_{jk}(V_j-V_k)^2
\; .
\end{equation}
The last term is always positive, so in order to satisfy $\Delta E\le 0$ one
must have that
\begin{equation}
{G^2}-\frac{4g}{3}\;G+\Delta^2\le 0
\; ,
\end{equation}
or
\begin{equation}
\left(G-\frac{2g}{3}\right)^2+\Delta^2 \le \left(\frac{2g}{3}\right)^2
\; ,
\label{eq:bounds}
\end{equation}
which imposes constrains to the allowed values of $G$ and $\Delta$. The
physical values are restricted to inside a circle of radius $2g/3$ and origin
at $(2g/3,0)$ in the $G-\Delta$ plane, as shown in
Fig.~\ref{fig:cond-circ}.  The boundary of the circle correspond to the cases
where $\Delta E=0$ when no energy is transfered to oscillator modes.
\begin{figure}
\includegraphics[width=6cm]{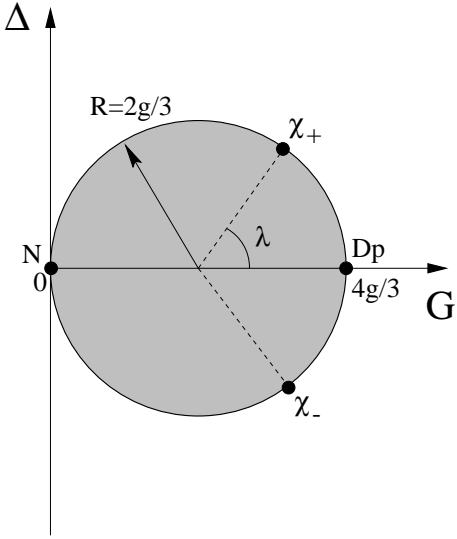}
\caption{Physical values of $G$ and $\Delta$, allowed from energy
  conservation, must lie inside the circle of radius $2g/3$ and origin at
  $(2g/3,0)$ in the $G-\Delta$ plane. The boundary of the circle
  corresponds to the case when no energy is transfered to oscillator modes
  upon scattering off the junction ($\Delta E=0$). The points corresponding
  to boundary conditions N, D$_\shbox{P}$,
 and $\chi_\pm$ lie on this boundary, and
  are are marked on the figure.}
\label{fig:cond-circ}
\end{figure}

\subsection{T-invariant case: $\Delta=0$}
If time-reversal is a symmetry in the problem, then $\Delta$ must vanish, so
that $G_{ij}=G_{ji}$. In this case, the allowed physical values of $G$ lie in
the segment that spans the diameter of the circle in the $\Delta=0$ axis, or
\begin{equation}
0\le G_S\le \frac{4g}{3}\frac{e^2}{h}
\; .
\end{equation}
The equality $\Delta E=0$ occurs when $G=0$ and when $G=4g/3$; these
cases correspond, respectively, to N and \DP boundary conditions (for a
summary of the conductances for different boundary conditions, see
section~\ref{sec:results}). Hence, for these two boundary conditions, no
energy is transfered into oscillator modes, and it is simply redistributed
among the zero modes of the three boson fields.

\subsection{T-broken case: $\Delta\ne 0$}
This is the general case, in which the physical values of $G$ and
$\Delta$ must lie within the circle, and the boundary corresponds to
the cases $\Delta E=0$. There are two particular cases, as shown in
Fig.~\ref{fig:cond-circ}, which correspond to the two chiral fixed
points $\chi_\pm$, when the angle $\lambda$ satisfies $\tan
\lambda=\pm g/\sqrt{3}$. In this case, the conductances are given by
$G_S=\frac{4g}{3+g^2}\frac{e^2}{h}$ and
$G_A=\sqrt{3}\,\Delta\frac{e^2}{h}=\pm g\,G_S$.


\section{Y-junctions attached to Fermi liquid leads}
\label{sec:FLL}
In this Section, we discuss the effective conductance of the junction
in the presence of Fermi liquid (FL) leads,
as introduced in Fig.~\ref{fig:deviceB}.
The FL leads can be thought of $g_{\rm lead}=1$
wires connected at
the endpoints of the three Y-junction wires with $g$ parameter. If the leads
and the three quantum wires are connected by a point contact, then one can
show that only one mode of the FL couples to the wire. This is similar
to the case of the Kondo problem, where only the $s$-channel couples to the
point-like impurity. In the case of the FL-quantum wire coupling at a point,
it is not the $s$-channel but instead some other channel that couples. 
~\cite{Chamon-Fradkin97}.

When we derived the conductance of the junction, we assumed that the wires
were subject to voltages $V_i$, $i=1,2,3$. If we have Fermi liquid leads,
what happens is that the reservoirs are at voltages $\bar V_i$, $i=1,2,3$,
and one must obtain the relation between the $\bar V_i$ and $V_i$. If that is
done, all we have left to do 
is to re-express the currents $I_i$ in terms of the
$\bar V_i$. In other words, we need to calculate the conductance
${\bar  G}_{ij}$ from the $G_{ij}$ we already computed:
\begin{equation}
  I_j = \sum_k G_{j k} V_k = \sum_k {\bar G}_{j k} {\bar V}_k
\; .
\end{equation}
The question is how to relate the $\bar V_i$ to the $V_i$. The answer is that
the voltage difference $\bar V_i-V_i$ across the lead/wire contact is
related to the current $I_i$ that flows through the contact:
\begin{equation}
  I_i = G_c \;(\bar V_i-V_i)
\; ,
\end{equation}
where the fixed point conductance for $g>1$ is
$G_c=2g/(g-1)\;e^2/h$. This can be derived by looking at the problem
of a single wire connected to FL leads, or in general for
two TLLs with $g_1>g_2$, as done in
Ref.~\cite{Chamon-Fradkin97}, in which case
$G_c^{-1}=(g_2^{-1}-g_1^{-1})/2 \;e^2/h$. For the case $g<1$, the fixed point
conductance is zero, and there is no transport between the leads and
the wires. Notice that if $g\to 1$ then $G_c\to \infty$, and hence
$\bar V_i=V_i$ for any finite current; this makes sense, as for if
both sides have $g=1$, there is no voltage drop across the lead/wire
interface.

It is easy to get ${\bar G}_{ij}$ now. We express $V_i$ in terms of $\bar
V_i$ and $I_i$:
\begin{equation}
V_i= \bar V_i - I_i/G_c
\; ,
\end{equation}
and then substitute in the formula for the current in terms of the $V_i$:
\begin{equation}
I_j = \sum_k G_{j k} V_k = \sum_k G_{j k} (\bar V_k - I_k/G_c)
\; ,
\end{equation}
or in matrix notation
\begin{equation}
{I}= {\bm G} \;{\bar{V}} - G_c^{-1} {\bm G} \;{I}
\; ,
\end{equation}
{}from which we obtain that
\begin{equation}
{I}= (\openone + G_c^{-1} {\bm G})^{-1}\;{\bm G} \;{\bar{V}} 
={\bar{\bm G}} \;{\bar{V}} 
\end{equation}
or, finally,
\begin{equation}
{\bar{\bm G}}=(\openone + G_c^{-1} {\bm G})^{-1}\;{\bm G} 
\; ,
\label{eq:GbarG}
\end{equation}
and more simply
\begin{equation}
{\bar{\bm G}}^{-1}={\bm G}^{-1}+ G_c^{-1} \openone
\; ,
\label{eq:RbarR}
\end{equation}
which can be interpreted physically as saying that the resistance tensors of
the Y-junction must be added to the resistances due to the lead/wire
interface resistances. [Notice, though, that this is a formal expression
since in practice the matrix $\bm G$ is singular because of the condition
$\sum_k G_{jk} =0$ -- hence one must use Eq.~(\ref{eq:GbarG}) instead of
Eq.~(\ref{eq:RbarR}).] We would like to point out that our expressions
Eqs.~(\ref{eq:GbarG},\ref{eq:RbarR}) lead to consistent results with those of
Refs.~\cite{Maslov-Stone,Safi-Schulz} when applied to the simpler two-wire
problem.


As an example, let us consider the effective conductance at the
chiral fixed points $\chi_\pm$, in the presence of the FL leads.
All we have to do is use the expression for the conductance tensor $\bm G$
and substitute into Eq.~(\ref{eq:GbarG}).
For the chiral fixed points $\chi_{\pm}$,
recall the conductance Eq.~(\ref{eq:GDGA}). 

Substitution of Eq.~(\ref{eq:GDGA}) into Eq.~(\ref{eq:GbarG}) leads to
\begin{equation}
  \bar G_{jk}^{\pm} = \frac{e^2}{h}\;[(3 \delta_{jk}  -1)
         \pm 1 \;\epsilon_{jk}]/2,
\label{eq:barGDGA}
\end{equation}
which is simply what one would obtain from the conductance tensor
Eq.~(\ref{eq:GDGA}) if we set $g=1$. This result is in agreement with
numerical results that find that the current vanishes in one of the leads
despite interactions, if the wires are connected to long non-interacting
segments~\cite{Meden2}. The scaling exponent for the corrections, however,
should still depend on $g$, and is given by $\Delta=4g/(3+g^2)$, which is
also consistent with the numerical findings obtained for small attractive
interactions. The sensitivity to the flux (and flow to $\chi_\pm$) should
also be a property of the interacting junction, not the FL leads.

The effective conductance $\bar{G}_{jk}$
at other fixed points can be obtained in similar manner.
The results are summarized in Sec.~\ref{sec:results}.
We find that the effective conductance tensor at each known
fixed points can be identified with a ``bare'' conductance tensor
of the junction of three Fermi liquids ($g=1$) at an appropriate
fixed point.
This is rather natural, because the system with finite (interacting) wires
and infinite FL leads would eventually be renormalized into a
junction of three FL leads in the low-energy limit.

\section{Open problems}
\label{sec:open}
In this paper we studied the problem of a junction of three quantum
wires connected by a ring, through which a magnetic flux can be
applied. The electron systems in the three leads were described in
terms of a Tomonaga-Luttinger liquid of spinless particles, with
dimensionless interaction parameter $g$. One of the main difficulties in
studying the problem with three wires is the necessity to include
Klein factors that ensure the proper fermionic statistics for
electrons in different leads.

We approached the Y-junction problem using different methods: mapping
to the dissipative Hofstadter model, boundary conformal field theory,
and delayed evaluation of boundary conditions. These methods gave
consistent results and allowed us to calculate the low-voltage and
low-temperature fixed point conductance tensor for the junction as a
function of the interaction parameter $g$. (See section
\ref{sec:results} for a summary of the results in the paper.)
In this concluding section, we would like to list a few of the open
directions that require further investigation.

The first one is the inclusion of electron spin in the problem. Even
in the case of tunneling between just two wires, the inclusion of the
spin degree of freedom leads to a rather rich phase diagram for the
charge and spin conductances as a function of the interaction
parameter~\cite{Kane}. For example, one finds situations where the
spin conductance can vanish while the charge conductance does not, and
vice versa, or systems that conduct both charge and spin, or that
conduct neither. In the case of the three-lead junction the situation
could, in principle, be much richer. One might wonder, for instance,
whether there can be fixed points where the junction can separate the
spin and charge of an electron incoming from one lead into different
outgoing leads.

It appears that the method of delayed evaluation of boundary
conditions may lend itself naturally to the study of the problem with
electron spin.  The reason is that this approach seems to be
generalizable to multiple species. In the case of three leads with
spin, one could try to extend the treatment laid down in this paper by
taking six branches altogether.

One of the major open problems concerning the Y-junctions is the
nature of the stable fixed point in the range $1<g<3$ when
time-reversal symmetry is preserved ($\phi=0,\pi$). So far, we have
not been able to identify the correct boundary condition for this
``mysterious'' ($M$) fixed point. We can, however, be sure that the
nature of this $M$ fixed point is rather different from $N,D$ and
$\chi_\pm$. For example, using the energy conservation arguments
presented in the scattering approach of section
\ref{sec:Econservation}, we can argue that energy is dissipated into 
oscillator modes upon scattering off the junctions, which was not the
case in the other four fixed points $N,D$ and $\chi_\pm$. We expect
that the $M$ fixed point should also describe the infrared stable
behavior of the resonant tunneling model of Nayak {\it et
al.}~\cite{Nayak}.

It is also clearly important to test our results by numerical methods,
which could also guide analytical work in understanding the $M$ fixed
point. One numerical method, used by Barnabe-Theriault {\it et
al.}~\cite{Meden2}, is an approximation based on the functional
renormalization group approach. Within this method, the chiral fixed
points we predicted were observed for small attractive interactions,
when time-reversal symmetry was broken by an external field. Another
fixed point, with maximal symmetric conductance for non-interacting
fermions, was found for the case of zero flux. It is not clear whether
this maximal conductance value is related to the conductance of the
$M$ fixed point once the renormalization effect of Fermi liquid leads
is taken into account (the interaction in the wires in the numerical
studies was turned off far away from the junction), or whether the
approximation scheme can only capture conductance values that are
equivalent to those of single particles scattering of a renormalized
barrier. It would be interesting to push this approach towards larger
values of the attractive potential ($g>3$), when the $D$ fixed point
becomes stable, since the physics in this case will be that of pair
tunneling, and hence values for the conductance greater than the
maximal one for single particle scattering should be attainable (see
\ref{sec:results} for a discussion).

Finally, the presence of the essential Klein factors in the Y-junction
is related to a Fermi minus sign problem which would presumably make
Monte Carlo methods difficult. This is to be contrasted with the two
wire case where the Klein factors can be trivially eliminated and the
resulting Coulomb gas can be very effectively studied by classical
Monte Carlo
\cite{Moon}. The Y-junction problem is perhaps  the simplest system 
with a fermion sign problem, which comes only from the junction. (In a
lattice formulation, the problem corresponds to three 1D systems
connected together by only one rung at one of the end points.) Hence,
it seems natural that this system, together with some of the
analytical results that we derive, may be used as a benchmark for
approximations that aspire to solve the fermion sign problem.

\section*{Acknowledgements}
The authors would like to thank V. Meden for useful correspondence and
for showing us his numerical results.
The work by MO was supported in part 
by a Grant-in-Aid for scientific research
and by the 21st Century COE program at Tokyo Institute of Technology
``Nanometer-Scale Quantum Physics'',
both from MEXT of Japan.
CC was supported in part by the NSF grant DMR-0305482.
IA was supported in part by NSERC of Canada,  
the Canadian Institute for Advanced Research and a 
JSPS invitation fellowship. 
The authors would also like to thank
the hospitality of the Physics Departments at BU, UBC and Tokyo Tech, 
and the Aspen Center for Physics, where parts of this work
were done.

\appendix

\section{Free fermion case}
\label{sec:free-fermion}
In this appendix we calculate the conductivity for the free fermion model
($g=1$). This is done by first calculating the $S$-matrix and then using the
Landauer formalism.  We consider both lattice and continuum free fermion
models.

\subsection{Tight binding model}
We first consider a tight-binding version of the continuum model that we
study using bosonization. $\psi_{n,j}$ annihilates a fermion on site $n$ on
wire $j$. Here $n=0,1$, $2$, ...$\infty$ and $j=1$, $2$, $3$, with the wire
index $j=0$ identified with $j=3$. The Hamiltonian is (setting the hopping
amplitude $t=1$):
\begin{eqnarray}
H&=&-\sum_{n=0}^\infty \sum_{j=1}^3(\psi^\dagger_{n,j}\psi_{n+1,j}+h.c.)
\nonumber\\
&&-(\tilde \Gamma /2) \sum_{j=1}^3[e^{i\phi /3}\psi^\dagger_{0,j}\psi_{0,j-1}+h.c.].
\end{eqnarray}
Writing a single-electron state as:
\begin{equation}
|\Psi \rangle=\sum_{n,j}\Phi_{n,j}\,\psi^\dagger_{n,j}|0\rangle,
\end{equation}
where $\Phi_{n,j}$ is the lattice wave-function, it is 
seen to obey the lattice Schr\"odinger equation:
\begin{widetext}
\begin{eqnarray} 
E\Phi_{n,j}&=&-[\Phi_{n+1,j}+\Phi_{n-1,j}]\qquad \qquad \qquad 
\qquad \qquad \qquad  (n\geq 1)\nonumber \\
E\Phi_{0,j}&=&-\Phi_{1,j}-(\tilde \Gamma /2)[e^{i\phi /3}\;\Phi_{0,j-1}+
e^{-i\phi /3}\;\Phi_{0,j+1}].
\label{SE}\end{eqnarray}
The scattering solutions of this equation take the form:
\begin{equation}
\Phi_{n,j}=A_{{\rm in},j}\;e^{-ikn}+A_{{\rm out},j}\;e^{ikn},
\label{ALR}
\end{equation}
for all $n\ge 0$ and $j=1,2,3$.  The corresponding energy, which follows from
the first line of the Schr\"odinger equation Eq.~(\ref{SE}) is:
\begin{equation}
E=-2\cos k.
\label{eq:latticeE}
\end{equation}
Noting that Eqs.~(\ref{ALR})~and~(\ref{eq:latticeE}) imply
\begin{equation}
\Phi_{1,j}+E\Phi_{0,j}=-(A_{{\rm in},j}\;e^{ik}+A_{{\rm out},j}\;e^{-ik}),
\end{equation}
we see that the second line of the Schroedinger equation Eq.~(\ref{SE}) gives
\begin{equation}
(\tilde \Gamma /2)\;[e^{i\phi /3}\;(A_{{\rm in},j-1}+A_{{\rm
out},j-1}) +e^{-i\phi /3}(A_{{\rm in},j+1}+A_{{\rm
out},j+1})]=A_{{\rm in},j}\;e^{ik}+A_{{\rm out},j}\;e^{-ik}
\;.
\end{equation}
Assembling the three components of the amplitudes, $A_{{\rm in},j}$ and
$A_{{\rm out},j}$ into vectors, $\vec A_{{\rm in}}$ and $\vec A_{{\rm out}}$,
this becomes:
\begin{equation}
[e^{-ik}-(\tilde \Gamma /2)\,M
]\;\vec A_{{\rm out}}=-[e^{ik}-(\tilde \Gamma /2)\,M 
]\;\vec A_{{\rm in}},
\end{equation}
\end{widetext}
where
\begin{equation}
M\equiv e^{i\phi /3}B^{-1}+e^{-i\phi /3}B,\label{M}
\end{equation}
and 
\begin{equation}
B\equiv \left(\begin{array}{lll}
0&1&0\\
0&0&1\\
1&0&0
\end{array}\right).
\label{B}
\end{equation}
Thus, the outgoing and incoming amplitudes are related by a 3-dimensional 
S-matrix:
\begin{equation}
\vec A_{{\rm out}}=S\vec A_{{\rm in}},
\end{equation}
where:
\begin{equation}
S=-[e^{-ik}-(\tilde \Gamma /2)\,M]^{-1}
[e^{ik}-(\tilde \Gamma /2)\,M].
\label{S}
\end{equation}
or, equivalently,
\begin{equation}
S=-e^{i2k}\,[1-e^{ik}(\tilde \Gamma /2)\,M]^{-1}
[1-e^{-ik}(\tilde \Gamma /2)\,M].
\label{SS2}
\end{equation}
\subsection{Continuum model}
We use the continuum model with  the Hamiltonian 
density:
\begin{eqnarray}
{\cal H}&=&i\sum_j\psi^\dagger_j {d\over dx}\psi_j
\\
&&-\delta (x)\sum_j\left\{
\Gamma [e^{i\phi /3}\psi^\dagger_j\psi_{j-1}+h.c.]+r\psi^\dagger_j\psi_j
\right\}.
\nonumber
\end{eqnarray}
Here we have ``unfolded'' the system so that all fields are left-movers.  We
do not explicitly write the subscripts $L$, and we have set $v_F=1$.  The
corresponding Schr\"odinger equation is:
\begin{widetext}
\begin{equation}
i{d\over dx}\Phi_j-\delta (x)\left\{\Gamma \;[e^{i\phi /3}\;\Phi_{j-1}
+e^{-i\phi /3}\;\Phi_{j+1}]+r\;\Phi_j\right\}=E\Phi_j.
\end{equation}
Writing the scattering states as
\begin{equation}
\Phi_j (x)=\left\{
 \begin{array}{lc}
  A_{{\rm in},j}\;e^{-ipx}& (x>0) \\
  A_{{\rm out},j}\;e^{-ipx}& (x<0)
   \end{array}
\right.
\end{equation}
with energy $E=p$, we obtain (for $E\to 0$):
\begin{equation}
i(A_{{\rm in},j}-A_{{\rm out},j})=\frac{1}{2}
\left\{\Gamma [e^{i\phi /3}\,(A_{{\rm in},j-1}+A_{{\rm out},j-1})
+e^{-i\phi /3}\,(A_{{\rm in},j+1}+A_{{\rm out},j+1})]+
r\;(A_{{\rm in},j}+A_{{\rm out},j})\right\}.
\end{equation}
\end{widetext}
Note that we have taken $\Phi_j(0)$ to be the average of its 
values at $0^+$ and $0^-$.  This can be rewritten:
\begin{equation}
[(r+2i)+\Gamma M]\;\vec A_{{\rm out}}=-[(r-2i)+\Gamma M]\;\vec A_{{\rm in}},
\end{equation}
where $M$ is the same matrix defined in the discussion of the 
tight-binding model, Eq. (\ref{M}).
Thus the S-matrix for the continuum model is:
\begin{equation}
S=-[(r+2i)+\Gamma M]^{-1}[(r-2i)+\Gamma M].
\end{equation}
This will be the same S-matrix as in the tight-binding model [Eq.~(\ref{S})],
for a particular value of $k$, if we choose $r$ and $\Gamma$ so that:
\begin{equation}
e^{ik}=-\frac{r-2i}{|r-2i|}
\end{equation}
and
\begin{equation}
\tilde\Gamma/2= {\Gamma \over |r-2i|} .
\end{equation}
It can be shown that the continuum limit of the tight-binding model, 
with the spectrum linearized around $k_F$, gives values of $\Gamma $ 
and $r$ satisfying this equation when $k=k_F$.  Note, in particular, 
that for the particle-hole symmetric case, $k_F=\pi /2$, $r=0$ 
and $\Gamma =\tilde \Gamma$. 

\subsection{The S-matrix}
An explicit expression for the S-matrix of Eq.~(\ref{SS2}) can be found by
straightforward algebra (we will drop the overall phase factor $e^{i2k}$
which is of no physical consequence to the conductances). This S-matrix can
be parameterized by the most general $Z_3$ symmetric form:
\begin{equation}
S=S_0+S_-B+S_+B^{-1}.
\end{equation}
It is convenient to define:
\begin{equation}
z_{\pm}\equiv (\tilde \Gamma /2)e^{i(k\pm \phi /3)}.
\end{equation}
The coefficients $S_i$ in the S-matrix then are given by:
\begin{widetext}
\begin{eqnarray} 
S_0&=&(z_+z_- + |z_+|^2 +|z_-|^2 +z_-^2z_+^*+z_+^2z_-^*-1)/D\nonumber 
\\ 
S_+&=&(z_-^* -z_+ -z_-^2 +z_-z_+^*)/D
\nonumber \\
S_-&=&(z_+^*-z_--z_+^2+z_+z_-^*)/D,
\label{Sa}
\end{eqnarray}
where:
\begin{equation}
D=1-3z_+z_--z_-^3-z_+^3.
\end{equation}
It can be readily seen that $S\to -e^{i2k}I$ at $\tilde \Gamma \to 0$ and 
$S\to -I$ at $\tilde \Gamma \to \infty$.
For $\tilde \Gamma /2=1$ a perfectly chiral S-matrix occurs, for 
some value of $\phi$,  
with $S_+$ or $S_-=0$.  In this case $|z_{\pm}|=1$.
Thus we see that:
\begin{equation}
z_-z_+S_+=(z_+-z_-z_+^2-z_-^3z_++z_-^2)/D=(1-z_+z_-)(z_++z_-^2)/D.
\end{equation}
Thus $S_+=0$ when $z_-^2z_+^*=e^{i(k-\phi )}=-1$ or
\begin{equation}
k-\phi = (2n+1)\pi ,
\end{equation}
for integer $n$. At this point, 
\begin{equation}
S_0=(z_+z_-+1+z_-^2z_+^*+z_+^2z_-^*)/D=0,
\end{equation}
and therefore only $S_-=-(z_-^*)^2\ne 0$. This corresponds to an electron
 incident on wire $j$ being transmitted with probability one to wire $j-1$.
 Conversely, $S_-=S_0=0$ when $\tilde \Gamma /2=1$ and
\begin{equation}
k+\phi = (2n+1)\pi ,
\end{equation}
so that an electron incident on wire $j$ is transmitted 
with probability one to wire $j+1$. 

In the time-reversal invariant cases, $\phi =0$ or $\pi$, $S_+=S_-$; 
an electron incident on wire $j$ has equal amplitude to be 
transmitted to wire $j-1$ or $j+1$. The maximum possible 
value of $|S_+|=|S_-|$ consistent with unitarity of the S-matrix is 
$|S_+|=2/3$, $|S_0|=1/3$. For $\phi =0$, one obtains
\begin{equation}
S_+={z^*-z\over (1+z)(1-2z)},
\end{equation}
with $z=(\tilde \Gamma /2)e^{ik}$. For any value of $k$, $|S_+|$ reaches 2/3
for some value of $\tilde \Gamma $ of order one.

\subsection{Conductance}
In the free fermion case, we obtain the conductance directly from the 
S-matrix by the Landauer formalism. We assume a thermal distribution 
of electrons heading towards the junction from distant reservoirs 
with different chemical potentials. The total current on wire $j$ 
is this incoming current minus the reflected current minus 
the current transmitted from the other 2 wires. Thus:
\begin{equation}
I_j=e\int {dk\over 2\pi }\,v(k)\left[(1-|S_{jj}|^2)\,n_F(\epsilon_k-\mu -eV_j)
-|S_{j,j+1}|^2\,n_F(\epsilon_k-\mu -eV_{j+1})-
|S_{j,j-1}|^2\,n_F(\epsilon_k-\mu -eV_{j-1})\right]
\end{equation}
\end{widetext}
where $n_F(\epsilon )$ is the Fermi distribution function. 
Using $v=(d\epsilon /dk)/\hbar$ and expanding to linear order in the $V_j$'s 
gives at $T=0$:
\begin{equation}
I_j={e^2\over h}\{ [1-|S_{jj}|^2]V_j-|S_{j,j+1}|^2V_{j+1}-
|S_{j,j-1}|^2V_{j-1} \},
\end{equation}
where the S-matrix is evaluated at $k=k_F$. 
Thus the conductance tensor is:
\begin{equation}
G_{jk}={e^2\over h}\left[ (3\delta_{jk}-1){|S_+|^2+|S_-|^2\over 2}+
{|S_+|^2-|S_-|^2\over 2}\epsilon_{jk}\right], 
\end{equation} where $S_{\pm}$
are given by Eq. (\ref{Sa}).  In particular, for the perfectly chiral
S-matrices discussed above, this reduces to: 
\begin{equation}
G_{jk}={e^2\over 2h}\left[ (3\delta_{jk}-1)\pm \epsilon_{jk} \right].
\end{equation}

\section{Review of boundary conformal field theory techniques}
\label{sec:BCFT}
Here we briefly review the boundary conformal field theory (BCFT) techniques,
developed largely by J. Cardy~\cite{Cardy-review}, which have been applied to
various quantum impurity problems.

We consider a general conformal field theory, such as a collection 
of free bosons or fermions, defined 
on the 1/2-line, $x\geq 0$,  with a conformally invariant boundary 
condition at $x=0$. In general two-point correlation functions 
will be affected by the boundary. While the correlation function 
of some operator, ${\cal O}(x,\tau )$ may behave as:
\be \langle {\cal O}(x,\tau ){\cal O}(x',\tau ' )\rangle =
{1\over [(x-x')^2+(\tau -\tau ')^2]^{\Delta}},\ee
in the bulk (i.e. in the absence of a boundary or far from the 
boundary), in the limit $x,x' \ll |\tau -\tau '|$ it behaves as:
\be \langle {\cal O}(x,\tau ){\cal O}(x',\tau ' )\rangle 
\propto {1\over |\tau - \tau '|^{2\Delta_B}}\ee
Here $\Delta$ is the bulk scaling dimension of the operator, ${\cal O}$ 
and $\Delta_B$ is its boundary scaling dimension.  In general, 
$\Delta_B$ will depend on the boundary conditions. The boundary 
conditions are always assumed to imply:
\be {\cal P}(t,0)=0,\label{Peq0}\ee
where:
\be {\cal P}(t,x)\equiv {\cal H}_R(t-x)-{\cal H}_L(t+x).\ee
${\cal P}$ is the momentum density and ${\cal H}_{L/R}$ are the 
left and right-moving parts of the Hamiltonian density.

We can determine all boundary scaling dimensions for an arbitrary 
conformally invariant boundary condition, $A$, denoted $\Delta_A$, from the 
finite size spectrum of the Hamiltonian on a strip of length $l$ 
with boundary conditions $A$ at both ends. This is done 
by the conformal transformation:
\be z=le^{\pi w/l},\ee
Here 
\be z=\tau + ix,\ee
covers the infinite half-plane and $w=u+iv$ the strip 
with $0<v<l$. The correlation function for two points 
at the edge of the strip can be obtained by this conformal 
transformation:
\bea \langle {\cal O}(u_1){\cal O}(u_2)\rangle &=&
\left\{{{\partial z\over \partial w}(u_1){\partial z\over \partial w}(u_2)
\over |z(u_1)-z(u_2)|^2}\right\}^{\Delta_A}\nonumber \\
&=& \left[ {2l\over \pi}\sinh {\pi \over 2l}(u_1-u_2)\right]^{-2\Delta_A}
.\eea
As $u_2-u_1\to \infty$, this approaches:
\be  \langle {\cal O}(u_1){\cal O}(u_2)\rangle \to 
\left({\pi \over l}\right)^{2\Delta_A}\exp \left[-\pi \Delta_A {(u_2-u_1)\over l}\right] .\ee
On the other hand, since $u$ is 
imaginary time, 
we may evaluate the correlation function on the strip 
by inserting a complete set of states:
\be \langle {\cal O}(u_1){\cal O}(u_2)\rangle
= \sum_n|\langle 0|{\cal O}|n\rangle_{AA} |^2\exp [-E_n(u_2-u_1)].\ee
Here $|0\rangle$ is the groundstate and $|n\rangle$ an arbitrary excited 
state for the strip Hamiltonian with b.c.'s $A$ at both ends. 
(The groundstate occurs here since the strip has infinite 
length in the imaginary time, $u$, direction.)
$E_n$ is  
the energy of the $n^{th}$ state (measured from the 
groundstate energy). As $u_2-u_1\to \infty$, the lowest 
energy excited state that can be created from the groundstate by 
the operator ${\cal O}$ dominates so we conclude that this state has 
energy:
\be E_1=\pi {\Delta_A\over l}.\label{fss}\ee
There is a one-to-one correspondance between boundary operators and 
the finite size spectrum with the boundary scaling dimensions 
and finite size energies related by Eq. (\ref{fss}).

Another very useful quantity to consider is the partition function $Z_{AB}$
with conformally invariant b.c.'s $A$ and $B$ at the 2 ends of a 
finite system of length $l$, at inverse temperature, $\beta$. This 
may be expressed in terms of the finite size spectrum on the strip as:
\be
Z_{AB}=tr \exp [-\beta H_l^{AB}]=\sum_n \exp [-\beta E_n^{AB}],
\label{ZABq}
\ee
where the $E_n^{AB}$ are the energies on the strip with b.c.'s $A$ and $B$ 
at the two ends. This picture is sometimes referred to as
``open string channel''.
On the other hand, we may interchange space and imaginary 
time and write instead:
\be Z_{AB}=\langle A|\exp [-lH_\beta^P]|B \rangle ,\ee
where $|A\rangle$ and $|B\rangle$ are {\it boundary states} that correspond 
to the b.c.'s $A$ and $B$ respectively. Here $H_\beta^P$ denotes 
the Hamiltonian with periodic b.c.'s on a ring of length $\beta$.
This point of view is sometimes referred to as ``closed string channel''.
All boundary states obey the condition:
\be {\cal P}(x)|A\rangle =0.\label{Peq0B}\ee 
The eigenstates of $H^P_\beta$ are of the form:
\be E^P_n = {2\pi \over \beta}\Delta_n,\label{fssB}\ee
where the $\Delta_n$ are bulk scaling dimensions. This follows, 
similarly to the correspondance between boundary scaling dimensions 
and boundary energies, 
from a conformal transformation from the infinite plane to the cylinder of 
circumference $\beta$. 

\section{Review of boundary conformal field theory for standard free bosons}
\label{sec:refboson}
\subsection{One component free boson}
A simple example of boundary dimensions and energies is given 
by the periodic boson. Thus we consider the bulk Lagrangian density:
\be {\cal L}= {g\over 4\pi} (\partial_\mu \varphi )^2,\ee
and assume that $\varphi (x,t)$ is a periodic variable so that 
$\varphi$ is identified as
\be
\varphi \leftrightarrow \varphi + 2\pi n \ \ (n\in \mathbf{Z}).
\label{eq:phiperiod}
 \ee 
\begin{widetext}
These boundary conditions determine the bulk mode expansion for $\varphi (x,t)$ 
on a circle of circumference $\beta$, $0<x<\beta$:
\be \varphi (t,x)=\hat \varphi_0+{2\pi \over \beta}\left[\hat P'x+{1\over g}\hat Pt\right]
+{1\over \sqrt{2g}}\sum_{n=1}^\infty {1\over \sqrt{n}}\left\{ 
a_n^L\exp \left[-inx_+\frac{2\pi}{\beta}\right]+a_n^R\exp \left[-inx_-\frac{2\pi}{\beta}\right]+ h.c.\right\}.\label{mode}\ee
Here:
$\hat P$ is the momentum operator conjugate to the constant term, $\hat \varphi_0$:
\be [\hat \varphi_0,\hat P]=i,\ee
and $\hat P'$ is another momentum operator.  They both have integer eigenvalues. 
The integer eigenvalues of $\hat P'$ follow from the periodic b.c.'s and 
the angular nature of $\varphi$:
\be \varphi (\beta ) =\varphi (0)+2\pi n .\ee
The integer eigenvalues of $\hat P$ follow from the fact that 
it is conjugate to an angular variable, $\hat \varphi_0$.  Explicitly, 
the wave-function, $\exp [-iP\varphi_0]$, with $P$ an eigenvalue of $\hat P$, 
must be single valued. 
\be x_{\pm} \equiv t\pm x,\ee
and $a_n^{L/R}$ are boson creation operators for the 
left and right moving finite momentum modes. This mode expansion is consistent 
with the equal-time commutation relations:
\be \left[ \varphi (x),{\partial\varphi (x')\over \partial t}\right]={2\pi \over g}i\delta (x-x'),\ee
which follow from the normalization of the Lagrangian. We may decompose $\varphi$ 
into left and right moving modes:
\be \varphi (t,x)=\varphi_L(x_+) +\varphi_R(x_-),\ee
and then write the dual field:
\be
\theta \equiv g(\varphi_L-\varphi_R).
\label{eq:thetadef}
\ee
This has the mode expansion:
\be
\theta (t,x)=\hat \theta_0+{2\pi \over \beta}[\hat Px+g\hat P't]
+\sqrt{g\over 2}\sum_{n=1}^\infty {1\over \sqrt{n}}\left\{ 
a_n^L\exp \left[-inx_+\frac{2\pi}{\beta}\right]
-a_n^R\exp \left[-inx_-\frac{2\pi}{\beta}\right]+ h.c.\right\}.
\label{eq:thetamode}
\ee
$\hat P'$ is the momentum conjugate to $\hat \theta_0$:
\be [\hat \theta_0,\hat P']=i,\ee
and we see from this mode expansion that $\theta (t,x)$ is also an angular variable:
\be \theta (t,x)\leftrightarrow \theta (t,x)+ 2\pi n \ \  (n\in \mathbf{Z}).\ee
\end{widetext}
We may equally well write the Lagrangian density in terms of $\theta$:
\be {\cal L}={1\over 4\pi g}(\partial_\mu \theta )^2.\ee
The bulk primary operators are $\exp \left\{ i[n\varphi (t,x)+m\theta (t,x)]\right\}$ with 
scaling dimension:
\be \Delta = {n^2\over 2g}+{gm^2\over 2}\ \  (n,m\in \mathbf{Z}).\ee
(All descendent operators are simply primaries multiplied by products of 
multiple derivatives of $\varphi$. These have dimensions equal to that 
of the corresponding primary plus a positive integer. )
Inserting the mode expansion of $\varphi$ into the Hamiltonian, gives the 
finite size spectrum with periodic b.c.'s on a circle of circumference $\beta$:
\be H={2\pi \over \beta}\left[ {{\hat{P}}^2\over 2g}+{{\hat{P'}}^2 g\over 2}
+\sum_{n=1}^\infty n(a_n^{L\dagger}a_n^L+a_n^{R\dagger}a_n^R)\right] .\ee
We see that the relationship, Eq. (\ref{fssB}) between the 
bulk scaling dimensions and finite size energies with periodic b.c.'s 
is obeyed. 

Now consider the Dirichlet (D) b.c. on $\varphi$:
\be \varphi (t,0)=\varphi_0,\ \ (\hbox{a constant, independent of $t$}),\ee
implying:
\be {\partial \varphi \over \partial t}(t,0)=0.\label{Dbc}\ee
We may calculate the dimensions of all boundary operators directly 
by recognizing that the D b.c.:
\be \varphi_R(t,0)=\varphi_0-\varphi_L(t,0),\ee
determines $\varphi_R(t,x)$ (for $x>0$) as the 
analytic continuation of $\varphi_L(t,x)$ to the negative $x$-axis:
\be \varphi_R(t,x)=\varphi_0-\varphi_L(t,-x),\ \  (x>0). \ee
The correlation function for $\varphi_L(t,x)$:
\be \langle \varphi_L(t,x)\varphi_L(0,0)\rangle =-{1\over 2g}\ln x_+ + \hbox{constant},\ee
is unaffected by the boundary. Thus we see that the boundary operators can be 
rewritten by the replacements:
\bea \varphi (t,0)&\to& \varphi_0 \nonumber \\
\theta (t,0) &\to & 2\varphi_L(t,0)-\varphi_0.\eea
Thus the non-trivial operators are:
\be \exp [im\theta (t,0)]\to \exp [2im\varphi_L(t,0)-im\varphi_0],\ee
of scaling dimension:
\be \Delta_D=gm^2.\label{Delta_D}\ee
Note that this is twice the {\it bulk} scaling dimension of $\exp [im\theta ]$. 
This factor of 2 arises from the b.c. The bulk scaling dimension of 
$\exp [im\theta ]$ consists of equal contributions of $gm^2/4$ from 
the left and right factors: $\exp [\pm i\varphi_{L/R}]$.  Upon imposing 
the D b.c. the right factor vanishes and the left factor has 
an extra factor of 2 in the exponent which quadruples the scaling dimension:
$gm^2/4\to gm^2$.  

\begin{widetext}
We may check the general  BCFT results 
on this simple example. Consider the finite size spectrum with 
the same D b.c. on both ends of the strip of length $l$ (with the same value of $\varphi_0$). 
The D b.c. essentially sets $\hat P =0$ and $a_n^R=-a_n^L$.  
The mode expansion of Eq. (\ref{mode}) becomes:
\be \varphi (t,x) \to \varphi_0+{2\pi \over l}\hat P'x+{1\over \sqrt{2g}}
\sum_{n=1}^\infty {1\over \sqrt{n}}\;2\sin \left({n\pi x\over l}\right)
\left[a_n^L\exp \left(-{in\pi t\over l}\right)+h.c.\right].\ee
$\hat P'$ must have integer eigenvalues due to the D b.c. and the periodic 
nature of $\varphi$. 
Thus the finite size spectrum with 
D b.c.'s is:
\be H_l^{DD} = {\pi \over l}\left[ g {\hat{P'}}^2+ \sum_{n=1}^\infty na_n^{L \dagger}a_n^L
\right] .\label{fssD}
\ee
We see that the general relation, Eq. (\ref{fss}) between the 
dimensions of boundary operators and finite size spectrum on 
a strip with the corresponding boundary conditions is obeyed. 

We may also find the corresponding boundary state. This must obey 
the operator equation:
\be \varphi (0,x)|D(\varphi_0)\rangle = \varphi_0
|D(\varphi_0)\rangle ,\ \  (\hbox{independent of $x$}).
\label{Dbc1BS}\ee
Note that this implies:
\be {\partial \varphi (0,x)\over \partial x}|D(\varphi_0)\rangle =0.\label{DbcBS}\ee
Note that, compared to the operator boundary condition of Eq. (\ref{Dbc}), 
$t$ and $x$ have been interchanged. This is a consequence of the 
interchange of space and time involved in going between the two 
interpretations of $Z_{AB}$. In the boundary state representation, 
the boundary corresponds to the circle $\tau =0$ and 
Eq. (\ref{DbcBS}) is the condition that $\varphi$ (when acting 
on the boundary state) be constant along 
the boundary. The D boundary state is:
\be |D(\varphi_0)\rangle =(2g)^{-1/4}\exp \left[ -\sum_{n=1}^\infty 
a_n^{L\dagger}a_n^{R\dagger} \right] \sum_{P=-\infty}^\infty  \exp [-iP\varphi_0]
|(0,P)\rangle.\label{DBS}\ee
Here $|(0,P)\rangle$ is the eigenstate of $\hat P'$ with eigenvalue $0$,
 the eigenstate of $\hat P$ with (integer) eigenvalue $P$, and 
the groundstate of all the harmonic oscillators. The condition 
Eq. (\ref{Dbc1BS}) follows using the explicit form of 
the wave-function $|(0,P)\rangle$: 
\be \langle  \varphi_0|(0,P)\rangle \propto \exp [iP\varphi_0].\label{wf}\ee
[We apologize for the confusing, but 
unfortunately standard, notation here. In Eq. (\ref{wf}), $\varphi_0$ 
is a {\it co-ordinate} whereas in Eq. (\ref{DBS}), $\varphi_0$ 
is a fixed number, corresponding to an eigenvalue of the co-ordinate.]
It is straightforward to calculate $Z_{DD}$ in 
the boundary state representation:
\be \langle D(\varphi_0)|\exp [-lH^P_\beta ]|D(\varphi_0)\rangle 
=(2g)^{-1/2}{1\over \eta (\tilde q)}\sum_P \exp \left[-l {2\pi \over \beta}
{P^2\over 2g}\right].
\label{ZDD}\ee
\end{widetext}
Here we have introduced the Dedekind $\eta$-function:
\be \eta (\tilde q)\equiv \tilde q^{1/24}\prod_{n=1}^\infty (1-\tilde q^n),\ee
and the convenient notation:
\be \tilde q \equiv e^{-{4\pi l \over \beta}}.\ee
The factor of $1/\eta (\tilde q)$ in Eq. (\ref{ZDD}) comes from 
the oscillator mode factor in the boundary state of Eq. (\ref{DBS}) 
and we have included the universal groundstate energy for free bosons
 (from the zero point motion/ Casimir effect):
\be E_0= -{\pi \over 6\beta}.\ee
The sum in Eq. (\ref{ZDD}) can also be written in terms of $\tilde q$:
\be Z_{DD}=(2g)^{-1/2}{1\over \eta (\tilde q)}\sum_P \tilde q^{\;P^2/4g}.
\label{ZDD2}\ee
In order to check that $Z_{DD}$ indeed gives the correct finite size 
spectrum with D b.c.'s we need to re-express it in terms of:
\be q\equiv e^{-{\pi\beta \over l}},\label{q}\ee
i.e. perform a modular transformation.  The modular transformation of 
the Dedekind $\eta$-function is:
\be \eta (\tilde q) = \sqrt{\beta \over 2l}\;\eta (q).\label{etamod}\ee
The modular transformation of the sum in Eq. (\ref{ZDD}) can be computed 
using the Poisson summation formula, i.e. the Fourier transform of 
the periodic $\delta$-function, $\delta_P(x)$:
%
%

\bea
\sum_{P \in \mathbf{Z}} \tilde{q}^{-P^2/(4g)}
&=&
\sum_{P \in {\mathbf{Z}}} \exp{\left[-{\pi lP^2\over g\beta}
\right]}
\nonumber \\
&=&\sqrt{g\beta \over l}\sum_{P' \in {\mathbf{Z}}}
\exp{\left[-{\pi {P'}^2g\beta \over l}\right]}
\nonumber \\
&=& \sqrt{g\beta \over l}\sum_{P' \in {\mathbf{Z}}} q^{\;g{P'}^2}.
\label{summod}
\eea
inserting Eqs. (\ref{etamod}) and (\ref{summod}) into Eq. (\ref{ZDD})
gives:
\be 
Z_{DD} = {1\over \eta (q)}\sum_{P'}q^{g{P'}^2}.\ee
Using the definition of $q$ in Eq. (\ref{q}) and the representation 
Eq. (\ref{ZABq}) for $Z_{AB}$, we extract the finite size energies with 
D b.c.'s at both ends of the strip:
\be E^{DD} = {\pi \over l}[g{P'}^2 + \hbox{integers}],\ee
in agreement with Eq. (\ref{fssD}).

\begin{widetext}
The Neumann (N) b.c. is:
\be {\partial \varphi \over \partial x}(t,0)=0.\ee
This is equivalent to $\partial \theta /\partial t = 0$ or equivalently:
\be \theta (t,0)=\theta_0.\ee
This implies:
\be \varphi_R(t,x)=-\theta_0/g+\varphi_L(t,-x),\ \  (x>0).\ee
Now the non-trivial primary boundary operators are:
\be \exp [in\varphi (t,0)]\propto \exp [2in\varphi_L(t,0)-in\theta_0/g],\ee
of dimension:
\be \Delta_N=n^2/g,\label{Delta_N}\ee
again twice the dimension of the corresponding bulk operator. 
The mode expansion is:
\be \varphi (t,x) \to \varphi_0+{2\pi \over lg}\hat Pt+{1\over \sqrt{2g}}
\sum_{n=1}^\infty {1\over \sqrt{n}}\;2\cos \left({n\pi x\over l}\right)
\left[a_n^L\exp \left(-{in\pi t\over l}\right)+h.c.\right],
\label{modeN1}\ee
with $\hat P$ having integer eigenvalues.  The corresponding spectrum 
can be read from after rewriting the Hamiltonian as:
\be H_l^{NN}={\pi \over l}\left[ {{\hat P}^2\over g}+
\sum_{n=1}^\infty na_n^{L\dagger}a_n^L\right].\label{EN1}\ee
The corresponding boundary state, obeying:
\be \theta (0,x)|N(\theta_0)\rangle =\theta_0|N(\theta_0)\rangle ,\ee
is:
\be
|N(\theta_0)\rangle =
\left( {g\over 2}\right)^{1/4}
\exp{\left[ \sum_{n=1}^\infty a_n^{L\dagger}a_n^{R\dagger} \right]}
\sum_{P'=-\infty}^\infty  \exp{[-iP'\theta_0]}
|(P',0) \rangle.
\label{DBS2}
\ee
\end{widetext}
\subsection{Multi-component free boson}

In some applications, including the subject of the present paper,
we need to consider a multi-component free boson field theory.
While Ref.~\cite{QBM} also contains a similar review,
here we summarize the basics of the multi-component free boson
field theory conforming to the normalizations of the present paper.
Let
\be
\vec{\varphi} = ( \varphi_1, \varphi_2, \ldots, \varphi_c)
\ee
be a $c$-component free boson field, with the Lagrangian density
\be
{\cal L}= {g\over 4\pi} (\partial_\mu \vec{\varphi} )^2.
\ee
At this point, the theory is just a collection of $c$ free bosons
which are independent of each other.
However, we will be interested in various possible boundary
interactions which couples different components.
In fact, we often introduce a multi-dimensional generalization
of the periodicity~(\ref{eq:phiperiod}),
which is often called  compactification.
Sometimes the compactification ties different components together,
so that they cannot be regarded as completely independent,
even before we introduce an interaction at the boundary.

The general form of the compactification would be given as
the identification
\be
\vec{\varphi} \leftrightarrow \vec{\varphi} + 2\pi \vec{R},
\label{vecphicom}\ee
where $\vec{R} \in \Lambda$ for a Bravais lattice $\Lambda$.
The different components would be completely independent
only if $\Lambda$ is rectangular. 
For a given Bravais lattice $\Lambda$, we can define
a reciprocal lattice $\Lambda^*$ so that 
\be
  \vec{K} \cdot \vec{R} \in \mathbf{Z}
\ee
for any vectors $\vec{K} \in \Lambda^*$ and $\vec{R} \in \Lambda$.
[Note that the lattices, $\sum_{i=1}^3n_i\vec R_i$ and 
$\sum_{i=1}^3m_i\vec K_i$ for arbitrary integers $n_i$ and $m_i$,
where the $\vec K_i$ are defined in Eq. (\ref{Kdef}) and the $\vec R_i$ in
Eq. (\ref{eq:Rdef}), are reciprocal.]
\begin{widetext}
Imposing  periodic boundary conditions in the space direction,
we can generalize the mode expansion~(\ref{mode}) to the present case as
\be
\vec{\varphi} (t,x)=
\hat{\vec{\varphi}}_0+
{2\pi \over \beta}\left[\vec{R}x+{1\over g}\vec{K} t\right]+
{1\over \sqrt{2g}}\sum_{n=1}^\infty {1\over \sqrt{n}}\left\{ 
\vec{a}_n^L\exp{\left[-inx_+\frac{2\pi}{\beta}\right]}+
\vec{a}_n^R\exp{\left[-inx_-\frac{2\pi}{\beta}\right]}+
\mbox{h.c.} \right\}.
\label{eq:modec}
\ee
Here we have replaced the zero-mode ``momentum'' operators
with their eigenvalues $\vec{R} \in \Lambda$ and
$\vec{K} \in \Lambda^*$.
The dual field $\vec{\theta}$ is defined similarly
to Eq.~(\ref{eq:thetadef}).
It has a similar mode expansion to the above,
generalizing Eq.~(\ref{eq:thetamode}).
As a result, $\vec{\theta}$ can be regarded as compactified as
\be
\vec{\theta} \leftrightarrow \vec{\theta} + 2\pi \vec{K},
\label{vecthetacom}\ee
where $\vec{K} \in \Lambda^*$.

The vacua of the oscillator modes
\be
 | (\vec{R}, \vec{K}) \rangle
\ee
are then labelled by
$\vec{K}$ and $\vec{R}$ which are the eigenvalues
of $\hat{\vec{P}}$ and $\hat{\vec{P}}'$.
\end{widetext}

Generalizing~(\ref{DBS}),
the Dirichlet boundary state corresponding to
$\vec{\varphi}=\vec{\varphi}_0$ at the boundary is given as
\be
 |D(\vec{\varphi}_0)\rangle =
(2g)^{-c/4} \left(V_0(\Lambda)\right)^{-1/2}
\sum_{\vec{K}\in \Lambda^*} \exp [-i\vec{K}\cdot \vec{\varphi}_0]
|(\vec{0},\vec{K})\rangle\rangle,
\label{eq:DBSc}
\ee
where $V_0(\Lambda)$
is the $c$-dimensional volume of the unit cell of
the Bravais lattice $\Lambda$,
and
\be
|(\vec{0},\vec{K})\rangle\rangle \equiv
\exp{\left[ -\sum_{n=1}^\infty 
\vec{a}_n^{L\dagger}\cdot \vec{a}_n^{R\dagger} \right]}
|(\vec{0},\vec{K})\rangle,
\ee
is the bosonic Ishibashi state.
The partition function for the cylinder with the same
Dirichlet boundary condition at both ends
(diagonal cylinder amplitude) is given,
generalizing~(\ref{ZDD}), as
\bea
&&Z_{DD}(\tilde{q}) =
\langle D(\vec{\varphi}_0)|\exp [-lH^P_\beta ]|D(\vec{\varphi}_0)\rangle 
\nonumber \\
&=& (2g)^{-c/2} \frac{1}{V_0(\Lambda)}
\left({1\over \eta (\tilde q)}\right)^c
\sum_{\vec{K} \in \Lambda^*}
\tilde{q}^{{\vec{K}^2/ (4g)}}
\label{ZDDc}
\eea
In order to find the spectrum of the boundary operators,
we must modular transform Eq.~(\ref{ZDDc}).
For this purpose, the multi-dimensional generalization
of Eq.~(\ref{summod}) is useful:
\begin{equation}
\left(\frac{1}{\eta(\tilde{q})}\right)^c
\sum_{\vec{K} \in \Lambda^*}
\tilde{q}^{\frac{1}{4g}\vec{K}^2}
= (2g)^{c/2}V_0(\Lambda)
\left(\frac{1}{\eta(q)}\right)^c
\sum_{\vec{R} \in \Lambda} q^{g \vec{R}^2}.
\label{eq:LatticeMod}
\end{equation}
Sometimes it is convenient to renormalize the lattices
as $\Lambda^*/\sqrt{g} \rightarrow \Lambda^*$ and 
$\sqrt{g} \Lambda \rightarrow \Lambda$,
to obtain the equivalent expression
\begin{equation}
\left(\frac{1}{\eta(\tilde{q})}\right)^c
\sum_{\vec{V} \in \Lambda^*}
\tilde{q}^{\frac{1}{4}\vec{V}^2}
= 2^{c/2} V_0(\Lambda)
\left(\frac{1}{\eta(q)}\right)^c
\sum_{\vec{W} \in \Lambda} q^{\vec{W}^2}.
\label{eq:LatticeMod-2}
\end{equation}

Using eq.~(\ref{eq:LatticeMod}), we find
\be
Z_{DD}(q) = \left({1\over \eta (q)}\right)^c
\sum_{\vec{R} \in \Lambda} q^{g \vec{R}^2}.
\label{eq:ZDDc-o}
\ee
In fact, the prefactor $(2g)^{-c/2} \frac{1}{V_0(\Lambda)}$
in the boundary state~(\ref{eq:DBSc}), which represents
the generally non-integer ``ground-state degeneracy''~\cite{gtheorem},
was chosen so that Eq.~(\ref{eq:ZDDc-o}) satisfies Cardy's
consistency condition.
Namely, the coefficient of Eq.~(\ref{eq:ZDDc-o}) must be unity
(or integer) to allow the interpretation as in Eq.~(\ref{ZABq}).
From Eq.~(\ref{eq:ZDDc-o}) we can read off the scaling
dimensions of
the boundary operators with  Dirichlet boundary
condition as
\be
\Delta_D = g \vec{R}^2 + \mbox{integer} ,
\ee
where $\vec{R} \in \Lambda$ as usual.  These are the dimensions 
of the boundary operators $\exp [i\vec \Theta \cdot \vec R]$, 
consistent with the compactification of Eq. (\ref{vecthetacom}).

Likewise, the Neumann boundary state corresponding to
$\vec{\theta}=\vec{\theta}_0$ at the boundary
is given as
\be
 |N(\vec{\theta}_0)\rangle =
\left(\frac{g}{2}\right)^{c/4} \sqrt{V_0(\Lambda)}
\sum_{\vec{R}\in \Lambda} \exp [-i\vec{R}\cdot \vec{\theta}_0]
|(\vec{R},\vec{0})\rangle\rangle,
\label{eq:NBCc}
\ee
where
\be
|(\vec{R},\vec{0})\rangle\rangle \equiv
\exp{\left[ +\sum_{n=1}^\infty 
\vec{a}_n^{L\dagger}\cdot \vec{a}_n^{R\dagger} \right]}
|(\vec{R},\vec{0})\rangle,
\ee
is the bosonic Ishibashi state.
The cylinder partition function for the Neumann boundary
condition at both ends reads
\be
Z_{NN}(\tilde{q}) =
(g/2)^{c/2} V_0(\Lambda)
\left({1\over \eta (\tilde q)}\right)^c
\sum_{\vec{R} \in \Lambda}
\tilde{q}^{{g \vec{R}^2/ 4}},
\label{eq:ZNNc}
\ee
in the closed string channel.
Modular transforming using Eq.~(\ref{eq:LatticeMod})
gives the same partition function in the open string channel
as
\be
Z_{NN}(q) = \left({1\over \eta (q)}\right)^c
\sum_{\vec{K} \in \Lambda^*} q^{\vec{K}^2/g}.
\label{eq:ZNNc-o}
\ee
The scaling dimensions of the boundary operators with 
Neumann boundary conditions are now given by
\be
  \Delta_N = \frac{\vec{K}^2}{g} + \mbox{integer},
\ee
where $\vec{K} \in \Lambda^*$.  
These are the dimensions 
of the boundary operators $\exp [i\vec \Phi \cdot \vec K]$, 
consistent with the compactification of Eq. (\ref{vecphicom}).

The discussions in this Appendix naturally
form a basis for the problem of the junction of three quantum wires,
which is mapped to the boundary problem of two-component 
free boson.
However, there is an important modification due to the
fermionic nature of the electrons.
This is discussed in Sec.~\ref{sec:twist}, before
applying the present formalism to the junction problem
in Sec.~\ref{sec:fpts}.

\section{Boundary conditions and the conductance}
\label{sec:BC_cond}
Physical properties of the multiple wire junctions can be related to the
boundary conditions imposed on the bosonic fields describing each
wire. In preparation for discussing conformally invariant boundary
conditions that correspond to RG fixed points for the 3-wire junction
problem, let us show below how the conductivity tensor
is extracted from the boundary conditions.

Let us start with a simple example first, that of a single quantum
wire, described by fields $\varphi(x,\tau)$ and $\theta(x,\tau)$ with
\begin{equation}
S=\frac{g}{4\pi}\; \int d\tau\;dx\;(\partial_\mu\varphi)^2
\end{equation}
or dual action
\begin{equation}
S=\frac{1}{4\pi g}\; \int d\tau\;dx\;(\partial_\mu\theta)^2
\;.
\end{equation}
As in Ref.~\cite{Kane}, we calculate the conductance within
the linear response theory.  Applying an AC electric field in an
interval $0<x<L$ and taking the DC limit, we obtain the Kubo formula
for the conductance of the wire as
\begin{widetext}
\begin{equation}
 G = \lim_{{\omega} \rightarrow 0_+} \frac{e^2}{h}\frac{1}{\pi {\omega} L}
	\int_{-\infty}^\infty d\tau\;e^{i {\omega}\tau} \int_0^L dx
	\langle T_\tau J(y,\tau) J(x,0)\rangle 
\;,
\end{equation}
where the current $J(x,\tau)=-i\partial_\tau\theta(x,\tau)$. The
current $J$ can be expressed in terms of chiral components as
$J=J_R-J_L$, with
$J_R=\partial\theta\equiv(\partial_x-i\partial_\tau)\theta/2$ and
$J_L=\bar\partial\theta\equiv(\partial_x+i\partial_\tau)\theta/2$. In
terms of the chiral currents, one has
\begin{eqnarray}
 G = \lim_{{\omega} \rightarrow 0_+} \frac{e^2}{h}\frac{1}{\pi {\omega} L}
	\int_{-\infty}^\infty d\tau\;e^{i {\omega}\tau} \int_0^L dx
&&\Big[
	\langle T_\tau J_R(y,\tau) J_R(x,0)\rangle 
	+\langle T_\tau J_L(y,\tau) J_L(x,0)\rangle 
\\
\;\;\;\;&&
	-\langle T_\tau J_R(y,\tau) J_L(x,0)\rangle 
	-\langle T_\tau J_L(y,\tau) J_R(x,0)\rangle 
\Big].
\end{eqnarray}
\end{widetext}

\subsection{Conductance of an infinite wire}

Let us consider first an infinite wire. In this case, the correlation
function between the different chiral components of the currents
$\langle J_R J_L \rangle$ vanishes.  Thus we need only the
``diagonal'' contributions.
\begin{equation}
\langle J_R(y,\tau) J_R(x,0) \rangle
=  \partial^2 \langle \theta(z) \theta(0) \rangle,
\end{equation}
where $z \equiv i \tau + (y-x)$.
Because 
\begin{equation}
 \langle \theta(z) \theta(0) \rangle = - \frac{g}{4\pi}\ln{|z|^2},
\end{equation}
we obtain
\begin{equation}
\langle J_R(y,\tau) J_R(x,0) \rangle
= - {g\over 4\pi} {1\over z^2} .
\end{equation} 
To calculate its contribution to the conductance, we first
perform the Fourier transformation.
A simple contour integral gives (for $\omega>0$)
\begin{equation}
\int_{-\infty}^{\infty} d \tau \;e^{i\omega\tau} {1\over (i\tau+u)^2}
= 2\pi \omega \;H(u) \;e^{-\omega u},
\end{equation}
where $u=y-x$ and $H(t)$ is the Heaviside step function. 
Combining with a similar calculation on the $\langle J_L J_L \rangle$
part, the DC conductance for the infinite wire reads
\begin{equation}
G = g\frac{e^2}{h}
\frac{1}{L}\int_0^L dx \;\big[ H(x-y) + H(y-x) \big]
= g\frac{e^2}{h} .
\label{eq:Gbulk}
\end{equation}
In this calculation, the length $L$ of the section on which the
voltage is applied does not affect the final result of the
conductance. Moreover, the location $y$ where one measures the
current does not matter, and it can even be outside of the region
where the voltage is applied. This is natural because in the DC
limit, the current would be uniform throughout the wire even if the
voltage is applied to a particular section of the wire.

However, the result~(\ref{eq:Gbulk}) should be treated with
caution.
As it has been discussed by several
authors~\cite{Tarucha,Maslov-Stone,Safi-Schulz},
the observed DC conductance per channel in a quantum wire
is generally the free electron value $G=e^2/h$ independent of
the TL parameter $g$, rather than
the renormalized value eq.~(\ref{eq:Gbulk}).
This is related to the fact that the voltage is applied
in a physical setting corresponds to the potential drop
between the two reservoirs to which the wire is connected.
The assumption of the electric field applied on the
finite interval of the wire does not reflect this.
Nevertheless, the calculation leading to eq.~(\ref{eq:Gbulk})
is useful as a starting point for various applications.
In fact, the physical result $G=e^2/h$ can be recovered within
the present approach, by modeling the reservoirs by Fermi liquid 
leads (TL liquid with $g=1$).

In this paper, we first ignore the issue of the reservoirs
and calculate the conductance simply using the present approach.
Later, in Sec.~\ref{sec:FLL} we will discuss the
physical DC conductance by including the Fermi liquid leads.

\subsection{Conductance of a semi infinite wire}
\label{sec:openwire}

As the simplest example involving a boundary, let us consider a
half-infinite wire with an ``open'' boundary at $x=0$ (through which
no current can flow.)  The wire is assumed to extend for $x>0$.  It
corresponds to the Dirichlet boundary condition on the field $\theta$
or equivalently the Neumann boundary condition on the field
$\varphi$. The boundary condition $\theta(0,\tau)=0$ is translated to
\begin{equation} 
	J_R(0) - J_L(0) = 0 .
\end{equation}
A convenient trick to respect this condition is to analytically
continue the right-mover current to $x<0$ and identify 
\begin{equation}
	J_L(x,\tau) \equiv J_R(-x,\tau)
\label{eq:NbcJ}
\end{equation}
for $x>0$.

As a general rule, correlation functions of the chiral operators are
not affected by the boundary condition.  Therefore the contributions
from $\langle J_R J_R\rangle$ and $\langle J_L J_L \rangle$ remains
the same as in eq.~(\ref{eq:Gbulk}).

Cross correlations between the different chiral components are
generally non-vanishing in the presence of the boundary, and
are determined by the boundary condition.
In the present case, all the left-mover current may be replaced by
the right-mover current (in fictitious, negative $x$ region) according
to eq.~(\ref{eq:NbcJ}).
As a result, 
\begin{eqnarray}
\langle J_R(y,\tau)J_L(x,0)\rangle 
&=& \langle J_R(y,\tau) J_R(-x,0) \rangle
\nonumber \\
&=& - \frac{g}{4\pi}\frac{1}{\big( i\tau +  (x+y)\big)^2}
\end{eqnarray}
Upon Fourier transform, this term contributes to the conductance a
term
\begin{equation}
G_{RL} = - g \frac{e^2}{h} \frac{1}{L}\int_0^L dx \;H(x+y)
	= - g\frac{e^2}{h}
\end{equation}
because $x,y>0$.
On the other hand, the corresponding one from $\langle J_L J_R\rangle $
vanishes because it involves the step function $H(-x-y)$ for
$x,y>0$.

Summing up all the contributions, we obtain the DC conductance
\begin{equation}
	G=0
\end{equation}
for the open wire. This is an intuitively obvious result.

\subsection{Conductance of a half-infinite wire attached to a superconductor}
\label{sec:wiresuper}

As another simple example, let us consider the half-infinite wire attached
to a superconductor at $x=0$. Again the wire is assumed to extend
to $x>0$.
Now the appropriate boundary condition on the boson field is given
by the Dirichlet boundary condition on $\phi$,
since it fixes the phase of a Cooper pair.
In terms of current, it is now given as
$J_R(0) + J_L(0) = 0$ and it is solved by identifying
\begin{equation}
	J_L(x,\tau) \equiv - J_R(-x,\tau)
\label{eq:DbcJ}
\end{equation}
for $x>0$.
The calculation exactly follows the previous case in Sec.~\ref{sec:openwire},
except that the contribution from the cross term $\langle J_R J_L\rangle$
flips the sign.
As a result, we find the DC conductance
\begin{equation}
	G = 2g \frac{e^2}{h},
\end{equation}
which is doubled from the bulk one.
This may be interpreted as a consequence of the (perfect) Andreev
reflection.

\subsection{Conductance of the Y-junction}

Now let us turn to our problem of the Y-junction.  We define three
half-infinite wires $1,2$ and $3$ which extends for $x>0$, and $x=0$
is represents the junction.  Each wire is bosonized separately and
described by the boson fields $\varphi_j$ and $\theta_j$. Currents
are also defined respectively.

We discuss the conductance matrix introduced in
Eq.~(\ref{eq:def-of-conduc-tensor}).
In the present approach, the conductance matrix is given as
\begin{equation}
G_{jk} = \lim_{\omega \rightarrow 0_+}
\frac{e^2}{h}\frac{1}{\pi  \omega L} \int_0^L dx
\int_{-\infty}^{\infty} \!\!\!d\tau\;e^{i\omega\tau}
	\langle J_j(y,\tau) J_k(x,0) \rangle,
\label{eq:YG1}
\end{equation}
where $J_j = -i \partial_\tau\theta_j$ is the current operator for wire $j$.

In analyzing the Y-junction, it is convenient to work on the rotated
basis defined in
Eqs.~(\ref{eq:newbasisphi},\ref{eq:newbasistheta}). We can express the
fields $\varphi_j$ and $\theta_j$ in this basis as
\begin{eqnarray}
\varphi_j&=&\frac{1}{\sqrt{3}}\Phi_0 
-\sqrt{\frac{2}{3}} (\hat z\times {\vec K}_j)\cdot \vec\Phi
\label{eq:vtr}
\\
\theta_j&=&\frac{1}{\sqrt{3}}\Theta_0 
-\sqrt{\frac{2}{3}} (\hat z\times {\vec K}_j)\cdot \vec\Theta
\;,
\label{eq:vtr2}
\end{eqnarray}
where the vectors ${\vec K}_j$ are defined in Eq.~(\ref{eq:Kvec}), and
recall that $\vec\Phi=(\Phi_1,\Phi_2)$ and $\vec\Theta=(\Theta_1,\Theta_2)$.

It also defines the transformation of the currents as
\begin{equation}
J_j = v_{j\mu} J_{\mu},
\end{equation}
where summation over $\mu$ is implicitly assumed,
\begin{equation}
	J^{\mu} \equiv - i \partial_\tau \Theta_{\mu}
\end{equation}
and the coefficients $v_{j\mu}$ can be read off from
Eqs.~(\ref{eq:vtr},\ref{eq:vtr2}):
\begin{equation}
v_{j\mu}=
\left\{
\begin{array}{ll}
1/\sqrt{3} & ,\; \mu=0 
\\
\sqrt{2/3} \;\epsilon_{\mu\nu} \;K^\nu_j & ,\;\mu=1,2   
\end{array}\right.\nonumber
\;.
\end{equation}

\begin{widetext}
{}From Eq.~(\ref{eq:YG1}), the conductance in this basis is given by
the new basis,
\begin{equation}
\label{eq:Y-cond}
G_{jk} = \lim_{\omega \rightarrow 0_+}
\frac{e^2}{h}\frac{1}{\pi  \omega L} \int_0^L dx
\int_{-\infty}^{\infty} \!\!\!d\tau\;e^{i\omega\tau}
\;
v_{j\mu} v_{k\nu}\; \langle J^{\mu}(y,\tau) J^{\nu}(x,0) \rangle,
\end{equation}
where the summation over $\mu,\nu$ is implicitly assumed.
\end{widetext}

Conservation of total charge implies that the ``center of mass'' boson
$\Phi^0$ is always subject to the Neumann boundary condition.  As a
consequence, the cross-terms $\langle J^0 J^{\mu}\rangle$ vanish for
$\mu\neq 0$.  Moreover, the contribution from $\langle J^0 J^0\rangle$
to the conductance also vanishes, because the calculation is identical
to that in the open wire. Therefore, the conductance of the Y-junction
is given by Eq.~(\ref{eq:Y-cond}) with the $\mu,\nu$ sum restricted to
over just $1,2$.

The boundary conditions will determine the
$\langle J^\mu J^{\nu}\rangle$ correlations. We will use the expression
Eq.~(\ref{eq:Y-cond}) explicitly in the calculations carried out for
the different boundary conditions we analyze in  Sec.~\ref{sec:fpts}.

\section{Junctions of three fractional quantum Hall edges}
\label{sec:fqh}
For the case $g=1$ and for flux $\phi=\pm \pi/2$, we showed that we could
understand the scattering in terms of an $S$-matrix that took
incoming/outgoing states cyclically from lead $1\to 2$, $2\to 3$, and $3\to
1$. We can gain a great deal of intuition beyond the $g=1$ case by studying
the case of tunneling in junctions of three edge states of fractional quantum
Hall (FQH) liquids, in particular those belonging to the Laughlin sequence
(filling fraction $\nu=1/(2m+1)$, $m$ odd). In this case, the dynamics of the
edge modes is described by chiral bosons, and the fractionally charged edge
quasiparticles are constructed from these chiral bosons~\cite{Wen91}.  We will
show here that one can still understand the scattering in junctions of three
FQH edges as cycling between the three leads.

One of the results below is that we can construct the Klein factors for the
multi-lead problem using the zero-modes of the chiral bosons, instead of the
Pauli matrices used by Nayak {\it et al.}~\cite{Nayak}.

The geometry we consider is shown in Fig.~\ref{fig:3edges}. The shaded areas
correspond to Hall liquids, and the white regions to vacuum.  Electrons
tunnel through the vacuum or the FQH liquid, while quasiparticles tunnel only
through the FQH liquid.

In order to understand the quasiparticle tunneling picture for the three
leads, we have to ensure that the quasiparticle operators for the three leads
respect the appropriate fractional statistics relations: this is the
generalization of the Klein factors needed in the fermionic problem of Nayak
{\it et al.}.

\vskip 1 cm

\begin{figure}
\includegraphics[width=7cm]{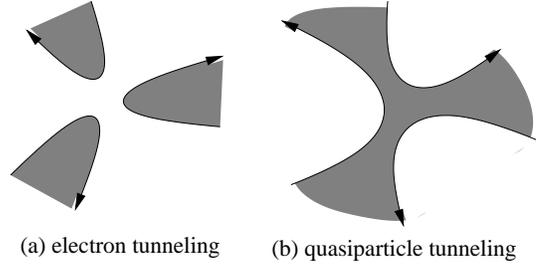}
\caption{Two tunneling geometries in a three lead FQH junction. (a) Electron
  tunneling geometry, where the Hall liquid is separated in three disjoint
  pieces (dark regions); tunneling occurs through the connected vacuum (white
  region), so only electrons can tunnel between the tips of the three FQH
  regions. (b) Quasiparticle tunneling geometry, where there is a single
  connected FQH liquid (dark region) and so fractionally charged
  quasiparticles can tunnel between the three closely spaced tips of the
  liquid.}
\label{fig:3edges}
\end{figure}

\subsection{Klein factors from the boson zero-modes}

Consider three right moving boson fields
\begin{equation}
\phi_i(t-x)=X_i+\frac{2\pi}{l}\;P_i\;(t-x)+\sum{\rm Oscillator\; modes}
\end{equation}
for $i=1,2,3$ labeling the three leads. The zero-modes satisfy
$[X_i,P_j]=i\;\delta_{ij}$. The Lagrangian is given by
\begin{equation}
{\cal L}=-\sum_{i=1}^3
\frac{1}{4\pi}\;\partial_x\phi_i
\left(\partial_t+\partial_x\right)\phi_i
\; .
\label{eq:fqh-lagrangian-deff}
\end{equation}
(In this appendix we adopt the standard normalization for edge modes,
see Ref.~\cite{Wen91}. This normalization is related to our standard
normalization throughout the rest of the paper by taking $4\pi\to2\pi$ in the Lagrangian Eq.~\ref{eq:fqh-lagrangian-deff}.)

The quasiparticle/electron operators are written as
\begin{equation}
\psi_i=e^{i\lambda X_i}\;e^{i\lambda \;\frac{2\pi}{l}P_i(t-x)}\;\tilde\psi_i
\end{equation}
where $\tilde\psi_i$ contains only oscillator
modes. ($\lambda=\sqrt{\nu^{-1}}$ for electrons and
$\lambda=\sqrt{\nu}$ for quasiparticles within the Laughlin sequence
$\nu=\frac{1}{2m+1}$.)  These $\psi_i$ have the correct statistics with
respect to themselves, but not among each other (they simply
commute). To fix this, define the Klein factors
\begin{equation}
\eta_i=e^{i\frac{\theta}{2}\sum_{j}\alpha_{ij}\,P_j}\;,
\end{equation}
which satisfy $[\eta_i,\eta_j]=0$, and
\begin{eqnarray}
\eta_i\;\psi_{j}&&=
\psi_{j}\;\eta_i\;\times
e^{-[\frac{\theta}{2}\sum_{k}\alpha_{ik}\,P_k,
\lambda\;X_j]}\nonumber\\
&&=\psi_{j}\;\eta_i\;\times
e^{\frac{i}{2}\theta\lambda\,\alpha_{ij}}\;.
\end{eqnarray}

If we define new edge operators 
\begin{equation}
\Psi_i=\eta_i\psi_i
\; ,
\end{equation}
we do not spoil the statistics between same edge operators as long as we
choose $\alpha_{ii}=0$ for $i=1,2,3$, so that $[\eta_i,\psi_i]=0$. Now
consider the statistical phase when two different edge vertex operators
$\Psi_i$ and $\Psi_{j}$ for $i\ne j$ (so that $[\psi_i,\psi_j]=0$) are
interchanged:
\begin{eqnarray}
\Psi_i\Psi_j
&&=\eta_i\;\psi_{i}\;\;\eta_j\;\psi_{j}\nonumber\\
&&=\Psi_j\Psi_i\;\times
e^{\frac{i}{2}\theta\lambda(\alpha_{ij}-\alpha_{ji})}\;.
\end{eqnarray}

It suffices to consider only anti-symmetric matrices so that
$\alpha_{ij}=-\alpha_{ji}$, with unit entries $\alpha_{ij}=\pm 1$. The
right statistical angle is picked up if we choose
$\theta\lambda=\pi\lambda^2$ or $ \theta=\pi\lambda$: for electrons
$\Psi^{el}_i\Psi^{el}_j=\Psi^{el}_j\Psi^{el}_i\times e^{\pm i\pi\nu^{-1}}$ and
for quasiparticles $\Psi^{qp}_i\Psi^{qp}_j=\Psi^{qp}_j\Psi^{qp}_i\times
e^{\pm i\pi\nu}$ when $i\ne j$.

Notice that there is still a sign ambiguity in choosing the three
$\alpha_{ij}=-\alpha_{ji}=\pm 1$ for $i\ne j$. To consistently fix the
correct signs, one must look, for example, at the relative phases as compared
to those obtained by closing the system into a single edge, as studied in
Ref.~\cite{Klein-wrap}. Doing so, we fix $\alpha_{12}=\alpha_{31}=1$ and
$\alpha_{23}=-1$.

\subsection{Tunneling terms and the connection with the Callan-Freed model}

Now we turn to the tunneling between the three edges:
$T_{ij}=\Psi^\dagger_i\Psi_j$. It is useful to also define
\begin{equation}
  \label{eq:2}
t_{ij}=\psi^\dagger_i\psi_j = e^{-i\lambda (\phi_i-\phi_j)}
\; ,
\end{equation}
which does not contain the Klein factors. In terms of the
 $\vec \Phi$ basis, defined in Eqs. (\ref{Phidef}) and (\ref{vecPhidef}), 
we can write six possible operators:
\begin{eqnarray}
&&t_{12}=e^{-i\lambda \vec K_3\cdot \vec\Phi},
\quad
t_{23}=e^{-i\lambda \vec K_1\cdot \vec\Phi},
\quad
t_{31}=e^{-i\lambda \vec K_2\cdot \vec\Phi},
\nonumber
\end{eqnarray}
with $t_{ji}=t_{ij}^*$ where the $\vec K_i$  
are defined 
in Eqs. (\ref{Kdef}). 
 The six operators
correspond to the six choices of $\pm \vec K_a$, $a=1,2,3$.

Now, the Klein factors are responsible for extra phases coming when
considering time-ordered products of the $T_{ij}$ operators instead of the
$t_{ij}$'s.

To illustrate this point, let us now calculate the phases obtained when
taking products of $T_{ij}$ operators along the elementary up and down
triangles in the lattice spanned by the vectors $\vec K_1$ and $\vec K_2$.

\begin{figure}
\vspace{.4cm}
\includegraphics[width=7cm]{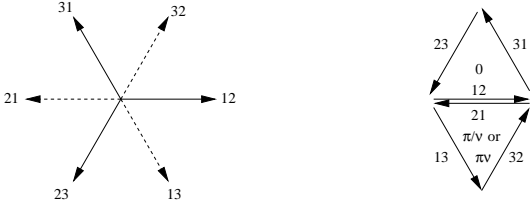}
\caption{The six possible tunneling 
between two edges are depicted on the left, in terms of vectors that span the
triangular lattice. On the right, the phases accumulated due to the Klein
factors are shown for elementary up and down triangles, according to
Eqs.~(\ref{eq:up-fqh},\ref{eq:down-fqh}).}
\label{fig:3tun}
\end{figure}

Using the commutation relations of the $\eta$ and $\psi$, one can show that
\begin{equation}
T_{12}T_{23}T_{31}=t_{12}t_{23}t_{31} \;,
\label{eq:up-fqh}
\end{equation}
so there is no extra phase due to the Klein factors when traversing the
elementary up triangles (see Fig.~\ref{fig:3tun}). On the other hand,
\begin{eqnarray}
T_{13}T_{21}T_{32}=t_{13}t_{21}t_{32}&\times& 
e^{i\theta\lambda [\alpha_{12}+\alpha_{23}+\alpha_{31}]}
\nonumber\\
=t_{13}t_{21}t_{32}&\times& 
e^{i\pi \lambda^2}
\; 
\label{eq:down-fqh}
\end{eqnarray}
when traversing down triangles.

This is equivalent to a flux flux $\pi\nu^{-1}$ in the electron tunneling
problem ($\pi\nu$ in the quasiparticle tunneling problem) for down triangles,
and a flux 0 for up triangles (See fig.~\ref{fig:3tun}).

Once we have identified the statistical phases for the two elementary
triangles in the lattice spanned by the vectors $\vec K_1$ and $\vec K_2$, we
can match order by order a Coulomb gas expansion of electron/quasiparticle
tunneling operators with a Coulomb gas expansion in the Callan-Freed model.
This procedure is explicitly carried out in Ref.~\cite{short-paper1} and in
full detail in section~\ref{sec:DHM}.

\subsection{Callan-Freed applied to the FQH junction}

In the case of Fig.~\ref{fig:3edges}(a), electron tunneling is irrelevant at
low energies, so the configuration is stable. Now, for
Fig.~\ref{fig:3edges}(b), quasiparticle tunneling is a relevant perturbation,
so the question is what is the strong coupling fixed point that is reached.
Here we focus on the ${\bf Z}_3$ symmetric case. Since we will concentrate on
the quasiparticle tunneling problem, we choose $\lambda=\sqrt{\nu}$.

Following Ref.~\cite{short-paper1} and section~\ref{sec:DHM}, we can match
the Coulomb gas expansion of the FQH tunneling problem to that of the
Callan-Freed problem model by choosing $\alpha,\beta$ in the dissipative
Hofstadter as follows:
\begin{equation}
\label{eq:alpha-fqh}
\frac{\alpha}{\alpha^2+\beta^2}=\nu
\end{equation}
and
\begin{equation}
\label{eq:beta-fqh}
4\pi\;{\cal A}
\;\frac{\beta}{\alpha^2+\beta^2}=2\pi n+\pi\nu
\; ,
\end{equation}
where ${\cal A} =\sqrt{3}/4$ is the area of the elementary
triangle of unit length, and $n$ is an integer. The first condition
Eq.~(\ref{eq:alpha-fqh}) is fixed by the scaling dimension $\nu$ of the
quasiparticle tunneling operator. The second condition
Eq.~(\ref{eq:beta-fqh}) is fixed by the statistics originating from the Klein
factors explained above.

We can then solve for $\alpha,\beta$, and using that the lattice dual to the
triangular lattice of unit length has lattice constant $2/\sqrt{3}$ (this
dual lattice contains the minima of the periodic potential in the problem,
and the distance between the nearby minima are the ``instantons''), we obtain
the scaling dimension $(2/\sqrt{3})^2\alpha$ (of the strong coupling
tunneling operators:
\begin{equation}
\label{eq:dim-fqh}
\Delta_n=\frac{4\nu}{3\nu^2+(2n-\nu)^2}
\; .
\end{equation}

As in Ref.~\cite{short-paper1}, there appears to be a whole family of minima
and their associated instanton operators labelled by different integer $n$.
Notice that the larger $|2n-\nu|$, the more relevant is the operator.
  
In what follows, let us consider the cases $n=\pm 1,0$. One can easily show
that, for $\nu< 1$, the choices $\pm 1$ lead to relevant operators
$\Delta_n<1$, so the associated strong tunneling fixed points are unstable.
Clearly, this will also be the case if $|n|>1$, since $\Delta_n$ decreases as
$|n|$ increases. So the only choice for a stable fixed point for $\nu< 1$ is
$n=0$. Notice that when $\nu=1$, there are two possible choices $n=0,1$ (see
\cite{short-paper1} and section~\ref{sec:DHM} for a discussion). 
In this case, we obtain a simple result,
\begin{equation}
\label{eq:1Delta-n0}
\Delta_0=\frac{1}{\nu}
\; .
\end{equation}
This result has a very simple physical interpretation: that the strong
quasiparticle tunneling regime of Fig.~\ref{fig:3edges}(b) leads to a dual
description similar to that in Fig.~\ref{fig:3edges}(a) where the interior of
the original single FQH region is pinched, leading to three disconnected FQH
leads, between which electrons can tunnel. Indeed, $\Delta_0=\frac{1}{\nu}$
is the scaling dimension of electron tunneling operators.


\end{document}